\documentclass[11pt]{article}

\usepackage{amsmath} 
\usepackage{graphicx} 
\usepackage{hyperref} 
\usepackage{natbib}
\usepackage[dvipsnames]{xcolor}
\usepackage{amsfonts}
\usepackage{mdframed}
\usepackage{multirow}
\usepackage{enumitem}
\usepackage{bbm}
\usepackage{subcaption}
\usepackage{overpic} 
\usepackage{nth}
\usepackage{bm}
\usepackage{setspace} 
\usepackage{booktabs}
\usepackage{array}
\usepackage{colortbl}
\usepackage{multirow}
\usepackage{makecell}
\usepackage{amsthm}

\usepackage[labelfont=bf]{caption}
\usepackage[margin=1in]{geometry}
\hypersetup{
  colorlinks = TRUE,
  citecolor=violet,
  linkcolor=violet,
  urlcolor=violet
} 

\theoremstyle{definition}
\newtheorem{proposition}{Proposition}

\newtheorem{assumption}{Assumption}

\newcommand{\starlanguage}{Significance Indicator: $^{***}$p$<$0.001, $^{**}$p$<$0.01, $^{*}$p$<$0.05.}
\newcommand{\aisub}{AI agent}
\newcommand{\persona}{prompt}
\newcommand{\Persona}{Prompt}
\newcommand{\NGamesS}{1,500}
\newcommand{\NParticipantS}{4,500}
\newcommand{\TotalGames}{883,320}

\usepackage{tikz}
\usetikzlibrary{positioning, shapes.geometric, arrows.meta}

\newcommand{\WorkingTitle}{General Social Agents}
\newcommand{\CROptTrainMAE}{0.2}
\newcommand{\CRBaselineTrainMAE}{0.42}

\newcommand{\CRTestAtheoryTotalProp}{32.5}

\newcommand{\CRTestBaselineGameAMAE}{0.52}
\newcommand{\CRTestBaselineGameBMAE}{0.29}

\newcommand{\CRTestBaselineTotalPropAtheo}{22.5}

\newcommand{\CRTestNOptBetterThanBaseline}{31}
\newcommand{\CRTestOptGameAMAE}{0.17}
\newcommand{\CRTestOptGameBMAE}{0.15}

\newcommand{\NNCRovel}{494}
\newcommand{\CRBaselineNovelMAE}{0.259}
\newcommand{\CRAtheoryNovelMAE}{0.264}
\newcommand{\CRTheoryNovelMAE}{0.206}
\newcommand{\CRPercImprove}{21}

\newcommand{\BaselineOneNineMass}{87}

\newcommand{\KlBasicCycleHuman}{1}
\newcommand{\KlBasicCostlessHuman}{1.26}

\newcommand{\KlHumanRawDist}{2.7}
\newcommand{\KlHumanOptDist}{0.3}
\newcommand{\KlPercDiffBasic}{89}

\newcommand{\KlHumanHistDist}{2.16}

\newcommand{\KlHumanMBDist}{2.36}

\newcommand{\KlHumanSameNumDist}{0}

\newcommand{\KlHumanRawCycleDist}{0.95}
\newcommand{\KlHumanOptCycleDist}{0.28}
\newcommand{\KlPercDiffCycle}{71}

\newcommand{\KlHumanRawCostlessDist}{0.93}
\newcommand{\KlHumanOptCostlessDist}{0.15}
\newcommand{\KlPercDiffCostless}{84}

\newcommand{\NTotal}{955}
\newcommand{\MaxPercImproveArad}{73}
\newcommand{\MinPercImproveArad}{53}

\newcommand{\KlHumanOptDistSeven}{0.16}

\newcommand{\BaselineBarLambda}{0.43}
\newcommand{\ExpBaselineBarLambda}{1.54}

\newcommand{\BaselineHatP}{0.715}

\newcommand{\NashBarLambda}{0.32}
\newcommand{\ExpNashBarLambda}{1.38}

\newcommand{\NashHatP}{0.622}

\newcommand{\CHBarLambda}{0.39}
\newcommand{\ExpCHBarLambda}{1.48}

\newcommand{\CHHatP}{0.64}

\newcommand{\UniformBarLambda}{0.2}
\newcommand{\ExpUniformBarLambda}{1.22}

\newcommand{\PureRandBarLambda}{1.44}
\newcommand{\ExpPureRandBarLambda}{4.21}

\newcommand{\MaxSymmNash}{10,051}
\newcommand{\PropSelectedPure}{59}

\newcommand{\CountUniqueEq}{467}
\newcommand{\CountPayoffDomEq}{328}
\newcommand{\CountRiskDomEq}{1,026}
\newcommand{\GamesConverged}{1,487}

\newcommand{\SampleSizeHuman}{4,249}
\newcommand{\SampleSizeGame}{1,490}

\newcommand{\OptHighestMatchRate}{24}
\newcommand{\OptTopThreeMatchRate}{53}

\newcommand{\OptGameAllMatchRate}{86}

\begin{document}  
\date{\today}
\title{\WorkingTitle{}\thanks{
Thanks to Tyler Cowen and the Mercatus Center for generous funding and intellectual support.
Thanks to 
Alex Moehring,
Daniel Rock,
David Holtz,
Drew Fudenberg,
Jenny Allen,
Jessica Hullman,  
John List,
Kehang Zhu,
Leland Bybee,
Michael Zhao,
Seth Benzell,
Sophia Kazinnik,
Soumitra Shukla, and
Steve Tadelis
for their time and helpful comments.
We are deeply grateful to Reanna Ishmael for software development support.
The experiments in this paper were preregistered on \url{https://aspredicted.org/} numbers $222695$, $231091$, and $241394$.
Author contact information, code, and data are currently or will be available at \url{http://www.benjaminmanning.io/}. 
Both authors have a financial interest in \url{https://www.expectedparrot.com/}. Horton is an economic advisor to Anthropic.
In preparing this paper, the authors utilized generative AI models extensively as tools to assist with editing and evaluation. 
The authors retain full responsibility for all content and conclusions presented herein.
}
} 
\author{
Benjamin S. Manning \\ MIT  \and   
John J. Horton \\ MIT  \& NBER
}
\date{\today}

\pagenumbering{gobble} 
\maketitle

\vspace{-0.5cm}
\begin{abstract}

\noindent Useful social science theories predict behavior across settings. 
However, applying a theory to make predictions in new settings is challenging: rarely can it be done without ad hoc modifications to account for setting-specific factors.
We argue that AI agents put in simulations of those novel settings offer an alternative for applying theory, requiring minimal or no modifications.
We present an approach for building such ``general'' agents that use theory-grounded natural language instructions, existing empirical data, and knowledge acquired by the underlying AI during training.
To demonstrate the approach in settings where no data from that data-generating process exists---as is often the case in applied prediction problems---we design a heterogeneous population of \TotalGames{} novel games.
AI agents are constructed using human data from a small set of conceptually related but structurally distinct ``seed'' games.
In preregistered experiments, on average, agents predict initial human play in a random sample of \NGamesS{} games from the population better than (i) a cognitive hierarchy model, (ii) game-theoretic equilibria, and (iii) out-of-the-box agents.
For a small set of separate novel games, these simulations predict responses from a new sample of human subjects \emph{better} even than the most plausibly relevant published human data.
\end{abstract}

\onehalfspacing
\newpage \clearpage
\pagenumbering{arabic}

\section{Introduction} 
\label{sec:introduction}

A general, low-cost method for accurately simulating human behavior with AI agents would have wide application in the social sciences \citep{Charness2025NextGen,jackson2025ai,promising2025LLM}.
Recognizing this potential, a growing literature explores whether large language models (LLMs) can simulate human responses in various settings.\footnote{\citep{argyle2022out,aher2022using,binz2023cognitive,brand2023using,park2023generative,
jackson2024turing,1000agents2024park,chang2024networks,Manning2024Automated,
capra2024llms,hansen2024simulating,kim2024homoeconomicus,
wang2025market,Duch2025Possum,zhu2025evidence,Broska2025Mixed,
bybee2025ghost,fish2025collusion}}
Across dozens of experiments, samples of these agents respond with remarkable similarity to humans---even when simulating novel studies that did not appear in the underlying LLM's training corpus \citep{hewitt2024predictingLLM,binz2025centaur,ValidityLLM2024,Tranchero2024LLMtheorizing,suh2025finetuningscaled}.
Yet within this literature, others find settings where the very same models are poor human proxies.\footnote{\citep{opinions2023Santukar,WhichHumans:2023,Cheng2023caricature,Gui2023Causal,gao2024caution}}
This inconsistency poses a challenge for AI simulations as robust and credible predictive models---particularly in settings where no prior human data exists.
The core challenge is not simply achieving a close match between AI and human responses in one setting, but building agents that will generalize reliably.

A natural starting point is improving the instructions given to agents.
These instructions, or ``\persona{s},'' are written descriptions given to the LLM specifying who it is, what it believes, or how it should behave and reason \citep{horton2023large}.
Such second-person instructions (e.g., \emph{``You respond as type-X person''}) can profoundly affect output distributions because advanced LLMs have been explicitly fine-tuned to follow instructions \citep{bai2022helpfulharmless,ouyang2022training}.
With an appropriate prompt (which can be massive),\footnote{To date, Google's Gemini 1.5 can accommodate 10 million tokens \citep{Gemini152024}. 
This is roughly equivalent to 15{,}000 pages, or about 20 copies of the Handbook of Experimental Economics \citep{Kagel1995Handbook}.} highly capable models can perform complex reasoning and mathematical tasks at levels sometimes better even than those of highly skilled humans.

Despite the existence of these powerful, steerable models, constructing agents whose behavior is similar to that of real humans in a wide variety of settings is nontrivial---even when human data are available to guide the search.
The set of possible \persona{s} is vast, ranging from simple combinations of social or demographic traits to complex programmatic instructions related to how humans make decisions \citep{zhu2025ExplPredicting,Jackson2025Mixture}.
As in other machine learning applications, the challenge is not only to avoid underfitting but also to guard against overfitting.
By iterating through enough \persona{s}, one can almost always find some arbitrary \persona{} that shifts the LLM's responses to closely match a given human distribution.
For example, an LLM instructed \emph{``you randomly offer between \$6 and \$9''} may perfectly reproduce a distribution of human responses in a \$20 dictator game, but such a \persona{} would be nonsensical for a \$5 dictator game.
In contrast, a \persona{} grounded in the underlying behavioral drivers---e.g., \emph{``you are self-interested but fair''}---can perform well in-sample and plausibly extend to a range of allocation games.
Standard data-driven approaches, such as a train-test split within a single dataset, cannot reliably distinguish between these two cases; the latter appears better only when tested in truly new settings.
If the goal is to predict behavior in settings with no prior human data, how should researchers construct \persona{s}?

In this paper, we build agents whose behavior in simulations usefully generalizes to what we see from humans across entire domains.
Our approach mirrors what researchers generally try to do in social science, but in reverse. 
Rather than testing a theory with empirical data, theory is embedded in agents (via natural language instructions), which are then used to generate candidate data.
The theory and agent composition is then optimized to reduce error with respect to real-world data from a domain where predictions are desired.
Human data from distinct but conceptually similar settings serve as held-out test sets or are incorporated into training to improve generalization.
We show that ``general'' agents constructed and validated in this way can improve the predictive power of \aisub{s} in novel settings.
Both steps are essential: without theoretical grounding, optimized \persona{s} may fail to meaningfully improve even in-sample predictions, and without cross-setting validation, they are prone to overfit.

The approach uses two kinds of data: (i) \emph{training data}---existing human data used to optimize AI simulations, and (ii) \emph{validation} or \emph{test data}---existing human data related to but distinct from the training data, used to test whether the optimized agents generalize.
The ultimate goal is to produce agents that can more accurately predict human behavior in novel \emph{target settings} where no prior human data exist, but that lie in the same broad domain as the training and validation settings.

The first step of the approach is to limit the ``space'' of \persona{s} to a subset motivated by some economic theory or causal mechanism relevant to the novel setting of interest.
This theory-grounding is analogous to constraining the functional form of the hypothesis class in machine learning.
Continuing with the dictator game example, where social preferences likely determine behavior, one might choose the candidate set of \persona{s} characterized by known drivers of the relevant preferences (e.g., \emph{``You are \{level\}}'' for all \emph{levels} $\in$ \{\emph{self-interested but fair, altruistic, selfish}\}).

The second step is to optimize over this set to best match the human training data.\footnote{There is an active literature on optimizing prompts \citep{khattab2024dspy}.}
All candidate \persona{s} are put in simulations of the settings that produced the training data (e.g., a \$20 dictator game).
Optimization effectively filters these candidate \persona{s} down to a final subset whose simulated responses are closest to the human training distribution.
We employ two optimization methods: (i) a selection method that identifies the optimal mixture of \persona{s} from the candidate set \citep{leng2024calibrate,Jackson2025Mixture,bui2025mixture}, and (ii) a construction method that optimizes numerical parameters embedded directly in the prompts.
If the final optimized set achieves a good in-sample fit, we have reason to believe they will generalize because they are also grounded in a relevant theory.
Poor in-sample fit suggests a mismatch between the theory and the training setting or an inadequate operationalization of that theory.
In this case, we revise the candidate \persona{s} and re-optimize.

Given strong in-sample fit, we assess whether the final set of \persona{s} generalize using a train-test split approach inspired by the principles of invariant risk minimization \citep{Peters2016invariant,Heinze2018invariant,arjovsky2020invariant}.
The final set of \persona{s} is placed in simulations of the settings that produced the human validation data, where we also expect the underlying theory to hold (e.g., optimize on a \$20 dictator game, but test on a \$5 dictator game).
To be clear, this means the validation set necessarily comes from a distinct data-generating process from the one that produced the training data. 
We then compare the predictions from these simulations to the relevant human distributions.
By construction, sets of \persona{s} with strong test performance---those that accurately predict the validation data---are then those that capture generalizable relationships predictive of human behavior across contexts.
Those that fail validation likely do not.
Consequently, if the novel target setting is governed by the same theory or causal mechanism used to construct the optimized \persona{s} (e.g., the target is a \$50 dictator game), we may gain confidence that they will better predict human responses in that setting.

We illustrate this approach and provide evidence of its efficacy with training and validation data drawn from experiments in the behavioral economics literature.
All simulations use \textsc{Gpt-4o} with the temperature set to 1, though the approach is agnostic to the choice of model and hyperparameters.\footnote{\textsc{Gpt-4o}, among the most widely used LLMs, and the temperature setting are defaults for the software used to run our simulations \citep{Horton2024EDSL}.}
We first apply the selection method to \citet{1120Arad2012}'s 11-20 money request game, where participants request an amount and receive a bonus if they choose exactly one less than their opponent. 
We use only the original dataset from the paper, which contained fewer than 200 observations.
This setting is appealing to study because optimal play depends not only on the focal agent's capabilities, but also on their beliefs about how others will reason.
Endowing \aisub{s} with distinct \persona{s} corresponding to varying degrees of strategic reasoning (the theoretical focus of \citeauthor{1120Arad2012}) produces a mixture that closely matches the original human data.

When we validate these optimized samples on distinct variants of the 11-20 money request game, they are better predictors of initial human play than baseline \aisub{s} with no additional instructions.
By contrast, scientifically meaningless or ``atheoretical'' \aisub{s} derived from historical figures, pseudo-scientific Myers-Briggs personality types, and those instructed to select particular numbers can sometimes match human distributions in one variant of the game but fail to generalize across others.

We next test the predictive power of the optimized agents in target settings where no prior human data exist.
To do so, we construct four new games and collect responses from samples of preregistered crowdsourced participants \citep{horton2011online} on Prolific. 
These games are derived from the original 11-20 game (and its variants), but adapted to other numeric ranges (1-10 and 1-7).
The optimized sample of theory-grounded \persona{s} produces responses that predict the new human data far better than the off-the-shelf baseline.
Prediction error is decreased by \MinPercImproveArad\%-\MaxPercImproveArad\% across the games.
Furthermore, these simulations predict the results of the new experiments in some games \emph{better} than the most plausibly relevant human data from \citeauthor{1120Arad2012}; in one case, the KL divergence is halved.
By contrast, the alignment of the atheoretical \persona{s}---which failed validation---with the new human data is often similar to or worse than the baseline AI.

What statistical guarantees does this approach afford?
Without a correctly specified causal model, no statistical procedure can guarantee performance in arbitrary new environments \citep{Pearl2009Causality}.  
Formal guarantees with existing data, like those required for prediction-powered inference \citep{Angelopoulos2023Prediction} and other related methods \citep{Egami2023Using,Hardy2025Population}, would require a strictly firewalled validation set: data never used in the construction of the underlying LLM \citep{ludwig2025llmapplied,sarkar2024lookahead,ML2017Sendhil,spiess2025causaltext}.
This is a tall order impossible to meet in practice, even with existing public weight LLMs, never mind private models. 
However, what we can guarantee is prediction performance over a \emph{pre-committed} family of settings. 

The setup is very similar to the theoretical framework of \cite{andrews2025transfer}, who study how well predictions from different economic and machine learning models transfer across economic domains.
The first step is to establish a clearly defined space of experimental settings---for example, variants of public goods games, dictator games, or other allocation games differentiated by instructions, parameters, or other structural features.
This idea is related to that in the literature on program evaluation and external validity, where population-level treatment effects can be estimated by randomizing over a defined set of possible samples \citep{Hotz2005PredictingEfficacy,Allcott2015Site,Dehejia2019External}.
From this space, a random subset of settings is drawn and randomly assigned to human subjects, who provide responses.\footnote{This is also similar to the integrative experiment designs of \cite{20Q2024Almaatouq}.}
By comparing these human responses with LLM-generated predictions, we can make externally valid estimates across the entire space.

A key distinction from the setups laid out in \citeauthor{Hotz2005PredictingEfficacy}, \citeauthor{Allcott2015Site}, and \citeauthor{Dehejia2019External} is that appropriate coverage does not require that the family of settings share a data-generating process or that the underlying samples of human subjects are from the same population.
Indeed, the variance of estimates naturally reflects how closely the held-out settings resemble those sampled.  
If all scenarios are variants of a single setting---e.g., a dictator game with different monetary amounts---performance is tightly bounded; if they span disparate domains---e.g., many types of allocation games---estimates are likely less precise.
Note that even if a setting inadvertently appears in the model's pre-training data, the random sampling procedure still accounts for its contribution---such cases may merely reduce prediction error.

We explore this approach to inference by constructing a population of \TotalGames{} novel strategic games.
These were inspired by the original 11-20 game but were modified along several dimensions, such as support of possible choices, the size and nature of the bonus, the manner in which money is allocated, etc. 
The population of games is the full factorial combination of these dimensions.
The resulting games differed substantively from each other; in some cases, they were cooperative, others zero-sum, many had a dominant strategy equilibrium, while some had no dominant strategies at all.
The vast majority are surely not in any LLM's training corpus to date.
From the population, we sample \NGamesS{} unique games and have \NParticipantS{} human subjects each play a game in a final preregistered experiment.
We take the theoretically-grounded level-$k$ agents---agents constructed months before the novel set of \TotalGames{} games existed---and evaluate their ability to predict the human responses.

The theory-grounded agents are significantly better predictors of initial human play than the baseline AI off-the-shelf.
In the average game, they assign $\ExpBaselineBarLambda$ times more probability to the actions actually taken by human subjects.
The optimized agents are similarly effective compared to two well-established theoretical benchmarks: (i) a symmetric Nash equilibrium calculated for each of the 1,500 games ($\ExpNashBarLambda$ times more probability), and (ii) the game-specific predictions from the cognitive hierarchy model of \cite{Camerer2004Cognitive} ($\ExpCHBarLambda$ times more probability).
Because the games were randomly sampled, the corresponding confidence intervals over the relative predictive power are externally valid for the much larger population.

The goal of AI simulations in this paper is to harness two extensive sources of information to create better predictive agents: (i) well-established theoretical models from the social sciences, and (ii) the immense knowledge about human behavior that LLMs have implicitly learned during training \citep{lindsey2025biology,ameisen2025circuit}.
In a sense, \aisub{s} are a general vessel through which theory can be flexibly applied to any setting.
Our approach aims to generate such agents by conducting a structured trial run across various settings, building evidence that theoretically-grounded \persona{s} generalize effectively to similar yet distinct environments.
These principles can be applied to any domain where relevant human data exists---even those with multiple interacting agents \citep{Manning2024Automated,qian2025strategicbargaining}.
One can imagine iteratively identifying novel scenarios where predictions are needed, finding the closest relevant existing data, and continuously optimizing and calibrating agents as needed.\footnote{LLMs could automatically construct candidate \persona{s} as input to such feedback loops \citep{Jackson2025Mixture}.}
Our main contribution is to systematize these ideas for constructing agents and then provide empirical evidence that they can substantially improve the predictive power of AI simulations in new settings.

The remainder of this paper is organized as follows.
Section~\ref{sec:approach} begins with a concrete example of identifying generalizable relationships and constructing \persona{s} that better predict human subjects in new settings.
We then describe the approach more generally and specify the optimization methods.
Section~\ref{sec:predict_new_AR} illustrates the approach and demonstrates its efficacy empirically with several experiments.
Section~\ref{sec:ext-valid} demonstrates how agents can be used in a pre-committed setting to provide externally valid statistical estimates of predictive accuracy at scale.
The paper concludes in Section~\ref{sec:conclusion}.

\section{Building \persona{s} that generalize}
\label{sec:approach}

This section explains the approach for constructing \aisub{s} that better approximate human response distributions in new settings.
We begin with an extended example of predicting behavior in a novel public goods game where no previous human data exists.
We then discuss the importance of validating across distinct data-generating processes and grounding candidate \persona{s} in relevant social science theories.
The section concludes with a summary of the approach and a brief description of the optimization methods used to construct agents.

\subsection{A motivating example}

Suppose we want to predict how people will share resources in a novel public goods game (as they do in \cite{alsobay2025integrative}).
In this new game, for which we have no previous data, three participants will be endowed with \$5 and can choose to contribute any portion of their endowment, which will be multiplied by 3 and then divided equally among all participants.
While we do not have any prior public goods game data, we do have human data from a related \$20 dictator game.
It is related in the sense that one might reasonably expect some generalizable features of human choice to affect allocations in both games.
The observed human offers from the dictator game are \{6,6,7,7,8,8,9,9\}.

Before LLMs, one might have tried to train a standard machine learning model (e.g., decision trees or regression) solely on these dictator game offers.
However, such models would struggle to transfer to the structurally distinct public goods game, as they cannot adapt across different game formats without retraining.
LLMs offer a more flexible alternative. 
Rather than training a new model from scratch, we can prompt 
an existing LLM to simulate responses by instructing it to behave as a human participant.
For instance, we might use a baseline system prompt: \emph{``You are a human.''}
We can then append game-specific prompts such as \emph{``You are playing a dictator game with \$20...''} or \emph{``You are playing a public goods game with \$5...''} without any additional training.

Suppose that, when prompted with the dictator game instructions, the LLM produces a response distribution \{3,3,3,3,3,4,4,4\}.
Although well within the allowable offers, this distribution clearly diverges from the observed human data  (\{6,6,7,7,8,8,9,9\}) and leaves us with little hope that it could effectively predict responses to the novel public goods game.
Thus, even though the model may capture some general aspects of human behavior as a baseline, i.e., a tendency to offer nontrivial amounts \citep{Henrich2001InSearch}, it does so imperfectly.

One might think this problem could be addressed by first randomly splitting the human sample of dictator game offers into training and testing sets.
And then identifying the best-performing \persona{s} on the training set, and validating their performance on the test set \citep{ML2017Sendhil,ludwig2025llmapplied}.
Indeed, an LLM instructed \emph{``You randomly choose numbers between 6 and 9''} could reasonably predict both training and testing sets for any split of the human data better than the baseline LLM with no persona.
However, such an approach will almost certainly fail to generalize beyond the specific data-generating process that produced the dictator game offers.
It has, in effect, still overfit.
Rather than overfitting to the training data, it has overfit to the whole data-generating process.

Now, consider a more theoretically grounded \persona{}: \emph{``You are self-interested but fair.''}
Suppose an LLM endowed with this \persona{} also generates offers in the 6-9 range for the \$20 dictator game.
Crucially, this \persona{} aligns with known causal factors that drive human sharing behavior more broadly, like behavioral parameters in empirically tested theoretical models (e.g., \cite{charness2002understanding}).
It is a flexible decision-making program that can be effectively applied to various types of allocation games.
Yet, without explicit knowledge of the causal drivers governing behavior in the new public goods game, nothing in the data produced by either \persona{} alone rules out the atheoretical random-number prompt.
This illustrates the core challenge.
We seek to construct and select \persona{s} that do not overfit to a single data-generating process, but instead capture the stable behavioral drivers relevant across settings.
Then, we might gain confidence that they will better predict human responses in novel target settings governed by the same drivers.

\subsection{Identifying generalizable behavioral relationships}

As our motivating example illustrates, a traditional train-test split does not adequately guard against overfitting to a single data-generating process. 
This approach is statistically valid only when training and testing samples are independently drawn from the same distribution \citep{statisticalvapnik1998}. 
However, our objective differs fundamentally: we seek \persona{s} that remain predictive even when the underlying data-generating process shifts.
By ``data-generating process,'' we refer specifically to the experimental setting (the environment in which behavior occurs), the population from which behavior is observed, and the outcomes they generate.
In econometric terms, two processes diverge when the given input covariates appear with different distributions of values or when the covariates themselves are entirely different random variables.
For example, a model trained on \$20 dictator games may fail on \$5 dictator games due to different stake values (covariate shift) or when moving from dictator to public goods games (structural shift).

Without direct training data from the novel target setting, no standard statistical procedure ensures predictive accuracy \citep{ben2010theory,Klivans2024Domain}. 
Theoretically, only a fully specified causal model could guarantee accurate predictions \citep{Pearl2009Causality}. 
In practice, however, constructing or inferring such causal models generally requires strong, often unverifiable assumptions, which are rarely feasible in complex social science contexts.

Instead, we propose leveraging principles from invariant risk minimization \citep{arjovsky2020invariant} to identify behavioral relationships that remain stable despite shifts in the data-generating process. 
Rather than splitting a single dataset, we deliberately choose training data from one data-generating process (or several related processes) and validate using data from a related but distinct process from those used for training.
\Persona{s} that consistently predict behavior across these distinct datasets likely capture generalizable, and possibly causal \citep{Peters2016invariant, Heinze2018invariant}, drivers of behavior.
As a result, these validated \persona{s} should be more effective in predicting human responses in novel but theoretically similar settings.

Returning to our motivating example, suppose we have additional human data from a \$5 dictator game, with observed offers \{1,1,1,2,2,2\}.
Although stakes differ, both the \$20 and \$5 dictator games likely share a common decision-making process: individuals give slightly less than half the available amount, a reasonable balance more similar to what is observed in humans \citep{Henrich2001InSearch}.
By using the  \$20 dictator game as training data and the \$5 dictator game for testing, we are implicitly filtering for \persona{s} based on their capacity to predict this proportional response.
The earlier atheoretical prompt \emph{``You randomly choose numbers between 6 and 9,''} still fits the \$20 game perfectly, but clearly fails validation on the \$5 game.
By contrast, the theory-grounded \persona{} \emph{``You are self-interested but fair,''} likely generalizes effectively across both settings, and most importantly, the novel public goods game.

This validation process becomes even more robust when multiple distinct but theoretically-related datasets are available. 
Imagine dictator-game responses from \$1, \$5, \$10, \$20, \$50, and \$100 games, all exhibiting offers slightly below half.
Optimizing over multiple settings simultaneously further reduces the risk of overfitting. 
With each new training data-generating process, it is less likely that an arbitrary prompt will generalize in-sample.\footnote{We do not offer empirical examples of optimizing over multiple training settings in the main text (only single settings in Section~\ref{sec:predict_new_AR}).
Appendix~\ref{app:predict_new_CR} provides such analyses, including an additional preregistered experiment.}
Thus, if the underlying relationship governing these settings also holds for a novel target setting, the validated \persona{s} should robustly predict behavior there as well.

Yet, even with an effective validation method, a fundamental practical challenge remains: identifying the initial set of candidate \persona{s}. 
While our motivating example made this step appear straightforward, selecting plausible \persona{s} in more complex settings can be far less obvious.
In a perfect world, an optimization procedure would take in a massive set of prompts and filter them down to a smaller set that generalizes.
However, this does not yet work in practice because many different prompts can be used to generate the same behavior in a given setting.
For example, the number of possible natural language prompts that get an LLM to respond with numbers between 6 and 9 is enormous.
If we were to start from a large pool of prompts, then the optimization procedure might endlessly produce sets of \persona{s} that fit in-sample but fail on the validation set.
For now, it is essential to ground the candidate \persona{s} with a principled starting point, which---we argue and later demonstrate empirically---economic and behavioral theories offer.

\subsection{Theory as guide towards generalizability}

A core function of economics is to construct models that capture causal and generalizable relationships that remain stable across environments.
For example, the idea that humans make reference-dependent utility choices is not specific to one particular economic environment, but has been shown to broadly apply to decision-making under uncertainty \citep{prospectkahneman1979}.
Conveniently, these are exactly the types of generalizable relationships that we would expect to better predict human behavior in new settings when supplied to an LLM.
Our approach is to narrow the search space of possible \persona{s} by grounding candidate \persona{s} in such theories.
By doing so, we might increase the likelihood of identifying \persona{s} that we have prior reason to believe reflect genuine, stable behavioral patterns.
Without doing so, we risk dramatically underfitting and failing even to accurately predict the training data.

Economic theories---and theories from other social sciences that embrace methodological individualism \citep{friedman1953methodology}---are well-suited to be cast into \persona{s}, since good theory rests on the choices and actions of individuals.
Specifically, we consider a prompt to be ``theoretically grounded'' when it instructs the LLM to make predictions about how a human would respond based on some underlying theory or model of human behavior.
For example, a utility function that trades off one's own payoff against inequality might become \emph{``You value your own earnings but dislike outcomes where you earn far more (or less) than others.''}
A highly capable and tool-using LLM could even be given that utility function and parameters directly (e.g., $u(x_{self}, x_{other})= x_{self} - | x_{self} - x_{other}|$).
A prospect-theoretic model reasonably maps to \emph{``You are risk averse in gains, risk seeking in losses, and probability weight the very likely and very unlikely outcomes''}---and of course these could be parameterized as well.

This definition is necessarily broad because what constitutes a theory is also broad.
One might consider good theories to be those that are parsimonious, predictive, and interpretable.
We view these as effective guides for constructing \persona{s} that are grounded in theory.
But just as there is no universal rule as to what makes a theory good, there is not a singular all-encompassing playbook for candidate \persona{s}.
Here, the researcher must leverage their expertise appropriately to determine which theories are most relevant to the domain where predictions are desired.
This should be relatively straightforward in practice.
One can even simply ask an LLM to construct a candidate set of prompts based on a given paper and then adjust the prompts accordingly.
For example, the linked notebook---which took 15 minutes to code---contains an example where we take PDFs of 5 well-known papers in the behavioral economics literature and generate 10 agents from each.\footnote{\url{https://expectedparrot.com/content/32751dae-7a69-430d-9f1a-0c3f50ccef5b}}

It is worth highlighting just how flexible such agents are as predictive models.
They can make predictions in response to \emph{any} setting described in natural language.
This is usually impossible for traditional machine learning and mathematical economic models, which operate within a fixed parameter space. 
One cannot introduce additional covariates to a regression model once it is trained. 
Similarly, a prospect-theoretic model for evaluating risky gambles cannot instantaneously incorporate other contextually relevant factors---the recent performance of financial markets, the demographic composition of the experimental sample, or even something as mundane as whether participants are making decisions before or after lunch. 
\aisub{s}, by contrast, can interpolate over such details through natural language instructions. 
Whether these agents actually leverage this flexibility effectively is, of course, an empirical question, but one that is easily testable.

Once a set of \persona{s} is defined, we can optimize them against an appropriate sample of relevant human decisions.
This may involve searching for an optimal mixture of \persona{s}, or adding continuous parameters directly in a single prompt and adjusting them until the simulated distribution aligns with the training data.
From a machine learning perspective, the candidate set of \persona{s} defines the functional form of our hypothesis class, with theory guiding this specification to effectively navigate the bias-variance tradeoff.\footnote{One could imagine an ``oracle'' \persona{} akin to a perfect program, describing every preference, heuristic, and belief update rule---allowing the LLM to accurately produce responses in all contexts for a person.
In effect, our approach is to identify portions of this oracle that are most relevant to the given training and testing data.
Increasing context windows suggest that such highly generalizable agents may be possible in the not-so-distant future.}
The set is then pruned (or tuned, depending on the method) to best fit the training data.

As an example, consider the extensive theoretical and empirical economic literature studying how people reason strategically \citep{STAHL1994,STAHL1995,1120Arad2012, Camerer2004Cognitive}.
Level-$k$ models, for instance, predict that individuals best respond based on their beliefs about others' levels of reasoning. 
In the $\frac{2}{3}$ guessing game, players aim to pick $\frac{2}{3}$ of the average number chosen by the group \citep{LkNagel1995,keynes1936general}. 
If we had human data from such a game, we could construct a series of \persona{s} that specify different levels of strategic thinking (e.g., \emph{``You are a level-0 reasoner,'' ``You are a level-1 reasoner,''} etc.).
We then identify the mixture that best fits the training distribution and test whether it generalizes to other variants of the game (e.g., different values than $\frac{2}{3}$).
Then, the optimized \persona{s} can be used to predict other related distributions in new settings.

Of course, constructing and validating these \persona{s} is not a perfectly specified problem.
There are various ways to translate a theory into natural language, and multiple theories may apply to a given setting.
The boundaries of a theory and the settings it plausibly governs are not always well-defined.
Nor can we guarantee that the data-generating processes of the training and testing sets are either sufficiently similar or sufficiently distinct to yield reliable validation.
Yet, as we will show empirically, these challenges can be overcome in practice, at least in some domains.

Ultimately, applying a \persona{}---or a set of \persona{s}---to a new environment requires an unavoidable inductive leap.
What we can do is try to make the leap explicit and interpretable.  
LLMs are extremely good at following explicit, well-defined instructions.
Because each \persona{} is a set of these flexible natural language instructions, researchers can both evaluate performance and reasonably assess its relevance for a new setting.\footnote{Such flexibility also makes \aisub{s} less susceptible to the Lucas critique \citep{Lucas1976Critique}---if at all.
Unlike traditional agent-based models or mathematical theories, which have hard-coded predetermined processes to generate predictions, LLMs can, as we will show, interpret new policies in context, reason through their implications, and adaptively formulate responses.
}
This mirrors how economic theory is used more broadly with real humans: it is tested against data, observed where it succeeds or fails, and cautiously extrapolated to settings just beyond current evidence.
When a new tariff is proposed, for example, the theory motivating the tariff is not known to be universally ``correct'' in a complex dynamic world ex ante.
It is a disciplined guess, shaped by assumptions and previous evidence, which the economist then uses to make predictions about the world.
Theory-guided \persona{s}, validated across environments, bring the same kind of disciplined reasoning to simulated \aisub{} predictions of human behavior.

\subsection{A brief summary of the approach}

We can summarize our approach into the following steps.

\begin{enumerate}[itemsep=1pt, topsep=0pt]
\item \textbf{Select Training and Testing Data.} Identify distinct samples of human-generated data that are presumably generated by the same mechanisms as the novel target setting(s) of interest. 
When possible, use multiple distinct datasets for both training and testing to increase confidence and minimize the possibility of selecting spurious \persona{s}.
\item \textbf{Propose Theory-Driven Candidate \persona{s}.} Generate a broad set of \persona{s} that are plausibly consistent with the proposed theory (or causal mechanisms if known) related to the training, testing, and novel settings.
\item \textbf{Optimize \persona{s} on Training Data.} Optimize the candidate \persona{s} to best match the training data. 
This might involve selecting a mixture of \persona{s} or adjusting trait parameters to minimize some statistical distance from observed responses.
Confirm that the optimized sample outperforms the baseline LLM off-the-shelf on the training data.
\item \textbf{Validate \persona{s} on Testing Data.} Apply the optimized \persona{s} to the testing data and evaluate their performance relative to the baseline LLM.
\end{enumerate}

Broadly speaking, our approach provides a disciplined ``trial run'' to confirm whether a given sample of \aisub{s} can reliably predict human behavior across multiple related settings.
Similar to applying economic models estimated from past data to inform predictions in new but structurally similar environments, the success of our approach depends critically on identifying and leveraging underlying stable behavioral relationships.

In Section~\ref{sec:predict_new_AR}, we demonstrate the approach empirically on a set of strategic games using training data from \cite{1120Arad2012}.
Appendix~\ref{app:predict_new_CR} provides another example using training data from \citep{charness2002understanding} and allocation games.
Both applications show that the approach substantially reduces prediction error relative to baseline LLM predictions in preregistered experiments with novel human samples.
We emphasize throughout that the two key elements of the approach---grounding candidate \persona{s} in economic or behavioral theories, and validating across distinct but related datasets---each independently address critical pitfalls.
Without theoretical grounding, optimized \persona{s} may fail to meaningfully improve even in-sample predictions; without validating across multiple related settings, optimized \persona{s} are prone to overfitting a single training context.

\subsection{Methods for optimizing \persona{s} in-sample}

We briefly describe two methods for optimizing \persona{s} with a given set of training data.
The first, the selection method, assumes we have a finite library of candidate \persona{s} and selects (or mixes) them to best fit the training data. 
We apply this method to the experiments presented in Section~\ref{sec:predict_new_AR} with data from \cite{1120Arad2012}.
An early version of this idea was suggested by \cite{horton2023large}, and several others have since explored applications \citep{leng2024calibrate,Jackson2025Mixture,bui2025mixture}.
The second, the construction method, parameterizes a prompt template with numeric trait dimensions and optimizes those parameters.
To the best of our knowledge, this method is novel.
An application of the method is presented in Appendix~\ref{app:predict_new_CR} using data from \cite{charness2002understanding}.

\paragraph{Selection Method.}
A finite set of unique candidate natural language \persona{s} is first specified. 
For each \persona{} $\theta \in \Theta$, the LLM is used to generate a predictive distribution $\hat P_\theta$. 
Let $P$ represent the observed ground-truth human distribution.
The objective is to solve
\[
\min_{\mathbf{w}} \; d \Bigl(P, \sum_{\theta \in \Theta} w_\theta \,\hat P_\theta\Bigr) \quad \text{subject to} \quad \sum_{\theta \in \Theta} w_\theta = 1, \quad w_\theta \ge 0,
\]
where $d$ is a chosen distance measure (e.g., KL divergence or the mean absolute distance between distributions). 
Once solved, these weights can be used to scale the appropriate mixture of \persona{s} (i.e., $\bm{\theta^{\star}}$) and applied to new settings.

\paragraph{Construction Method.}
Alternatively, a \persona{} template can be parameterized by numeric trait variables. 
This is best illustrated with an example.
Suppose $\phi_1$ and $\phi_2$ capture degrees of self-interest and inequity aversion, respectively. 
The \persona{} could be:
\begin{quote}
\singlespacing
\vspace{-.5cm}
$\theta(\phi_1, \phi_2)$ = \emph{``You weigh your own payoff with weight $\{\phi_1\}$, and you dislike creating disadvantageous inequality at level $\{\phi_2\}$. Please respond accordingly.''}
\end{quote}
$\hat P_\theta$ denotes the distribution induced by the LLM under parameter vector $\theta = (\phi_1, \phi_2)$. 
Given an observed human distribution $P$, the optimal parameters $\bm{\theta^{\star}}$ are found by solving $\min_{\theta} \; d \bigl(P, \hat P_\theta\bigr)$.
In practice, this can be solved using any derivative-free optimization algorithm, such as Bayesian optimization or evolutionary algorithms.
Note that one need not be limited to a single prompt; optimization can be solved using multiple templates or sets of agents with different values for the same template.

\paragraph{Measuring Performance.}  
Let $\hat P_{\bm{\theta^{\star}}}$ denote the LLM's distribution of responses under $\bm{\theta^{\star}}$, and $\hat P_0$ the baseline LLM off-the-shelf without any additional prompting.
Training loss is $d(P, \hat P_{\bm{\theta^{\star}}})$ and validation loss is defined analogously on the held-out test setting. 

To assess whether $\bm{\theta^{\star}}$ generalizes, we compare predictive fit against the baseline $\hat P_0$.  
This is an appealing reference measure because many distance metrics between distributions (like the KL divergence) are not easily interpretable.
Furthermore, improvement over the baseline provides direct evidence that designing and optimizing \persona{s} is a worthwhile endeavor in the first place.\footnote{Strong baseline performance would simply reflect the LLM's already high off-the-shelf predictive capacity.}

Relative improvement can be computed as a difference-in-distances: $d\bigl(P, \hat P_{0}\bigr) \;-\; d\bigl(P, \hat P_{\bm{\theta^{\star}}}\bigr)$.
A positive difference indicates that the \persona{s} provide better predictive power than the baseline.
When we optimize over a set of candidate \persona{s}, we seek those that yield this positive difference on both the training and testing data---that would suggest that the set generalizes.

The same framework applies across multiple training and evaluation settings, with optimization performed by averaging distances if needed.  
It also allows direct comparison of optimized \persona{s} to alternative \persona{} sets or benchmark models (e.g., game-theoretic equilibria) using the same measurement tools.  

\section{Predicting behavior in novel strategic games}
\label{sec:predict_new_AR}

Thus far, we have argued that theory-grounded \persona{s}, validated on distinct but related datasets, offer an approach for predicting human responses in entirely novel settings.
To empirically test the efficacy of our approach, we now turn to predicting human behavior in a set of strategic games adapted from \cite{1120Arad2012}'s (AR) study of strategic reasoning.

We begin by briefly reviewing the level-$k$ model of strategic reasoning that originally motivated AR. 
This model provides plausible theoretical underpinnings linking the training, testing, and novel games in this section.
We then describe the structure of the original 11-20 money request game and outline our procedure for constructing theory-grounded \aisub{s}: optimizing their parameters on human data from AR's original experiment, and validating their predictive performance on distinct but related variants also studied by AR.
Finally, we introduce an entirely new set of games adapted from AR's original design but featuring distinct numeric ranges and a novel participant sample recruited from Prolific. 
Agents are evaluated on their ability to predict human behavior in these never-before-seen games. 

Although this section focuses on strategic reasoning games, the same procedure can be applied in any setting.
In Appendix \ref{app:predict_new_CR}, we replicate the process using the allocation games from \cite{charness2002understanding}, which explore social preferences.
Two other important differences in that section are that we: (i) optimize across multiple related settings rather than a single training setting, and (ii) utilize the construction method to build agents.
We also present an additional preregistered experiment with human subjects to validate these agents in new settings.

\subsection{\cite{1120Arad2012}'s 11-20 money request game}
\label{sec:11-20_app}

The level-$k$ model posits that players differ in how many steps ahead they consider when forming their strategies \citep{LkNagel1995,STAHL1994,STAHL1995}.
The model defines different types of players, from level-$0$ to level-$k$. 
Level-$0$ players use some predefined arbitrary decision rule, while level-$k$ players ($k \geq 1$) best respond assuming others are level-$(k-1)$ reasoners.
Such a model highlights the idea that players' behavior depends not only on their decisions but also on their beliefs about how other people think.

To measure the distribution of level-$k$ thinkers in human populations, AR developed the 11-20 game.
The instructions are:
\begin{quote}
\vspace{-.5cm}
\singlespacing
\emph{You and another player are playing a game in which each player requests an amount of money. The amount must be (an integer) between 11 and 20 shekels.  Each player will receive the amount he requests. A player will receive an additional amount of 20 shekels if he asks for exactly one shekel less than the other player. What amount of money would you request?}
\end{quote}
This \emph{basic} version of the game clearly maps players' levels of reasoning to their choices.
A natural starting point is to choose 20 shekels.
This maximizes the guaranteed payment and provides an obvious starting point for level-0 thinking.
Next, level-1 players, anticipating level-0 players, will choose 20, best respond by requesting 19 shekels to earn the bonus. 
Level-2 players, expecting level-1 behavior, choose 18 shekels, and this pattern continues down to the minimum of 11.
More generally, a player choosing $(20-k)$ shekels plausibly reveals themselves as a level-$k$ thinker.

Interestingly, the 11-20 game does not have a pure strategy Nash equilibrium.
The best response to any choice greater than 11 is to undercut the opponent by 1.
But if the other player chooses 11, also selecting 11 is strictly dominated by every other strategy.
The top row of Table~\ref{tab:AR_results} shows the unique symmetric mixed strategy Nash equilibrium for the game, with most of its density lying between 15 and 17 (levels 3 to 5).

\begin{table}[h!]
\caption{Original results from \cite{1120Arad2012}}
\vspace{-.5cm}
\begin{center}
\begin{minipage}{1 \linewidth}
\label{tab:AR_results}
\footnotesize
\setlength{\tabcolsep}{7pt}
\renewcommand{\arraystretch}{1.25}
\begin{tabular*}{\textwidth}{@{\extracolsep{\fill}}lrrrrrrrrrr}
\hline
\textbf{Shekels Requested} & \textbf{11} & \textbf{12} & \textbf{13} & \textbf{14} & \textbf{15} & \textbf{16} & \textbf{17} & \textbf{18} & \textbf{19} & \textbf{20} \\
\textbf{\boldmath{Level-$k$}} & \boldmath{$L9$} & \boldmath{$L8$} & \boldmath{$L7$} & \boldmath{$L6$} & \boldmath{$L5$} & \boldmath{$L4$} & \boldmath{$L3$} & \boldmath{$L2$} & \boldmath{$L1$} & \boldmath{$L0$} \\
\hline
\emph{Nash Eq. Prediction (\%)} & \emph{0} & \emph{0} & \emph{0} & \emph{0} & \emph{25} & \emph{25} & \emph{20} & \emph{15} & \emph{10} & \emph{5} \\
Basic ($n = 108$) (\%)& 4 & 0 & 3 & 6 & 1 & 6 & 32 & 30 & 12 & 6 \\
Cycle ($n = 72$) (\%)& 1 & 1 & 0 & 1 & 0 & 4 & 10 & 22 & 47 & 13 \\
\hline
\emph{Nash Eq. Prediction (\%)} & \emph{0} & \emph{0} & \emph{0} & \emph{10} & \emph{15} & \emph{15} & \emph{15} & \emph{15} & \emph{15} & \emph{15} \\
Costless ($n = 53$) (\%)& 0 & 4 & 0 & 4 & 4 & 4 & 9 & 21 & 40 & 15 \\
\hline
\end{tabular*}
\end{minipage}
\end{center}
\begin{footnotesize}
\begin{singlespace}
\vspace*{-0.05in}
\emph{Notes:} This table reports the empirical PMFs for three versions of the 11-20 game from \citeauthor{1120Arad2012}. 
In the basic version of this game, two players each request a number between 11-20 shekels and they receive that amount. 
If a player requests exactly one less than their opponent, they win their request plus a 20 shekel bonus. 
In the cycle version, players also receive a 20 shekel bonus if they select 20 and their opponent selects 11. 
The costless version is identical to the basic version, except that players receive 17 shekels if they select any amount other than 20.
The basic and cycle versions of the game share a unique symmetric mixed strategy Nash equilibrium, which is shown in the first row.
The unique mixed strategy Nash equilibrium for the costless version is shown in the 4th row.
\end{singlespace}
\end{footnotesize}
\end{table}

Table~\ref{tab:AR_results} also shows that when AR tested this game on pairs of college students, they deviated significantly from the Nash equilibrium (see row titled Basic).
Most notably, 73\% of participants chose between 17 and 19 shekels, whereas only 45\% of the density for the mixed strategy Nash is on these values.

Finally, Table~\ref{tab:AR_results} provides results from two additional variants of the game from AR.
Both variants of the game still involve two players selecting numbers between 11 and 20.
They differ from the basic version in their payouts (see Appendix~\ref{app:instruct} for the full instructions).
In the \emph{cycle} version, players can earn a bonus of 20 shekels by undercutting the other's request by exactly one or by selecting 20 when the other selects 11. 
This version has the same unique mixed strategy Nash Equilibrium as the basic game because the only affected strategy profiles are (20,11) and (11,20), which are outside the equilibrium support.
However, if there is a distribution of level-$k$ thinkers---as opposed to perfectly rational game-theoretic best-responders---such a change might significantly impact play.

The \emph{costless} version has the same bonus structure as the basic game, but with a different payout for choices below 20.
Here, requesting 20 yields 20 shekels outright, while choosing a lower amount guarantees 17 shekels plus a 20-shekel bonus if the lower request is exactly one less than the other player's. 
It is comparatively ``costless'' to continue undercutting.
This game induces a symmetric mixed strategy Nash equilibrium that is more uniform across the choice set than the basic and cycle versions.

The human responses from these game variants, also from a similar sample of college students, are noticeably shifted towards 18-20 shekels---even with the basic and cycle sharing the same equilibrium.
AR attributes this to the increased salience and payoff of selecting a higher number.
AR concluded that the collective results from these three experiments are best explained by a mix of strategic types consisting of level-0, level-1, level-2, level-3, and random choosers.

The empirical human distributions from these three games comprise our training (the basic version) and validation (the costless and cycle versions) data.
The games are distinct enough such that the human response distributions are different, but still all likely well-explained by similar underlying human choice processes.

\subsection{Optimizing \aisub{s} in-sample}
\label{sec:apply_mixture}

\begin{figure}[h!]
\begin{center}
\begin{minipage}{\linewidth}
\caption{Response distributions for the basic version of the 11-20 game} 
  \vspace*{-0.15in}
\label{fig:compare_agents_train}
\centering
\includegraphics[width=.8\textwidth]{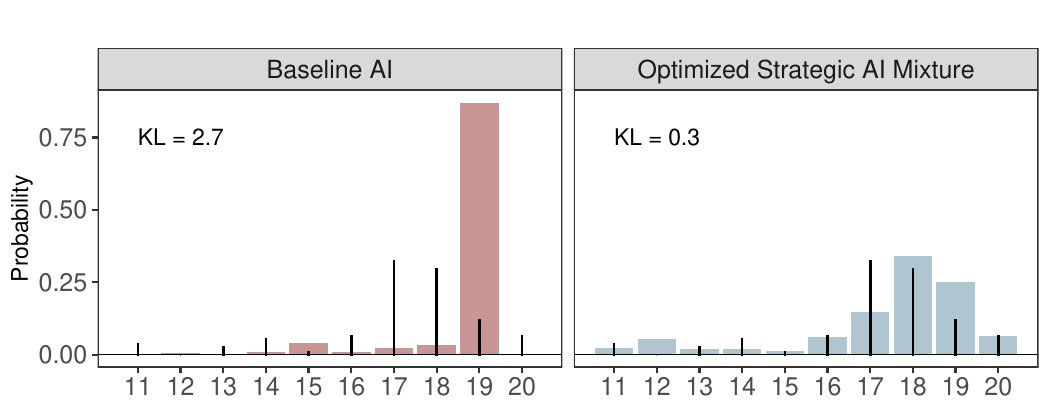}
\end{minipage}
\end{center}
\begin{footnotesize}
\begin{singlespace}
\vspace*{-0.15in}
\emph{Notes:} This figure displays empirical PMFs for three samples playing the basic 11-20 money request game: human subjects from \citeauthor{1120Arad2012} (the vertical black lines in both panels), the off-the-shelf baseline (left panel), and responses from our selected \aisub{s} based on the weights (right panel).
The KL divergence between the human and AI distributions is reported in the upper left corner of each panel.
\end{singlespace}
\end{footnotesize}
\end{figure}

We begin by eliciting the baseline AI's response distribution ($\hat P_0$).
We prompt \textsc{Gpt-4o} to play the basic version of the game 1,000 times without any additional instructions, setting the temperature to 1. 
Figure~\ref{fig:compare_agents_train} displays the results. 
The left panel shows the empirical PMF of the baseline AI responses (red), along with the empirical human distribution $P$ from AR's original experiment (vertical black lines). 
The baseline AI almost exclusively selects 19 shekels (\BaselineOneNineMass\%), demonstrating limited variability. 
Using the forward KL divergence as $d(\cdot,\cdot)$ with the human distribution as the reference, the distance between these distributions is $d(P,\hat P_0) =\KlHumanRawDist$.

Such a poor baseline result is unsurprising.
In a related exercise, \cite{gao2024caution} also elicit responses from various LLMs playing the same 11-20 game.
Even after applying diverse prompting strategies, fine-tuning, and endowing agents with distributions of demographic traits, they similarly find that LLMs strongly index on choosing 19 shekels.
While this outcome is not inherently problematic---19 shekels lies within the support of both the Nash equilibrium and the empirical human distribution---it underscores a limitation: demographic prompts (and the other techniques they employ) alone provide little leverage in predicting strategic human reasoning.
They do not provide any reliable, flexible program for the LLM to follow, nor do they allow for heterogeneity within the simulated sample.
In contrast, the level-$k$ model implies that optimal predictions require explicitly accounting for how individuals reason about others' decisions.
And because we know from AR that there is likely a distribution of reasoning levels in their human sample, we should apply various levels of reasoning to construct a heterogeneous sample of agents.
This may, in turn, better reflect the distribution of human strategic reasoning processes.

To operationalize the level-$k$ reasoning explicitly, we construct a set of natural language \persona{s} ($\Theta_{Strategic}$) corresponding to varying levels of strategic reasoning.
These candidate \persona{s}, listed in Table~\ref{tab:persona}, specify how far ahead each \aisub{} reasons about the opponent's decisions, effectively encoding beliefs about others' strategies.

We elicit response distributions $\hat P_{\theta}$ for each candidate \persona{} $\theta \in \Theta_{Strategic}$ by prompting \textsc{Gpt-4o} 100 times per \persona{}.\footnote{We then elicit each \aisub{'s} responses using Chain-of-Thought prompting, which encourages step-by-step reasoning before producing a final answer \citep{weiCOT2024}. We implement this through two sequential prompts. \textbf{Prompt 1}: \emph{\{\texttt{11-20 game instructions}\}. Reason out a few settings according to your personality and how others might respond.} \textbf{Prompt 2}: \emph{\texttt{\{11-20 game instructions}\}. You previously had the following thoughts: \{\texttt{Response to prompt 1}\}. What amount of money would you request?}. This procedure creates the agents (and their response distributions) over which we then optimize.}
We then employ the selection method to identify the optimal mixture of \persona{s} that minimizes the absolute difference between the CDFs implied by the empirical distribution of the human responses ($P$).
This distance can be minimized using simple nonlinear programming techniques.
The weights $\mathbf{w}^*$ corresponding to the optimal mixture appear in the second column of Table~\ref{tab:persona}.\footnote{See \url{https://www.expectedparrot.com/content/6f58d11f-98cc-4de5-bb89-edcf78042d79} for the agents.}

Most mass concentrates on two \persona{s}: one that reasons between levels 1 and 3 (47\%), and another varying more broadly from levels 0 to 5 (34\%). 
The remaining weight falls on more extreme behaviors (random choices or the safest guaranteed option). 
This aligns closely with AR's findings, whose human subjects predominantly exhibited level-0 through level-3 reasoning or made random choices.

\begin{table}[h!]
\begin{center}
\caption{Proposed \aisub{} \persona{s} and resulting mixture weights from the selection method} 
\label{tab:persona}
\footnotesize
\setlength{\tabcolsep}{7pt}
\begin{tabular}{p{0.85\textwidth}c}
  \hline
 \textbf{Persona} & \textbf{Weight} \\
\hline

You are generally a 0-level thinker---picking the option with the most guaranteed money. & 0.065 \\ 
  You vary between a 0 and 1-level thinker. & 0.000 \\ 
  You vary between a 1 and 2-level thinker. & 0.000 \\ 
  You vary between a 0, 1, and 2-level thinker. & 0.000 \\ 
  You vary between a 0, 1, 2, and 3-level thinker. & 0.000 \\ 
  You vary between a 1, 2, and 3-level thinker. & 0.469 \\ 
  You vary between a 0, 1, 2, 3, and 4-level thinker. & 0.013 \\ 
  You vary between a 0, 1, 2, 3, 4 and 5-level thinker. & 0.339 \\ 
  You randomly pick between lower numbers because you think that's the best way to win. & 0.114 \\ 
  You are Homo Economicus. & 0.000 \\ 
   \hline 
\end{tabular}

\end{center}
\begin{footnotesize}
\begin{singlespace}
\vspace*{-0.05in}
\emph{Notes:}
This table shows the set of \persona{s} $\Theta_{Strategic}$ used as input to the selection method.
The right column shows the optimized mixture weights $\mathbf{w}^*$ that minimize the absolute difference between the CDFs of the human distribution $P_s$ and the distribution of responses from the \aisub{s}.
Prepended to all the \persona{s} is: \emph{You are a human being with all the cognitive biases and heuristics that come with it}. 
We also include an explanation in the prompts for $k$-level reasoning for all \persona{s} besides the random one: \emph{A k-level thinker thinks k steps ahead. A 0-level thinker thinks 0 steps and would, therefore, just select the maximum amount that guarantees money.}
\end{singlespace}
\end{footnotesize}
\end{table}
  
Using these weights, we generate the sample $\bm{\theta}^*$ of 1,000 \aisub{s} by assigning each agent to one of the 10 \persona{s} with probability equal to its corresponding weight in Table~\ref{tab:persona}. 
The resulting empirical distribution of responses $\hat P_{\bm\theta^*}$ produced by this sample of 1,000 agents $\bm{\theta}^*$ appears in the right panel of Figure~\ref{fig:compare_agents_train} (blue).
The improvement over the baseline AI is substantial: $d(P, \hat P_{\bm\theta^*}) = \KlHumanOptDist$ is \KlPercDiffBasic\% smaller than $d(P, \hat P_0) = \KlHumanRawDist$, demonstrating a strong in-sample fit.
  
\subsection{Validation using game variants}

To validate these agent sets, we elicit their response distributions to the costless and cycle versions of the game.
Both games maintain our fundamental requirement of similar but distinct data-generating processes. 
They involve reasoning well-explained by level-$k$ thinking, but feature payoff structures and incentives that differ from those of the basic game.
Most important, humans do not play these games in the same way as the basic version.
The forward KL divergence between the basic and costless game is $\KlBasicCostlessHuman$, and between the basic and cycle game is $\KlBasicCycleHuman.00$ (see Table~\ref{tab:AR_results} for the human response distributions for all three games).
These divergences are substantial: each is more than three times larger than the divergence between the human and optimized agent distributions in the right panel of Figure~\ref{fig:compare_agents_train}.
Accurate predictions on these distinct games would indicate that our optimized \aisub{s} capture generalizable patterns rather than merely replicating the original training scenario.

\begin{figure}[h!]
\begin{center}
\begin{minipage}{\textwidth}
\caption{Response distributions for the cycle and costless versions of the 11-20 game} 
\label{fig:compare_agents_test}
\vspace*{-0.15in}
\centering
\includegraphics[width=.9\textwidth]{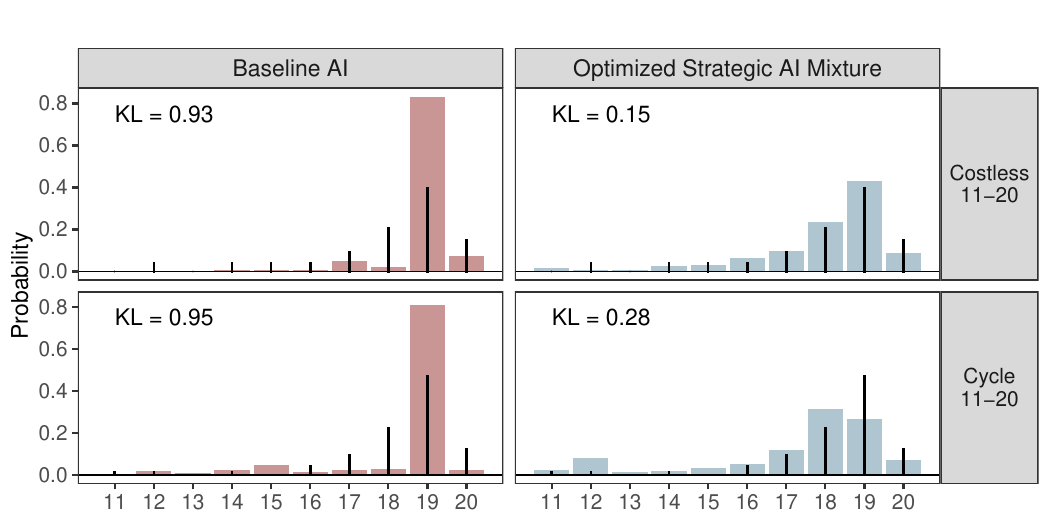}
\end{minipage}
\end{center}
\begin{footnotesize}
\begin{singlespace}
\vspace*{-0.05in}
\emph{Notes:} This figure displays empirical PMFs for the costless (top row) and cycle (bottom row) variants of the 11-20 game. 
The columns correspond to the baseline (red), the optimized \aisub{s} (blue).
Within each panel, the empirical PMFs from \citeauthor{1120Arad2012} are imposed in black.
The KL divergence between each human and the AI response distribution is displayed in each panel.
For both variants of the game, the selected \aisub{s} (blue) are far closer to the human distribution than the baseline (red), even though the selected \aisub{s} are constructed using only the basic version of the game.
\end{singlespace}
\end{footnotesize}
\end{figure}

We elicit responses from all 1,000 \aisub{s} in our optimized sample $\bm{\theta}^*$ for both the costless and cycle versions of the game. As a benchmark for evaluating relative performance, we also elicit responses from the baseline AI 1,000 times per game. 
Figure~\ref{fig:compare_agents_test} presents these results, comparing the empirical distributions from AR's original human experiments (black lines) with those generated by the baseline AI (red) and the optimized mixture of theory-grounded \aisub{s} (blue).

Consistent with the basic version of the game and \cite{gao2024caution}, the baseline AI overwhelmingly selects 19 shekels in both variants, a considerable divergence from AR's human subjects. 
In contrast, $\bm{\theta}^*$ is a far better predictor of both validation settings. 
The costless game in particular shows substantial improvement, with the KL divergence decreasing by \KlPercDiffCostless\% ($d(P, \hat P_{\bm\theta^*}) = \KlHumanOptCostlessDist$ vs. baseline $d(P,\hat P_0) = \KlHumanRawCostlessDist$).
It is almost perfectly predictive of the human responses when comparing the empirical PMFs.
In the cycle game, the KL divergence between the optimized AI and human responses is also reduced substantially, by \KlPercDiffCycle\% relative to the baseline ($d(P, \hat P_{\bm\theta^*}) = \KlHumanOptCycleDist$ vs. $d(P,\hat P_0) = \KlHumanRawCycleDist$).
Taken together, these results are interpreted as evidence that $\bm{\theta}^*$ has effectively generalized to the validation data---data that was not used to construct its mixture.
We therefore gain confidence that these agents may better predict entirely new settings that call for similar strategic reasoning.

\subsection{Optimizing among atheoretical \persona{s}}
\label{sec:arbitrary_select}

We now generate sets of arbitrary, atheoretical \aisub{s} using the basic version of the game, which ultimately fail validation on the cycle and costless versions.
These agents will offer a substantial contrast with $\bm{\theta}^*$ when applied to entirely novel games in the next subsection.
This exercise also highlights two potential pitfalls of AI simulations addressed by our approach: (i) without theoretically motivated candidate \persona{s}, optimization may fail to even improve predictive power in-sample over the baseline, (ii) atheoretical candidate sets can be optimized to effectively match particular samples of human data---even when such samples are obviously overfitting.
The former is addressed by using samples of \aisub{s} grounded in plausible theory, and the latter is addressed by using training and testing data from distinct settings (or training and testing both across many different settings).

To illustrate these points concretely, we introduce three new sets of candidate \persona{s}, none having any plausible relationship to strategic reasoning or the choices made in the different 11-20 games.
These are shown in Table~\ref{tab:bad_personas} in the appendix.
The first set consists of historical figures ($\Theta_{Hist}$);\footnote{Cleopatra, Julius Caesar, Confucius, Joan of Arc, Nelson Mandela, Mahatma Gandhi, Harriet Tubman, Leonardo da Vinci, Albert Einstein, Marie Curie, Genghis Khan, Mother Teresa, Martin Luther King, Frida Kahlo, George Washington, Winston Churchill, Mansa Musa, Sacagawea, Emmeline Pankhurst, and Socrates.} the second has the 16 Myers-Briggs personality types ($\Theta_{MB}$); and the third set comprises 10 ``Always Pick `N''' agents ($\Theta_{N}$), each of which is instructed to exclusively select a given integer from 11 to 20.\footnote{These agents all take the form of \emph{``You always pick N''} for $N \in \{11,\ldots, 20\}$.}
We apply the exact same selection procedure used in Section~\ref{sec:apply_mixture} to find optimized weights for each set, using only the human data from the basic version of the 11-20 game.

Table~\ref{tab:bad_personas} also shows the resulting weights.
For the historical figures, nearly all weight (89.1\%) collapses onto Julius Caesar and a small remainder on Confucius (10.1\%).
In the Myers-Briggs set, all weight concentrates on ENFP.\footnote{ENFP is Extraversion, Intuition, Feeling, Perceiving (\url{wikipedia.org/wiki/Myers-Briggs_Type_Indicator}).} 
While Julius Caesar is historically renowned for his strategic military prowess, it is unclear how a generic reference to his name translates into a meaningful \persona{} for this game. 
Likewise, Myers-Briggs constructs are widely considered to be pseudo-scientific and meaningless.

\begin{figure}[h!]
\begin{center}
\caption{Response distributions for the 11-20 games with atheoretical \aisub{s}}
\vspace*{-0.12in}
\label{fig:bad_1120}
\begin{subfigure}{\textwidth}  
\refstepcounter{subfigure}\label{fig:compare_bad_1120}
\centering
\begin{overpic}[width=.9\textwidth]{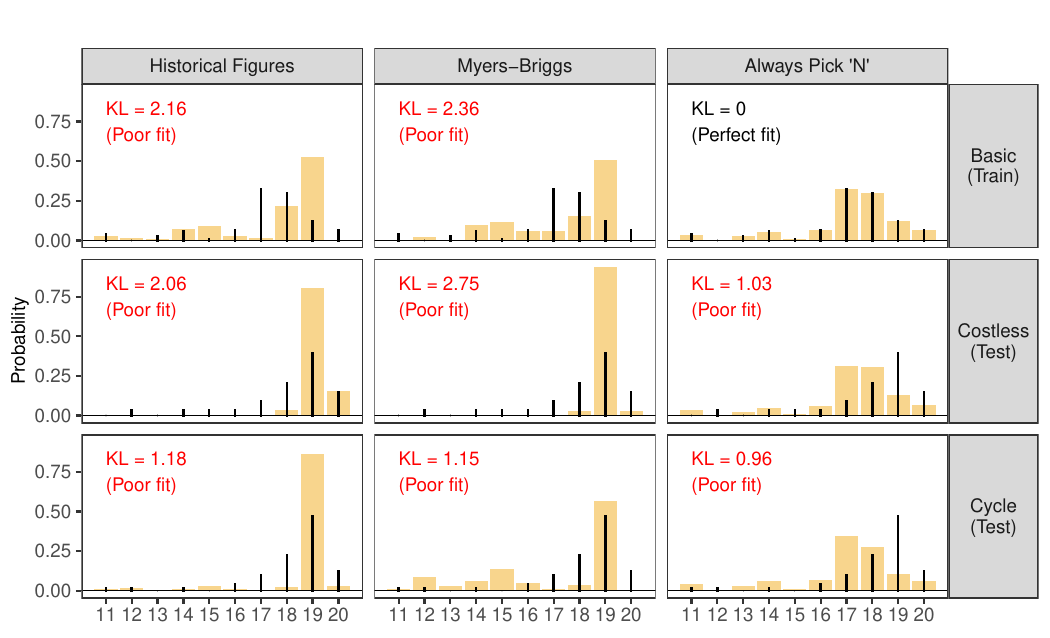}
\put(-1,55){\large\bfseries (a)}
\end{overpic}
\end{subfigure}
\begin{subfigure}{\textwidth}
\refstepcounter{subfigure}\label{fig:distances-11-20}
\centering
\begin{overpic}[width=.9\textwidth]{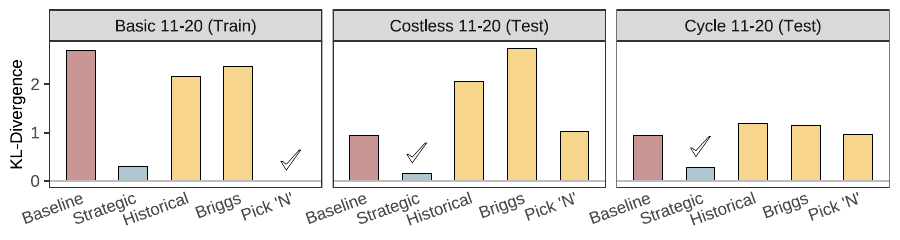}
\put(-1,25){\large\bfseries (b)}
\end{overpic}
\end{subfigure}
\end{center}
\vspace*{-0.2in}
\begin{footnotesize}
\begin{singlespace}
\textit{Notes:} (a)  shows empirical PMFs for three variants of the 11-20 game from \citeauthor{1120Arad2012}, compared to selected atheoretical \aisub{} samples optimized using only the basic version. Rows correspond to game variants, and columns correspond to \aisub{} types, with human data superimposed in black. Historical figures and Myers-Briggs subjects poorly match human distributions across all variants. 
The ``Always Pick `N''' set matches human data perfectly in-sample but fails to generalize.
(b) shows KL divergence between human and AI responses for games from \citeauthor{1120Arad2012}. 
The lowest KL divergence in each panel is indicated by a checkmark. 
Only the strategically-selected \aisub{s} consistently improve over the baseline in all games.
\end{singlespace}
\end{footnotesize}
\end{figure}

Figure~\ref{fig:compare_bad_1120} shows that these two selected samples do not even offer a good in-sample fit.
Each row corresponds to a different variant of the game---the top row is the basic.
The columns represent different \aisub{} types, with the empirical PMFs from AR superimposed in black.
The KL divergence between the distributions in each panel is shown in the top left of each panel.
After optimization, the in-sample KL divergence between the selected \aisub{s} and the humans in AR is $d(P, \hat P_{\bm\theta^*_{Hist}}) =\KlHumanHistDist$ and $d(P, \hat P_{\bm\theta^*_{MB}}) =\KlHumanMBDist$ for the historical figures and Myers-Briggs, respectively.
These are not much better than the baseline $d(P, \hat P_0) =\KlHumanRawDist$ and far worse than the strategic \aisub{s} $d(P, \hat P_{\bm\theta^*}) = \KlHumanOptDist$.

The core issue is that these ``theories'' are bad: historical personas and pseudo-scientific personality types are not causally related to how humans play these games, which means the \persona{s} cannot generate meaningful variation in the dependent variable.
The hypothesis classes are not flexible enough, and the optimization produces a predictive model that severely underfits.
To make an analogy, if $x$ covaries with $y$, then $y = mx + b$ may effectively fit a range of $(x,y)$ pairs, but $y = b$ cannot.

When validated out-of-sample on the costless and cycle variants (Figure~\ref{fig:compare_bad_1120}, bottom rows), these atheoretical personas perform even worse relative to the baseline. 
Both historical figures and Myers-Briggs types simply default to selecting 19 shekels, severely diverging from the shifted human response distributions.

The third atheoretical set (`Always Pick `N''') initially appears successful, achieving a perfect in-sample fit ($d(P, \hat P_{\bm\theta^*_{N}}) = \KlHumanSameNumDist$).
However, this is misleading---such personas offer no flexibility. 
Each agent always selects its assigned integer, between 11 and 20, regardless of setting changes, clearly overfitting to the training data.
Unsurprisingly, when validated on the human data from the new variants, these largely fail to improve over the baseline, merely reproducing their training distribution and failing to capture shifts in human responses (rightmost column of Figure~\ref{fig:compare_bad_1120}). 

Figure~\ref{fig:distances-11-20} succinctly compares these results along with the optimized mixture of strategic \aisub{s} and the baseline, marking the best-performing sample in each setting. 
Only $\bm{\theta}^*$ consistently outperforms the baseline and generalizes across all settings.
All three atheoretical samples are strictly worse than the baseline on both validation games.

These results underscore the importance of theory-driven candidate \persona{s} and validation across related but distinct settings.
We next show that failure to pass validation bodes poorly for predicting responses in new settings.

\subsection{Predicting the new games}
\label{sec:predict_new_games}

We now introduce the four novel strategic games.
To the best of our knowledge, these games are not in the LLM's training corpus, thus providing a stringent testing ground for the \aisub{s} we have explored so far.
Three of the games parallel AR's games in strategic structure but modify the implementation: participants choose between 1 and 10 (rather than 11 and 20) and earn points instead of shekels.
The instructions for the ``basic'' version of this 1-10 game highlight these differences:
\begin{quote}
\vspace{-.5cm}
\singlespacing
\emph{You are going to play a game where you must select a whole number between 1 and 10.
You will receive a number of points equivalent to that number.
After you tell us your number, we will randomly pair you with another player who is also playing this game. They will also have chosen a number between 1 and 10.
If either of you selects a number exactly one less than the other player's number, the player with the lower number will receive an additional 10 points.}
\end{quote}
We adapted the costless and cycle variants similarly.
The fourth ``1-7 game'' introduces an entirely new variant with a restricted choice set (see Appendix~\ref{app:instruct} for the full instructions for all games). 

Of the 1,000 participants we recruited from Prolific, \NTotal{} passed the validation check and were randomly distributed across the four games.
To ensure incentive compatibility, participants were paid \$1 for completion and had a 10\% chance of receiving the dollar value of the points they earned.
We preregistered our complete experimental design, including all prompts for both the baseline and selected \aisub{s}---the latter using only the weights optimized on the basic 11-20 game.
We have $\bm{\theta}^*$, $\bm{\theta}^*_{Hist}$, $\bm{\theta}^*_{MB}$, $\bm{\theta}^*_{N}$, and the baseline play each game.
All \aisub{} responses are elicited using \textsc{Gpt-4o} with the temperature set to 1.
All \aisub{} samples played these games \emph{before} the human subjects' data was collected.

\begin{figure}[h!]
\begin{center}
\caption{Analysis of novel 1-10 games: response distributions and KL divergences}
\vspace*{-0.1in}
\label{fig:all_110}
\begin{subfigure}{\textwidth}  
\refstepcounter{subfigure}\label{fig:all_110_a}
\centering
\begin{overpic}[width=\textwidth]{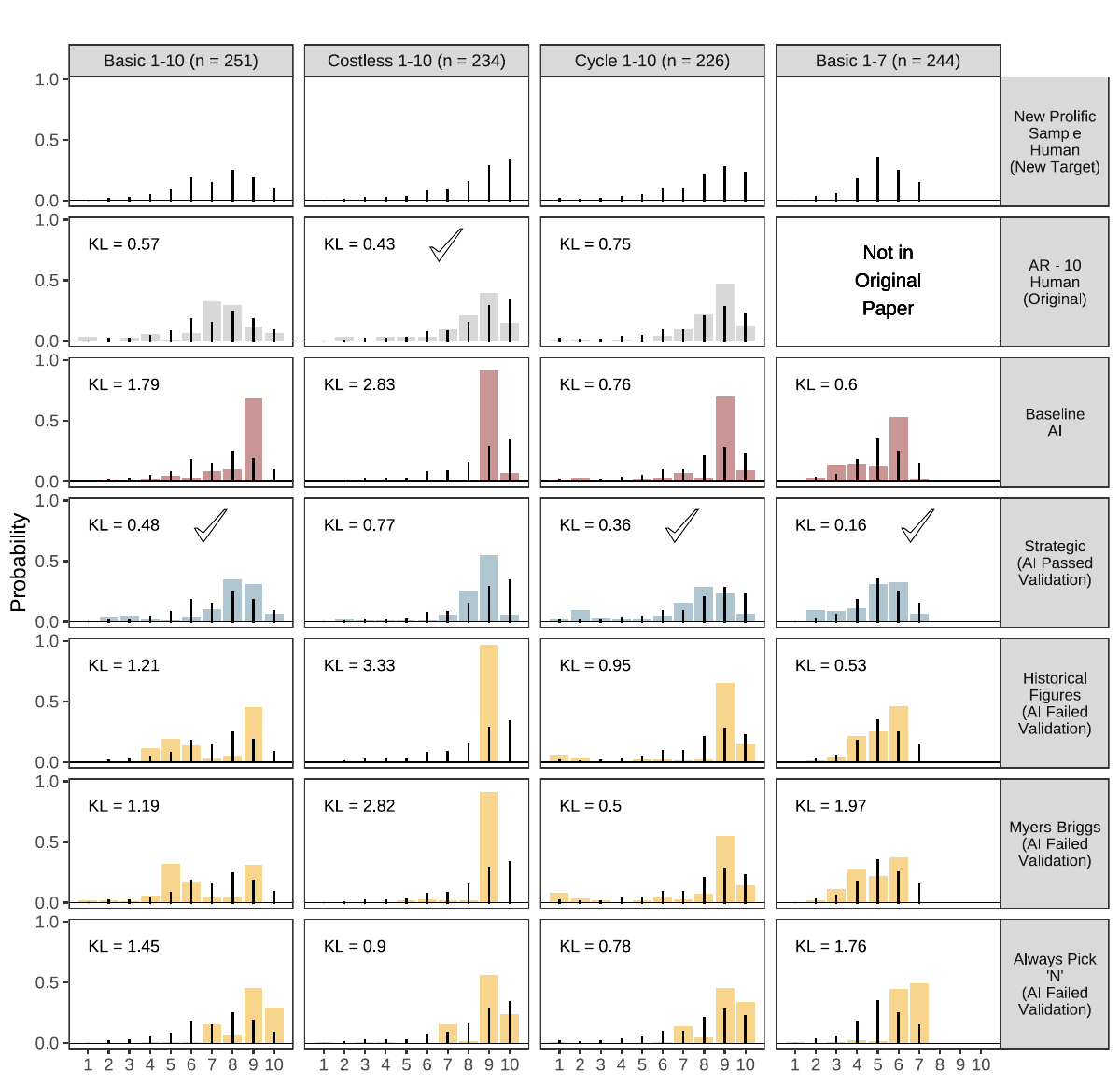}
\put(-1,92){\large\bfseries (a)}
\end{overpic}
\end{subfigure}
\begin{subfigure}{\textwidth}
\refstepcounter{subfigure}\label{fig:all_110_b}
\centering
\begin{overpic}[width=\textwidth]{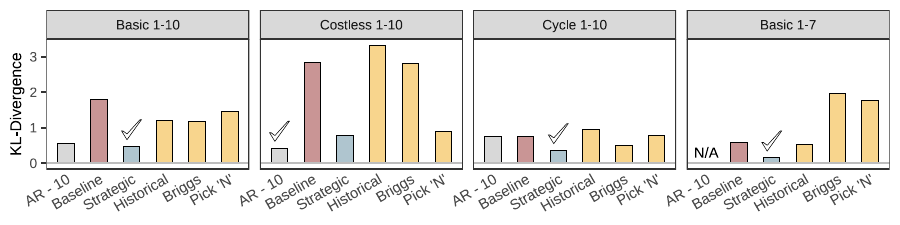}
\put(-1,23){\large\bfseries (b)}
\end{overpic}
\end{subfigure}
\end{center}
\vspace*{-0.2in}
\begin{footnotesize}
\begin{singlespace}
\textit{Notes:} (a) plots human and AI response distributions for the four games. 
Prolific data are in black superimposed on all other distributions. \citeauthor{1120Arad2012} data occupy the second row, shifted down by 10 to fit the 1-10 format.
The remaining rows show the various \aisub{} samples named in the right-hand column. 
(b) Reports the KL divergence between Prolific data and each other distribution.
The minimum in each panel is flagged by a checkmark.
\end{singlespace}
\end{footnotesize}
\end{figure}

Figure~\ref{fig:all_110_a} presents the response distributions of initial play for all subject samples---both human (top two rows) and AI (remaining rows)---across the four novel games. 
The responses from AR are all shifted down by 10 for ease of comparison---20 is now 10, 19 is 9, etc.
Response distributions from the Prolific sample (thin black bars) are superimposed on all other samples.

The Prolific responses differ notably from AR's original results.
Specifically, in the basic 1-10 game, the Prolific sample's distribution is more uniform, with a modal choice at 8 rather than AR's mode at 7; in the costless variant, Prolific responses peak at 10 instead of AR's 9 and 8; and the cycle variant yields a more uniform distribution relative to AR's original sample. 
The newly introduced 1-7 game lacks an analogous comparison in AR, but the modal response is 5, and most other choices are on 4, 6, and 7.

Assuming the difference is not due to sampling variation, the gap between the Prolific data and \citeauthor{1120Arad2012} is somewhat surprising. 
Both sets of games share the same fundamental strategic structure. 
Although the games differ in their Nash equilibria (which we know is not how actual humans play anyway), a fixed set of $k$-level reasoners would play them identically: level-0 would naturally choose the highest number (10), level-1 the second-highest (9), and so on. 
The divergence may therefore reflect setting-specific factors or differences in beliefs about others' reasoning---precisely the kinds of factors that otherwise make theory difficult to apply.
In the end, even highly relevant human data proves to be an imperfect predictor.

Moving down Figure~\ref{fig:all_110_a}, the remaining rows show response distributions from various AI samples.
Whether or not the AI sample successfully passed validation on the costless and cycle games is indicated below the sample name.
The baseline AI (red) generally provides a poor fit to the Prolific data, frequently concentrating choices on 9. 
The notable exception is the 1-7 game, where it disperses responses across several choices.

Critically, the optimized sample of strategically-motivated \aisub{s} robustly generalizes to these novel settings.
This is the only sample of \aisub{s} that was successfully validated on all of the data from AR's original games (the atheoretical samples each failed to improve in at least one of these games).
Figure~\ref{fig:all_110_b} reports the KL divergence between the Prolific data and each AI-generated distribution: the strategic agents consistently outperform the baseline AI by at least \MinPercImproveArad\% in every variant (Appendix Figure~\ref{fig:dist_compare_1_10} shows the same comparisons for several alternative distance metrics). 
Especially notable is the strategic agents' near-perfect alignment in the entirely novel 1-7 game (KL divergence = \KlHumanOptDistSeven). 
Indeed, the strategically motivated AI even outperforms AR's original human data in predicting our Prolific participants' responses in the basic and cycle games.

In stark contrast, atheoretical \aisub{s} fail to generalize.
All arbitrary samples predict the human responses no better than the baseline in at least two of the four games.
The  ``Always Pick `N''' agents are particularly nonsensical as they are solely instructed to select integers between 11 and 20, 
highlighting a severe case of overfitting to a particular data-generating process.

Overall, these results underscore our central claim: carefully identifying theoretically-grounded candidate \persona{s} and validating their predictive utility in related but distinct contexts can substantially enhance predictive accuracy in novel, unseen settings.
Only agents subject to this approach generalized to the novel strategic games. 

\section{External validity in pre-committed novel settings}
\label{sec:ext-valid}

We now turn to making guarantees about inference in novel settings.
This is made possible when we have a pre-committed family of settings from which we can randomly sample.
In particular, this framework allows us to compare the relative accuracy of two models in predicting human responses even for unsampled settings---where we have no previous data---within the same general domain.

The setup is related to that of \cite{Hotz2005PredictingEfficacy}, \cite{Allcott2015Site}, and \cite{Dehejia2019External}, where treatment effects from various ``sites'' are used to evaluate the external validity of a given intervention at the population level.
However, they assume a common underlying intervention---analogous to a single setting in our framework---across all sites.
When there are heterogeneous interventions, only special instances with strong additional assumptions allow for appropriate inference.
The following, which is very similar to the theoretical framework in \cite{andrews2025transfer}, requires no such assumptions.

Let $X=\{x_{1},\dots,x_{M}\}$ denote the pool of candidate settings for which we wish to make predictions.
$P(y|x)$ denotes the true human response distribution for $y \in Y$---the set of allowable responses.
We define a predictive distribution for some flexible model $\theta$---an LLM, an economic model, etc.---as $\hat P_{\theta}(y|x)$.
For a given setting $x$, the expected log-likelihood that the human distribution could have been produced by $\theta$ is
\[
\ell(x;\theta) = \mathbb{E}_{y\sim P(\cdot\mid x)}\bigl[\log\hat P_{\theta}(y|x)\bigr].
\]
Then the comparative predictive power of two models $\theta'$ and $\theta''$ can be measured via
\begin{equation} \label{eq:ratio}
  \Lambda(x) = \ell(x;\theta') - \ell(x;\theta'').
\end{equation}
A positive $\Lambda(x)$ means that $\theta'$ assigns more probability mass to the human responses than $\theta''$ for $x$.
Averaging over all settings in $X$ yields the population estimand
\begin{equation} \label{eq:avg-ratio}
\bar{\Lambda} = \mathbb{E}_{x\sim\pi}\bigl[\Lambda(x)\bigr]
\end{equation}
where $\pi$ is some distribution over $X$.
A positive $\bar{\Lambda}$ is interpreted as evidence that $\theta'$ is, on average, more predictive of human behavior than $\theta''$ across the entire population of $X$.

Suppose we observe a sample $S=\{s_{1},\dots,s_{n}\}\subset X$ and, for each
$s \in S$, independent human responses $y_s=(y_{s,1},\dots,y_{s,m_s})$.
Identification to estimate equations~\ref{eq:ratio} and \ref{eq:avg-ratio} from these samples requires the following assumptions.

\begin{assumption}  \label{ass:rand-settings}
(Unconfounded settings).
The observed settings $S=\{s_{1},\dots,s_{n}\}$ are randomly sampled from distribution $\pi$ with full support over $X$ such that $\pi(x)>0$ for all $x \in X$.
\end{assumption}

\begin{assumption}\label{ass:rand-humans}
(Random assignment and within-setting independence).
Humans are randomly assigned to settings in $S$.
Human responses within a setting are independent draws from $P(y \mid s)$.
\end{assumption}

\begin{assumption} \label{ass:response-support}
(Positivity and finite second moment).
Whenever $P(y\mid x)>0$, then $\hat P_{\theta'}(y\mid x)>0$ and $\hat P_{\theta''}(y\mid x)>0$. 
Moreover, $\mathbb{E}_{x\sim\pi}\bigl\{\mathbb{E}_{y\sim P(\cdot\mid x)}\bigl[(\log\hat P_{\theta'}(y\mid x) - \log\hat P_{\theta''}(y\mid x))^{2}\bigr]\bigr\} < \infty$.
\end{assumption}

The first two assumptions are basically identical to the assumptions of \emph{unconfounded location} and \emph{random assignment} in \citeauthor{Hotz2005PredictingEfficacy}.
They are also very similar in spirit to Assumption 1 in \citeauthor{andrews2025transfer}.
The first part of Assumption~\ref{ass:response-support} is similar to the covariate overlap assumption in causal inference; without it, some observed responses would have $\log 0$.
The second portion of Assumption~\ref{ass:response-support} is a standard finite second-moment condition.

For every setting $s \in S$, define the sample analogue to equation~\ref{eq:ratio} as:
\begin{equation} \label{eq:sample-ratio}
\hat\Lambda_{s}
\;=\;
\frac1{m_s}
\sum_{j=1}^{m_s}
\Bigl[\log\hat P_{\theta'}(y_{s,j}\mid s)
-\log\hat P_{\theta''}(y_{s,j}\mid s)\Bigr].
\end{equation}
Aggregate across settings to produce the sample analogue to equation~\ref{eq:avg-ratio}:
\begin{equation} \label{eq:sample-avg-ratio}
\bar\Lambda_{S}
\;=\;
\frac1n\sum_{s \in S}\hat\Lambda_{s}.
\end{equation}

\begin{proposition}[Unbiasedness and asymptotic normality]
\label{prop:sample-valid}
Suppose Assumptions~\ref{ass:rand-settings}--\ref{ass:response-support} hold.
Then
\[
\mathbb{E}[\bar{\Lambda}_{S}] = \bar{\Lambda}
\quad \text{and} \quad
\sqrt{n}\,(\bar{\Lambda}_{S}-\bar{\Lambda})
\xrightarrow{d} \mathcal{N}\bigl(0,\sigma^{2}\bigr),
\]
where $
\sigma^{2}
= \operatorname{Var}_{x\sim\pi}\!\bigl[\Lambda(x)\bigr]
\;+\;
\mathbb{E}_{x\sim\pi}\!\Bigl[\tfrac{1}{m_x}\,V_x\Bigr]$ and 
$V_x
= \operatorname{Var}_{y \sim P(\cdot \mid x)}
\bigl[\log\hat P_{\theta'}(y \mid x)
-\log\hat P_{\theta''}( y\mid x)\bigr].$\footnote{\emph{Proof.} (Unbiasedness).
For any setting $s$, Assumption~\ref{ass:rand-humans} implies $\mathbb{E}\bigl[\hat\Lambda_s \mid s\bigr]=\Lambda(s).$
Hence $\mathbb{E}\bigl[\bar\Lambda_S \mid S\bigr] = \frac{1}{n}\sum_{s \in S}\Lambda(s).$
Taking expectation over the i.i.d.\ draw of the settings (Assumption~\ref{ass:rand-settings}) yields $\mathbb{E}[\bar\Lambda_S]=\bar\Lambda.$
(Asymptotic normality).
The random variables $\{\hat\Lambda_s\}_{s \in S}$ are i.i.d. across settings with finite variance $
\operatorname{Var}(\hat\Lambda_s)
= \operatorname{Var}_{x\sim\pi}[\Lambda(x)]
+ \mathbb{E}_{x\sim\pi}
\bigl[{\textstyle\frac{1}{m_x}}\,V_x\bigr]
\;<\;\infty,$
where the decomposition follows from the law of total variance and the independence of human draws within each setting.
Because the moment condition in Assumption~\ref{ass:response-support} guarantees $V_x<\infty$, the central-limit theorem gives $\sqrt{n}\bigl(\bar\Lambda_S-\bar\Lambda\bigr)
\;\xrightarrow{d}\;
\mathcal N\bigl(0,\sigma^{2}\bigr).$
(Regularity).
By Assumption~\ref{ass:response-support},
$\log\hat P_{\theta'}(y\mid x)$ and $\log\hat P_{\theta''}(y\mid x)$ are finite, so all moments used above exist.}
\end{proposition}

Notably, this does not rely on a large sample size of humans for any particular setting.
The asymptotic variance in Proposition~\ref{prop:sample-valid} decomposes into two conceptually distinct parts.
The first term, $\operatorname{Var}_{x\sim\pi}\bigl[\Lambda(x)\bigr]$, reflects heterogeneity in model performance across settings.
The second term, $\mathbb{E}_{x\sim\pi}\bigl[\frac{1}{m_x}V_x\bigr]$, is sampling noise that arises because we estimate $\Lambda(x)$ with a finite number $m_x$ of human draws. 
As long as $m_x\ge 1$, this component is finite. 
Consequently, once a few human observations are obtained per setting, the precision of $\bar\Lambda_S$ is governed primarily by $n$ 
because each setting contributes only a single observation $\hat\Lambda_s$.
This also means there is no within-cluster correlation left to adjust for, so the standard sample variance estimator ($\hat{\sigma}^{2}\;=\;\frac{1}{n-1}\sum_{s\in S}\!\bigl(\hat{\Lambda}_{s}-\bar{\Lambda}_{S}\bigr)^{2}$) across settings already yields valid standard errors without needing to adjust for clustering.
As such, standard $z$-tests or Wald confidence intervals follow immediately when estimating if $\bar\Lambda_S$ is significantly different from zero.

Crucially, this construction imposes \emph{no assumption} that all settings share a single data-generating process.
The settings can be an arbitrarily eclectic mixture---public goods games, dictator games, or entirely unrelated tasks.   
Inference remains valid even when the model's performance differs sharply across sub-domains; the variability of estimates widens (or narrows) in proportion to the observed heterogeneity.

The remainder of this section is devoted to implementing the above framework on a pre-committed family of strategic games.
This set comprises \TotalGames{} novel and unique permutations of \citeauthor{1120Arad2012}'s money request game.
We randomly sample \NGamesS{} of these games for human subjects and \aisub{s} to play in a preregistered experiment.
We use this data to estimate the relative capacity of different \aisub{s} to predict human responses at scale.
In particular, we return to the strategic sample of optimized level-$k$ \aisub{s} presented in Table~\ref{tab:persona} from Section~\ref{sec:predict_new_AR}.
We compare these agents' ability to predict human responses across the \NGamesS{} games to the baseline AI, a cognitive hierarchy model, and symmetric Nash equilibria.
Because the games (and human subjects) are randomly sampled from the population according to a known distribution, confidence intervals over the comparisons are valid over the \TotalGames{} games.

\subsection{A pre-committed family of strategic games}

The pre-committed family of games generalizes the original 11-20 money request game by parameterizing it into six independent variable components.\footnote{\cite{alsobay2025integrative} similarly generate various public goods games and \cite{zhu2025capturing} do so for over 2,000 2-by-2 strategic games.}
Each symmetric game preserves the core idea behind the original: two players simultaneously select a whole number between two bounds, earning guaranteed points based on their individual choice plus a potential bonus determined by both players' choices.
The six parameters---lower bound, upper bound, gap to achieve the bonus, bonus size, rule to award guaranteed points, and bonus rule---are detailed in Table~\ref{tab:game_parameters}.
The table's upper portion enumerates the possible values for the first five parameters, while the bottom section presents the eleven possible bonus rules. 

\begin{table}[h!]
\begin{center}
\caption{Game parameters and possible values} 
\label{tab:game_parameters}
\scriptsize
\setlength{\tabcolsep}{7pt}
\begin{tabular*}{\textwidth}{@{\extracolsep{\fill}}p{1.8cm}p{9cm}p{4cm}}
\toprule
\textbf{Parameter} & \textbf{Possible Values} & \textbf{Description} \\
\midrule
\emph{Lower Bound} & The minimum number players can select & $\{1, 2, \ldots, 20\}$ \\
\emph{Upper Bound} & lower bound + $\{5, 6, \ldots, 20\}$ & Max number players can select \\
\emph{Bonus Size} & Points awarded when bonus condition is met & $\{1, 2, \ldots, 20\}$ \\
\emph{Gap} & Difference parameter used in certain bonus rules & $\{1, 2, 3, 4\}$ \\
\emph{Points Rule} & Rules to award guaranteed points (\# is number participants select) & \{\# - 2, \# - 1, \#, \# + 1, \# + 2, costless - 2\} \\
\midrule
\multicolumn{1}{@{}l}{\textbf{Bonus Rule}} & \multicolumn{1}{l}{\textbf{Bonus awarded to player when...}} & \multicolumn{1}{l}{\textbf{Mutual vs. Competitive}} \\
\midrule
\emph{Gap Low} & they select exactly \{\textit{gap}\} less than the opponent & Competitive \\
\emph{Gap High} & they select exactly \{\textit{gap}\}  more than the opponent & Competitive \\
\emph{More Than} & they select a number \{\textit{gap}\} more than the opponent's number & Competitive \\
\emph{Gap Abs.} & the absolute difference from the opponent's number equals \{\textit{gap}\} & Mutual\\
\emph{Equal} & they select the same number as the opponent & Mutual\\
\emph{Unequal} & they select a different number than the opponent & Mutual\\
\emph{Sum Even} & the sum of both numbers is even & Mutual\\
\emph{Sum Odd} & the sum of both numbers is odd & Mutual\\
\emph{Coord. Low} & both players select the lower bound & Mutual\\
\emph{Sum Upper} & the sum of both numbers equals the upper bound & Mutual (or not achievable)\\
\emph{Less Upper} & the sum of both numbers is less than upper bound & Mutual (or not achievable)\\
\bottomrule
\end{tabular*}
\end{center}
\begin{footnotesize}
\begin{singlespace}
\vspace*{-0.05in}
\emph{Notes:} This table shows the possible values for the six parameters of the pre-committed family of games.
The rightmost column indicates whether the bonus is mutually achievable (both players can receive it together) or competitive (only one player can receive it, if at all).
The ``Sum Upper'' and ``Less Upper'' bonus rules are not achievable when the lower bound exceeds half the upper bound. 
The naive Cartesian product of all parameter values yields 
$20 \times 16 \times 4 \times 20 \times 6 \times 11 = 1{,}689{,}600$ combinations. 
However, seven of the eleven bonus rules do not use the \textit{gap} parameter; for these rules, varying the gap value produces mechanically identical games. 
Collapsing such duplicates leaves \TotalGames{} unique games in the final population $X$.
\end{singlespace}
\end{footnotesize}
\end{table}

To illustrate how these parameters translate into actual games, consider the following example.
If the lower bound is 5, upper bound is 14, gap is 6, bonus size is 10, points rule is \# - 2, and bonus rule is the Gap Abs. rule, participants see:
\begin{quote}
\vspace{-.5cm}
\singlespacing
\emph{You are going to play a game where you must select a whole number between \textbf{5 and 14}. A player will receive a number of points equivalent to \textbf{that number minus two}. After you tell us your number, we will randomly pair you with another Prolific worker who is also playing this same game. They will also have chosen a number between 5 and 14. Both players will receive \textbf{an additional 10 points} if their requested numbers \textbf{differ from each other by exactly 6}. What number would you request?}
\end{quote}

The full factorial product of this parameterization yields $1{,}689{,}600$ games.
However, many of these games are mechanically identical because seven of the bonus rules do not use the \textit{gap} parameter.
Accounting for these duplicates, we have \TotalGames{} unique games in total---the pool of candidate settings $X$.
This family includes the original 11-20 game as a special case (lower bound 11, upper bound 20, bonus 20, gap 1, points rule \#, first bonus rule).
Besides this and the other games in Section~\ref{sec:predict_new_AR}, to the best of our knowledge, all of these games are novel.
They cannot be found in \textsc{Gpt-4o}'s training data---the model we use to generate AI responses.

Notably, the games exhibit dramatic variation in strategic difficulty. 
With bonus rules such as ``Unequal,'' the vast majority of strategy profiles lead to the bonus.
Conversely, for rules like ``Sum Upper'' or ``Less Upper'' bonuses are sometimes unattainable (when the lower bound exceeds half the upper bound). 
The games also vary in their incentive alignment between players.
As indicated in the rightmost column of the bottom panel of Table~\ref{tab:game_parameters}, some bonus rules are mutually achievable, while others are competitive---only one player can receive the bonus.
This heterogeneity creates a particularly stringent test of agents' predictive power, as successful generalization demands flexibility along multiple dimensions.

To construct $S$, we randomly sampled \NGamesS{} games from $X$. 
The intended design was uniform sampling for $\pi$ across all \TotalGames{} unique games. 
A very minor miscalculation in the deduplication process caused small deviations from uniformity: the seven bonus rules that do not use the \textit{gap} parameter were each sampled with probability $\approx 0.086$, while the four gap-using bonus rules were each sampled with probability $\approx 0.010$.\footnote{This miscalculation was only noticed after the experiment. As such, the preregistration (aspredicted \#241394) states that we were sampling from a larger number of games. 
The only difference is that the correct number is \TotalGames{}. The random sample of \NGamesS{} was still chosen before the experiment and is available here: \url{www.expectedparrot.com/content/db984e24-2810-4b21-be4e-91efde378e21}---the same link given in the preregistration.}
For points rules, the ``costless'' variant was sampled with probability $\approx 0.095$, and each of the remaining rules with probability $\approx 0.18$. 
All other parameters were sampled uniformly. 
Consequently, the estimand in this section is technically the expected relative predictive power of the models over the \TotalGames{} games under this slightly non-uniform distribution.
However, robustness checks will later show that the relative predictive power of the models is not particularly sensitive to the points rule or bonus rule, indicating that this minor departure from uniformity has no substantive effect on our conclusions.
The results are therefore likely to hold for many distributions $\pi$ over $X$.

\subsection{Eliciting \aisub{} responses}

We generate AI responses for each of the \NGamesS{} games in the set $S$ using two distinct samples of \aisub{s}. 
As a baseline sample, we prompt \textsc{Gpt-4o} at temperature 1 to independently play each game 100 times, without providing any additional instructions. 
For the optimized strategic sample, we use the same 10 \persona{s} from Table~\ref{tab:persona}, which were optimized using human experimental data from \citeauthor{1120Arad2012}. 
To generate this strategic sample, we proportionally scale the optimized persona weights to create a total of 100 \aisub{s}, each of which plays each game exactly once using \textsc{Gpt-4o} (the ``strategic'' sample of \aisub{s} hereinafter).\footnote{A very small fraction (less than 0.1\%) of the \aisub{} responses were invalid due to stochasticity inherent to the LLM at temperature 1. Following our preregistered analysis plan, we discard these invalid responses without resampling.}

This procedure produces an empirical distribution for both the strategic level-$k$ sample $\hat P_{\bm\theta^*}$ and the baseline AI $\hat P_{0}$ for every game $s \in S$.
These samples correspond exactly to those described in Section~\ref{sec:predict_new_AR} ($\hat P_{\bm\theta^*}$ and $\hat P_{0}$, respectively, in the notation of that section), differing only in the number of agents---here, each distribution is generated with 100 agents rather than 1,000.
To be clear, these agents were constructed \emph{months} before they played any of the \NGamesS{} games.
In total, the elicitation procedure produces approximately 300,000 individual \aisub{} responses.

\subsection{Predictive benchmarks}
\label{sec:nash-selector}

A key limitation of our statistical framework is that comparing the predictive power of different AI simulations provides no absolute benchmark for how well these samples predict human responses in general. 
Thus, we require a suitable theoretical or statistical benchmark for a more comprehensive analysis.
However, due to the scale and heterogeneity of our games, several appealing benchmarks are impractical.

Ideally, we would apply the standard level-$k$ model from Section~\ref{sec:predict_new_AR} to generate predictive distributions across these games. 
Unfortunately, no existing mechanical method reliably identifies which choices correspond to specific levels of reasoning across such diverse contexts. 
For example, the ``obvious'' choice for a level-$0$ player is not always the highest number---particularly when bonuses are large or the bonus rule involves selecting number 11. 
Consequently, higher-level reasoning does not follow the intuitive progression featured by the original game in \citeauthor{1120Arad2012}.

Another seemingly attractive alternative would be to follow the approaches of \cite{FudenbergLiang2019PredictPlay}, \cite{hirasawa2022using}, or \cite{andrews2025transfer}, which involve training a bespoke supervised machine learning model on past game data to predict responses. 
However, this method is infeasible for our purposes because of two problems. 
First, it relies heavily on having access to a large and representative dataset, which we currently do not possess.\footnote{The combined approach of \cite{zhu2025capturing}---using machine learning to flexibly estimate parameters in a well-defined behavioral model and then using that model to predict responses---is infeasible for the same reason.} 
Second, as discussed in Section~\ref{sec:approach}, traditional machine learning models (e.g., regression, decision trees, or generic neural networks) are not well-suited for predicting human behavior in novel settings because they cannot adapt to structural differences across settings.
Even with substantial representative data, if a game had even one new parameter value (e.g., a new bonus rule or an upper bound above the highest from the training data), the model would need to be retrained from scratch.

We employ two complementary benchmarks that each address a different part of the above limitations.
First, we use the cognitive hierarchy model of \cite{Camerer2004Cognitive}.
Throughout the economics literature, it is one of the most empirically accurate predictors of human responses in one-shot strategic games.
It can also produce predictions for all the structurally distinct games in $S$, but does require some prior information, which we discuss in the next subsection.
Second, we use Nash equilibrium with Harsanyi-Selten selection, which requires no empirical calibration, but, along with most equilibrium concepts, has not been particularly successful at predicting initial human play for static games in the literature. 

The challenge of finding an ideal benchmark actually highlights a unique strength of \aisub{s}: they can generate plausibly accurate predictions in virtually any setting without extensive training data or pre-specified behavioral parameters.
To our knowledge, no other well-known benchmarks are both likely to have predictive accuracy and flexible enough to be applied to such a wide range of structurally distinct games.

\subsubsection{Cognitive hierarchy model as a benchmark}

The cognitive hierarchy model of \cite{Camerer2004Cognitive} is a well-established behavioral model for predicting human behavior in strategic games. 
It has a similar structure to the level-$k$ model from \citeauthor{1120Arad2012}, assuming a distribution of players with different reasoning levels from zero through $k$. 
However, there are three key differences. 
First, rather than assuming level-$0$ players choose the highest or most obvious choice, it assumes they choose uniformly at random. 
Second, level-$k$ players best respond to the mixture of all lower-level players (levels $0$ through $k-1$), not just level-$(k-1)$. 
Specifically, each level chooses a pure strategy best response, and if multiple pure strategies are equally optimal, players at that level mix uniformly between them. 
For example, in the original 11-20 money request game, the cognitive hierarchy model assumes level-$0$ players choose uniformly at random, level-$1$ players best respond to this uniform distribution, while level-$2$ players best respond to the weighted mixture of level-$0$ and level-$1$ players, and so on.
Third, the number of level-$k$ players is assumed to follow a Poisson distribution ($f(k) = \frac{\tau^k e^{-\tau}}{k!}$).

The key limitation of this model in our setup---shared by the aforementioned supervised machine learning approaches---is that it requires a pre-specified value for $\tau$ to parameterize the distribution of different-level players.
Since our goal is to make predictions in the 1,500 games where we have no prior human data, we have no compelling value to choose for $\tau$ other than that from the literature.
In their meta-analytic validation across many strategic games and subject pools, \cite{Camerer2004Cognitive} find that $\tau = 1.5$ is often an accurate predictor of human response distributions.
As such, we adopt $\tau = 1.5$ as our ex ante parameter value.
This estimate implies the modal player is level-$1$, with 90\% of mass between levels $0$ and $3$.

\begin{figure}[h!]
\begin{center}
\begin{minipage}{\textwidth}
\caption{Predictive distributions from the cognitive hierarchy model for all 1,500 games in $S$} 
\label{fig:ch_plot}
\vspace*{-0.1in}
\includegraphics[width=\textwidth]{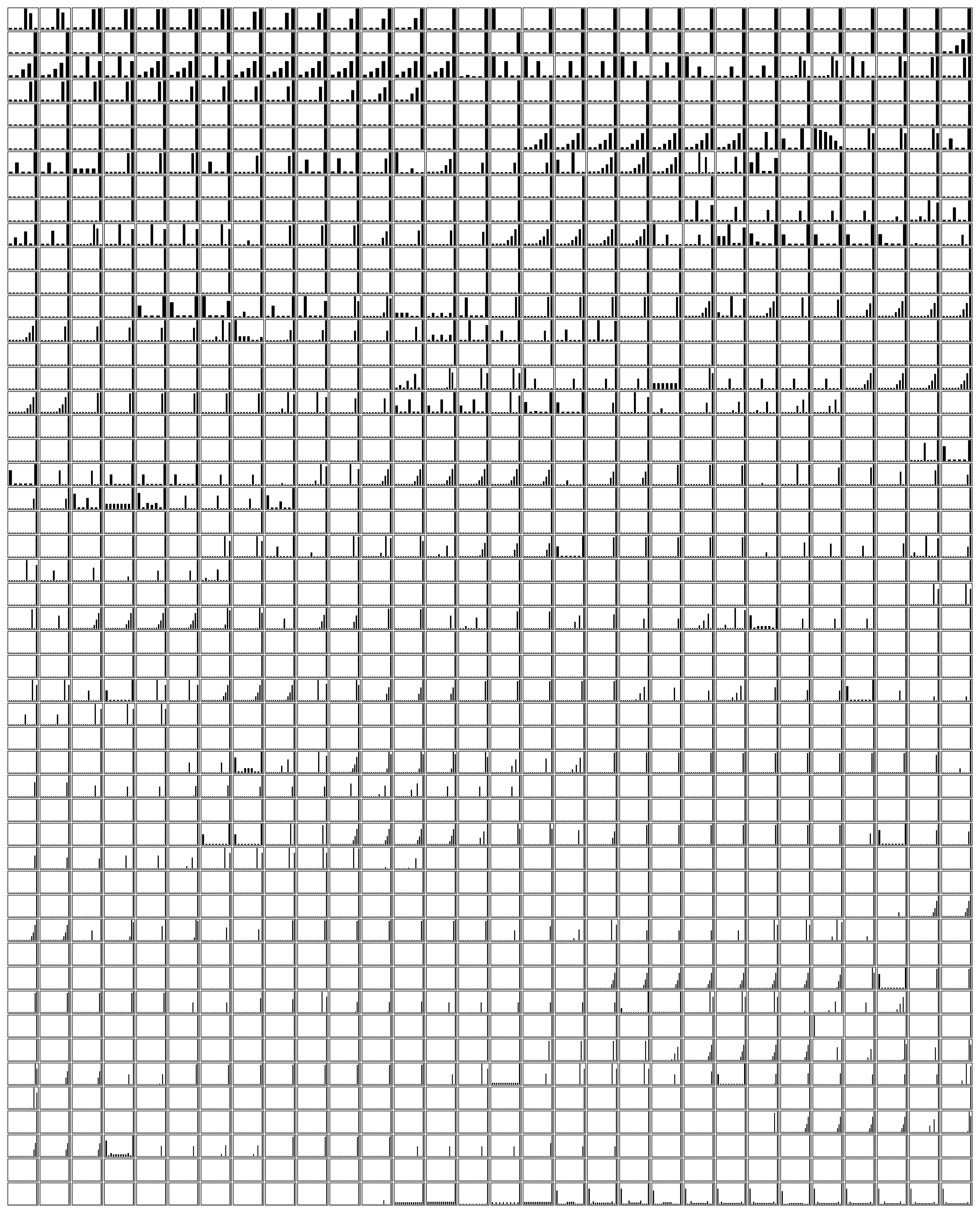} 
\end{minipage}
\end{center}
\begin{footnotesize}
\begin{singlespace}
\vspace*{-0.05in}
\emph{Notes:} This figure shows the predictive distributions of \cite{Camerer2004Cognitive} Poisson cognitive hierarchy model ($\tau = 1.5$) across all 1,500 sampled games in $S$, with each panel corresponding to one game.
\end{singlespace}
\end{footnotesize}
\end{figure}

With this as the distribution over reasoning levels, we mechanically calculate the cognitive hierarchy model's predictive distribution for each game in $S$.
Since this is mechanical for a given game, it does not affect statistical inference for $\bar\Lambda_{S}$.
Figure~\ref{fig:ch_plot} illustrates the diversity of predictive distributions across games.
Each panel corresponds to a predictive PMF for one of the 1,500 games, which are ordered top-left to bottom-right by their variance.
The x-axes are scaled freely, so games with different action spaces (e.g., 5, 11, or 20 options) span the same horizontal width. 
The y-axes are also scaled freely, ensuring that distributions with small probabilities remain visible.

From the modest natural-language permutations in Table~\ref{tab:game_parameters}, we obtain a striking range of predictive distributions from the cognitive hierarchy model.
Some spread probability across many options, while others concentrate only on the extremes—the highest and lowest actions.
A few resemble uniform distributions, others exhibit sharp spikes on a small set of actions, and still others increase monotonically with the action number.

\subsubsection{Harsanyi-Selten Nash equilibria as a benchmark}

The second benchmark is symmetric Nash equilibria.
Unlike the cognitive hierarchy model, this solution concept does not require any previous data to generate predictions.
It is also flexible: symmetric Nash equilibria exist for any symmetric two-player game with a finite number of actions \citep{nash1951noncooperative}.
They can be computed systematically across all game types in our dataset.
Furthermore, in our setting, all games are played independently by participants (AI and human), making symmetry a natural assumption that suggests a ``consistent common belief'' across the population \citep{STAHL1994}.

Since many games have multiple symmetric Nash equilibria, we require a systematic method to select a single equilibrium prediction. 
Indeed, one game in $S$ has $\MaxSymmNash$ symmetric equilibria. 
Unfortunately, there is no universally agreed-upon criterion for selecting the ``optimal'' symmetric equilibrium across all games \citep{camerer2003behavioral, tadelis2013game}.

We employ a slightly modified version of the equilibrium selection procedure developed by \citet{HarsanyiSelten1988}, which provides a principled approach grounded in stability and focal-point considerations. 
It guarantees the selection of a unique Nash equilibrium for every game in our dataset and can be slightly modified to ensure symmetry. 
The procedure also prioritizes equilibria that are stable in the sense of \citet{schelling1960strategy}, specifically favoring equilibria that are either payoff dominant (maximizing joint welfare) or risk dominant (robust to strategic uncertainty). 
This is appealing because many of the games---particularly those with bonus rules ``Equal'' or ``Coord. Low''---often have clear equilibria that are both payoff and risk dominant, which humans often choose \citep{camerer2003behavioral}. 

The Harsanyi-Selten procedure operates through a multi-stage filtering process, progressively narrowing the set of candidate equilibria. 
The procedure first identifies all Nash equilibria, then applies filters based on Pareto efficiency, symmetry requirements, and risk dominance, and finally employs tracing methods to resolve any remaining ties. 
See the pseudocode in Appendix~\ref{app:nash-selector} for details on the implementation.

We first use open-source software \citep{gambit2025} to calculate all Nash equilibria for the \NGamesS{} games in set $S$. 
We then apply the Harsanyi-Selten procedure to each game's set of equilibria, producing a single equilibrium distribution for \GamesConverged{} of the games.\footnote{
In fewer than 1\% of games, degeneracy issues prevented the code from converging.
Following our preregistration, we discard these games in our analysis when comparing the equilibria to the strategic \aisub{s}.}
Of these \GamesConverged{} games, \CountUniqueEq{} have unique symmetric equilibria.
The selection procedure was unnecessary in these cases. 
Among the remaining games with multiple symmetric equilibria, the procedure selects payoff-dominant equilibria---those Pareto superior to all alternatives---in \CountPayoffDomEq{} games, and risk-dominant equilibria in \CountRiskDomEq{} games. 
Finally, $\PropSelectedPure$\% of the equilibria selected by the Harsanyi-Selten procedure have pure strategies, with the remainder employing mixed strategies.

Similar to the predictions from the cognitive hierarchy model, the equilibrium distributions are highly diverse across games.
This can be seen in Figure~\ref{fig:eq_plot} in the appendix, which is of the same format as Figure~\ref{fig:ch_plot}.

\subsection{Eliciting human responses}

We collected human data from a sample of \NParticipantS{} Prolific workers using a custom online survey platform, which allows us to generate any game programmatically \citep{Horton2024EDSL}.
This human data supplies $y_{s}$ for each game $s \in S$, with which we can then estimate the relative predictive power of the different AI models and benchmarks.
The entire experimental design, all \aisub{} responses, and the statistical analysis in this section were preregistered before collecting the human subjects' data.\footnote{The sole exploratory analysis outside our preregistration is the comparison to the cognitive hierarchy model, added in response to suggestions we received after posting the paper online.}
Each Prolific worker was randomly assigned one of the \NGamesS{} sampled games such that each game had approximately three human players.

The survey flow began with a very simple attention check.
Participants were then shown the rules of their assigned game.
This was followed by a comprehension check, which asked participants to calculate the correct number of points for a hypothetical outcome of their game.
They were then asked to make their strategic choice.
Participants received a fixed payment of \$0.50 for completing the survey.
To align incentives with the game structure, 1\% of participants were randomly awarded performance-based bonus payments, with each point earned in their assigned game converted to US dollars at a 1:1 rate.
These bonuses were substantial, averaging \$23 across those who received them, with one participant earning \$48.

After a preregistered filtering based on the first attention check, removing participants who timed out on our platform, or those who selected a final number outside of the range of their game or did not select a whole number, our final sample size was \SampleSizeHuman{}, each playing one of the  \SampleSizeGame{} unique games.
These \SampleSizeGame{} games comprise the sample $S$ we use for analysis.

\subsection{Estimation}
\label{sec:estimation}

We estimate the relative predictive power of the different AI samples and the theoretical benchmark in three steps.
These are: (i) construct smoothed predictive distributions for each benchmark (besides the cognitive hierarchy model) in every sampled game; 
(ii) evaluate the strategic sample of \aisub{s} compared to each other model with per-game log-likelihoods and their paired differences;
(iii) attach sampling-error bounds that are externally valid for the full population of nearly one million games.
We address these steps in turn.

Many of the Harsanyi-Selten equilibria---mainly those pure strategies---and to a lesser extent the samples of \aisub{} place zero probability on strategies that humans sometimes take.
For these models, the product of all likelihoods would be zero, making the log-likelihood $-\infty$ and violating Assumption~\ref{ass:response-support}.
Dropping these games would heavily bias the results away from any pure strategy equilibria or the samples of \aisub{} where the agents mostly pick a single action---even when most people do select that action. 

To address this, we follow the convention in game theory where players are assumed to follow the equilibrium strategy with probability $1-\varepsilon$ and choose uniformly at random the remaining $\varepsilon$ of the time.
Mathematically, this is: $\tilde P_{\theta}(y\mid s) = (1-\varepsilon)\,\hat P_{\theta}(y\mid s) + \frac{\varepsilon}{K_{s}}$, where $K_{s}$ is the number of feasible actions in game~$s$.
Setting $\varepsilon=0.2$ implies that players follow their model 80\% of the time and choose uniformly at random the remaining 20\%.
We apply this smoothing to all models except the cognitive hierarchy model, which already incorporates random play and full support through its level-0 players (approximately 22\% when $\tau = 1.5$)---a feature specifically designed to capture this behavior \citep{Camerer2004Cognitive}.
Further smoothing is therefore unnecessary for this benchmark.

This additional smoothing is not without both empirical and theoretical support \citep{mckelvey1992experimental,mckelvey1995quantal}.
In the original 11-20 money request game, \citeauthor{1120Arad2012} estimate that 32\% of participants choose uniformly at random in their best fitting model.
In \citeauthor{STAHL1994}, at most one out of 40 participants is best explained by random choosing.
Across these and other studies, the evidence suggests that between 20-30\% random play captures observed behavior well.
We therefore report all main results with $\varepsilon = 0.2$ but provide robustness checks with $\varepsilon \in \{0.05, 0.1, 0.3\}$ in Appendix~\ref{app:figs}.

For each game $s$, we calculate Equation~\ref{eq:sample-ratio} five times.
In all five cases, $\theta'$---the numerator of the log-likelihood ratio $\hat \Lambda_s$---is the predictive distribution for the strategic sample of \aisub{s}.
The denominator $\theta''$ is one of five benchmarks: (i) the baseline AI, (ii) the Poisson cognitive hierarchy model (with $\tau = 1.5$), (iii) the Harsanyi-Selten Nash equilibria, (iv) a uniform distribution over all possible strategies, or (v) a randomly selected pure strategy distribution.
To be clear, this means all comparisons are made with respect to the strategic sample and $\hat \Lambda_s > 0$ implies the strategic sample is the best predictor of initial play for game $s$.
We then take the average across the games to estimate $\bar \Lambda_S$ from Equation~\ref{eq:sample-avg-ratio} for each benchmark.

Proposition \ref{prop:sample-valid} holds for each of these sample averages.
Games were drawn via a known and fully supported distribution over the population $X$ (Assumption \ref{ass:rand-settings}).
Human respondents were randomly assigned these games, and their answers were independent (Assumption \ref{ass:rand-humans}).
Smoothing guarantees the first part of Assumption \ref{ass:response-support}, and the possible human responses form bounded, discrete distributions---so the required second-moment condition is satisfied.
This means that confidence intervals must cover appropriately, and the results are externally valid for the population of all \TotalGames{} games.

We report bootstrapped confidence intervals for the five $\bar\Lambda_S$ values and provide robustness checks with Wilcoxon and random-sign permutation tests.
We also report the proportion of games for which the strategic \aisub{} is the best predictor---i.e. $\sum_{s \in S} \mathbf 1\{\hat \Lambda_s > 0\} / |S|$---with its exact Clopper-Pearson 95\% interval.
Such intervals are valid following a nearly identical argument leading to Proposition~\ref{prop:sample-valid}.

\subsection{Results}

\begin{figure}[h!]
\begin{center}
\begin{minipage}{\textwidth}
\caption{Predictive power of strategic \aisub{s} compared to other models ($\varepsilon = 0.2$)} 
\label{fig:ll_ratios}
\vspace*{-0.15in}
\includegraphics[width=\textwidth]{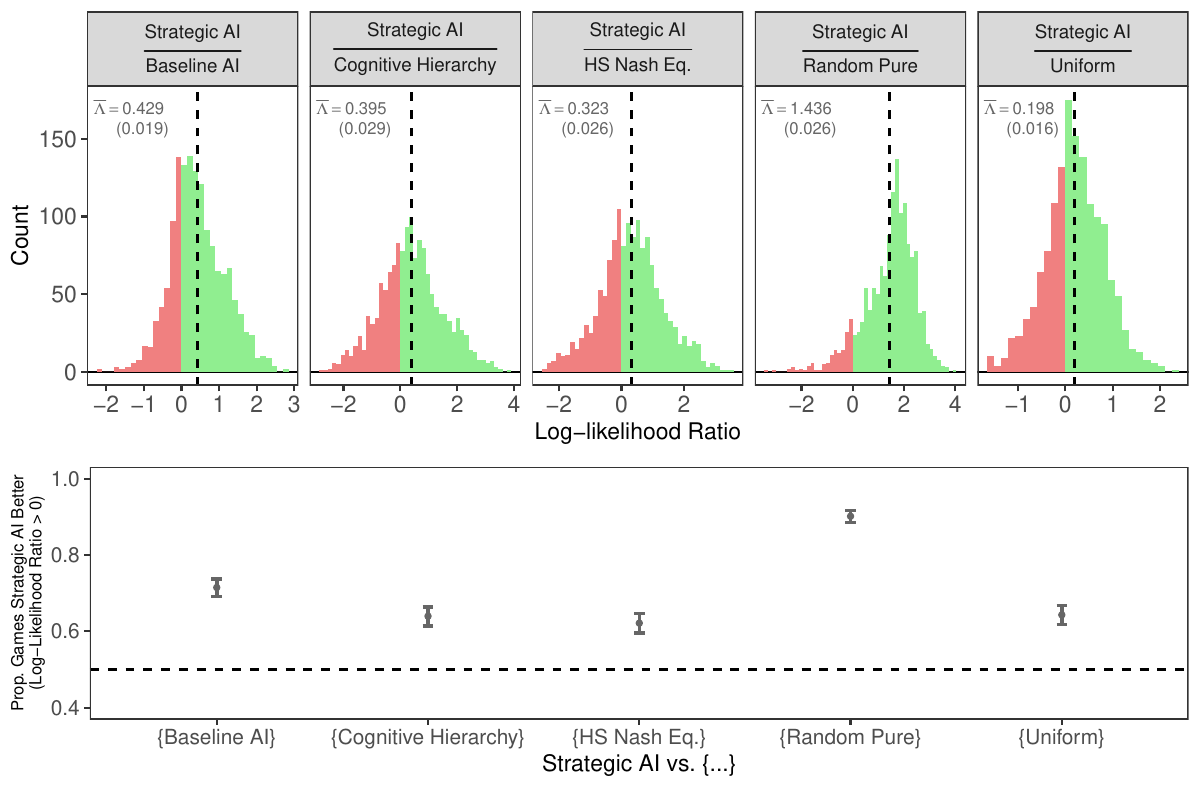} 
\end{minipage}
\end{center}
\begin{footnotesize}
\begin{singlespace}
\vspace*{-0.05in}
\emph{Notes:} The top panel shows the distribution of the log-likelihood ratios for each benchmark.
The vertical black line indicates the mean, which is also provided in the upper left corner of each panel.
The respective bootstrapped standard errors are in parentheses.
Green indicates ratios greater than zero, where the optimized strategic AI has more predictive power, and red is the converse.
The bottom panel shows the proportion of games for which the sample of strategic \aisub{s} is the best predictor of initial play. 
Error bars are 95\% Clopper-Pearson confidence intervals
\end{singlespace}
\end{footnotesize}
\end{figure}

The top panel of Figure~\ref{fig:ll_ratios} shows the estimation results.
Each panel provides the histogram of the game-by-game log-likelihood ratios ($\hat \Lambda_s$) for each comparison.
The vertical dashed black lines indicate the means of the log-likelihood ratios ($\bar \Lambda_S$), which are also given in the upper left corner of each panel.
Bootstrapped standard errors are in parentheses.
Green indicates ratios greater than zero, where the optimized AI has more predictive power, and red is the converse.

With $\varepsilon = 0.2$, the strategic sample of \aisub{} is, on average, significantly more predictive than any of the benchmarks ($p < 0.001$ for all comparisons).
These differences are substantial.
Starting with the leftmost panel, across all human observations in the dataset, the strategic sample of \aisub{s} achieves an average per-observation likelihood ratio of $e^{\BaselineBarLambda} = \ExpBaselineBarLambda$ in favor of the model, relative to the baseline~AI. 
In other words, the likelihood of the observed human data under the strategic \aisub{} model is, on average, $\ExpBaselineBarLambda$ times larger per observation than under the baseline.

Moving to the right, the corresponding average likelihood ratios are $e^{\CHBarLambda} = \ExpCHBarLambda$ compared to the cognitive hierarchy model and $e^{\NashBarLambda} = \ExpNashBarLambda$ compared to the Harsanyi-Selten-selected equilibria.
To be clear, this means that the strategic \aisub{s} outperform the cognitive hierarchy model by a larger margin than they outperform the Harsanyi-Selten Nash equilibria.
This is notable because throughout the literature, the cognitive hierarchy model has been a strong predictor of initial play in strategic games.
It has been explicitly shown to ``explain why equilibrium theory predicts behavior well in some games and poorly in others'' \citep{Camerer2004Cognitive}.
Of course, the cognitive hierarchy model might be more accurate with different values for $\tau$, but we had no way of knowing what these were ex ante---besides those from the literature---without the data sampled from the population we wanted to predict.
Finally, the strategic \aisub{s} show a large gain compared to the random pure-strategy benchmark $e^{\PureRandBarLambda} = \ExpPureRandBarLambda$, and a more modest but still significant advantage over the uniform distribution $e^{\UniformBarLambda} = \ExpUniformBarLambda$.

The bottom panel of Figure~\ref{fig:ll_ratios} shows the proportion of games for which the sample of strategic \aisub{s} is the best predictor---i.e., has a positive log-likelihood ratio ($\sum_{s \in S} \mathbf 1\{\hat \Lambda_s > 0\} / |S|$).
The results are consistent with the top panel.
The theory-grounded sample of strategic \aisub{s} better predicts the human responses in more games than any other model.
This proportion is large and significant for the baseline ($\BaselineHatP$).
It is smaller, although still substantial, for both the cognitive hierarchy model ($\CHHatP$) and the Harsanyi-Selten Nash equilibria ($\NashHatP$).
Because each game has only a small number of human observations ($m_s$ between 1 and 5), sampling noise in $\hat\Lambda_s$ mechanically attenuates this proportion toward 0.5, almost surely making these estimates highly conservative.
Empirical Bayes shrinkage methods \citep[see][]{walters2024empirical} could partially correct for this, but we leave the unadjusted proportions for simplicity.
This issue does not affect the sample averages $\bar\Lambda_S$ reported above, which remain unbiased.

Tables in Appendix~\ref{app:robustness_checks} provide the same statistical analyses for $\varepsilon \in \{.05, 0.1, 0.3\}$, respectively.
The above results are robust to these additional sensitivity checks.
$\bar \Lambda_S > 0 $ for all comparisons, and the proportions of games for which the optimized \aisub{} is the best predictor are all greater than 50\%.

Beyond relative comparisons, the strategic \aisub{s} demonstrate impressive absolute predictive accuracy.
Without any smoothing, \OptHighestMatchRate\% of human respondents selected the strategy for which the optimized \aisub{} assigns the most density.
Given that the number of possible strategies per game varied evenly between 5 and 20, this is notable.
Furthermore, \OptTopThreeMatchRate\% of human respondents selected one of the top three strategies for which the strategic \aisub{} assigns the most density.
And maybe most surprisingly, for \OptGameAllMatchRate\% of games, all human respondents selected a strategy within the support of the strategic \aisub{}.

\subsubsection{Predictive power by bonus rule}

In this subsection, we briefly analyze whether the predictive power of the strategic \aisub{} is sensitive to the different types of games from Table~\ref{tab:game_parameters} (and therefore different possible distributions for $\pi$).
In particular, we focus on the bonus rules, although the results are similar for all dimensions of the parameter space.
As a reminder, different bonus rules change the fundamental structure of the game.
For some, many strategy profiles led to the bonus (e.g., ``Unequal''), and others only one (``Coord. Low'').
Broadly speaking, the rules can be categorized into two types: mutually achievable and competitive.
In mutually achievable games, both players receive the bonus when the bonus condition is met.
In competitive games, only one player can receive the bonus.
This distinction roughly corresponds to the classic game-theoretic categories of coordination versus zero-sum games.

\begin{figure}[h!]
\begin{center}
\begin{minipage}{\textwidth}
\caption{Relative predictive power of strategic \aisub{s} ($\varepsilon = 0.2$) by bonus rule} 
\label{fig:bonus_rule_prop_plot}
\vspace*{-0.15in}
\includegraphics[width=\textwidth]{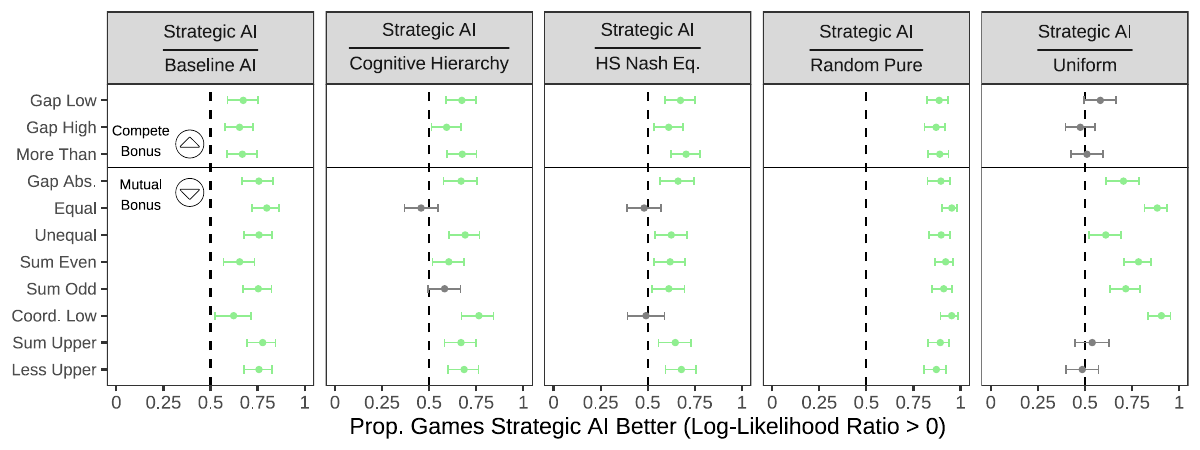} 
\end{minipage}
\end{center}
\begin{footnotesize}
\begin{singlespace}
\vspace*{-0.05in}
\emph{Notes:} This figure shows the proportion of games for which the strategic \aisub{s} is the best predictor of initial play for each bonus rule (on the y-axis).
The vertical dashed line corresponds to a 50-50 split, where there is no difference between the strategic \aisub{} and the reference model in that panel.
Bonus rules above the horizontal solid black line are mutually achievable (both players can receive the bonus), and those below are competitive (only one player can receive the bonus).
Green indicates that strategic \aisub{} significantly outperforms the reference model in more than 50\% of games with that bonus rule, and grey indicates no significant difference. 
Error bars show 95\% Clopper-Pearson confidence intervals.
The bonus rules are as follows:
\textbf{Gap Low}---selecting exactly \{\textit{gap}\} less than the opponent;
\textbf{Gap High}---selecting exactly \{\textit{gap}\}  more than the opponent;
\textbf{Gap Abs.}---when absolute difference equals \{\textit{gap}\} ;
\textbf{More Than}---when difference exceeds \{\textit{gap}\} ;
\textbf{Equal}---for matching opponent's number;
\textbf{Unequal}---for selecting different number than opponent;
\textbf{Sum Even}---when sum of both numbers is even;
\textbf{Sum Odd}---when sum of both numbers is odd;
\textbf{Sum Upper}---when sum equals the upper bound;
\textbf{Less Upper}---when the sum is less than the upper bound;
\textbf{Coord. Low}---when both players select the lower bound.
\end{singlespace}
\end{footnotesize}
\end{figure}

Figure~\ref{fig:bonus_rule_prop_plot} shows the proportion of games for which the strategic \aisub{} is the best predictor of initial play by bonus rule ($\sum_{s \in B} \mathbf 1\{\hat \Lambda_s > 0\} / |B|$ where $B \subset S$ is the set of games with a given bonus rule).
The figure is of a similar structure as the bottom panel of Figure~\ref{fig:ll_ratios}, with each panel showing a different benchmark comparison with $\varepsilon = 0.2$.
The y-axis shows the bonus rule, and the x-axis is the proportion of games for which the log-likelihood ratio is greater than zero.
Green indicates that strategic \aisub{} significantly outperforms the reference model in more than 50\% of games with that bonus rule, and grey indicates no significant difference.
Bonus rules above the horizontal solid black line correspond to games where both players can achieve the bonus simultaneously, while those below are competitive, permitting only one player to do so.

The strategic \aisub{} sample weakly dominates all other benchmarks; it never significantly underperforms for any bonus rule.
It outperforms the baseline and the random pure-strategy benchmark for every rule, the cognitive hierarchy model and the Harsanyi-Selten Nash equilibria for all but two rules, and the uniform benchmark for 7 of the 11 rules.
There is no apparent distinction between mutually achievable and competitive bonus rules.

In Appendix~\ref{app:robustness_checks}, we provide analogous plots for alternative values of $\varepsilon \in \{0.05, 0.1, 0.3\}$ and for the various points rules in Table~\ref{tab:game_parameters}.
We also report regressions of other game parameters (lower bound, upper bound, bonus size, gap) on the log-likelihood ratio of the strategic \aisub{} relative to the benchmarks ($\hat \Lambda_s$).
The results are similar to the ones presented here.
The strategic \aisub{s} are almost universally superior across different slices of the parameter space.
They are therefore likely to generalize to alternative distributions for $\pi$ across the population of games.

\section{Conclusion}
\label{sec:conclusion}

The great promise of \aisub{s} lies in their potential to accurately predict human behavior in novel settings.
Realizing this capability could transform social science research and public policy.
It could provide the social science equivalent of a lab bench in the physical sciences: an accurate, scalable playground to test ideas before large-scale and expensive implementation with humans.\footnote{This could be even more powerful given the often-observed researcher inability to predict results of their own experiments \citep{predictscience2019DellaVigna,milkmanMegastudies2021,Gandhi2023Hypothetical,Gandhi2024EffectSize,Duckworth2025ZearnMega}.}
Yet, with the current state-of-the-art foundation models, \aisub{s} are not yet reliable enough to be used in this way out of the box.

In this paper, we explored an approach to address this shortcoming.
Our approach relies on two key principles: (i) grounding candidate \aisub{s} in theories expected to drive human behavior in the target setting, and (ii) optimizing and then validating \aisub{s} in distinct but related settings presumed to be well-explained by the same theory.
Without theoretical grounding, optimized \persona{s} may fail to meaningfully improve even in-sample predictions.
Without validation across distinct but related datasets, optimized \persona{s} are prone to overfit a particular data-generating process.
Just as economists carefully extend established theories to novel policy contexts---relying on accumulated empirical validation rather than absolute certainty---optimizing theoretically-grounded \aisub{s} to match samples of human data, and then validating them in distinct but related settings, provides a principled foundation for predicting humans in new settings.

The improvements in predictive power yielded by this approach are substantial. 
In four novel and preregistered strategic games derived from \citeauthor{1120Arad2012}'s 11-20 money request game, theory-grounded \aisub{s}, optimized and validated through our methodology, reduced prediction errors by approximately \MinPercImproveArad-\MaxPercImproveArad\% compared to baseline AI predictions. 
Remarkably, these theory-grounded agents predicted initial play in some of the games better than the original human data from \citeauthor{1120Arad2012}. 
Importantly, our results are not confined to games involving strategic reasoning.
In Appendix \ref{app:predict_new_CR}, we apply the same procedure to the allocation games from \citeauthor{charness2002understanding}, and, using data from a preregistered experiment with entirely novel human responses, we find substantial improvements.

Although this approach provides no statistical guarantees for arbitrary novel settings---indeed, no procedure can guarantee predictive power in entirely novel domains without a fully specified and correct causal model---we demonstrated that we can make externally valid inferences within a pre-committed family of settings.
Using novel data from \SampleSizeHuman{} participants playing \SampleSizeGame{} games randomly sampled from a heterogeneous population of \TotalGames{} strategic games, we found that the strategic level-$k$ agents generalized effectively across this broad domain.
In \OptGameAllMatchRate\% of games, all human subjects chose actions within the support of the optimized \aisub{} responses. 
These general agents substantially outperformed the baseline AI off-the-shelf, a cognitive hierarchy model, and the Harsanyi-Selten equilibria.

The results were also robust to several alternative specifications and were consistent across different game structures.
Because the games were randomly sampled under the assumptions stated in Section~\ref{sec:estimation}, these results are externally valid for the entire population of \TotalGames{} games.
While we can only guarantee validity within this specific population and sampling distribution, the strong performance of theoretically motivated \aisub{s} across such a diverse set of strategic games suggests that similar approaches would likely generalize to alternative distributions and even other strategic games.

Looking ahead, researchers could leverage this approach by identifying broad domains where they need predictions and can obtain training and validation data from representative subsets---even if there is substantial structural variation within the domain.
Indeed, a key advantage of \aisub{s} is that they can make predictions in response to any setting expressed in natural language without any prior data.
This is often not possible with traditional machine learning methods and economic models.

Another exciting research direction would be to automate the theory-prediction-testing loop.
Such a system would start with novel human data and relevant experimental settings, iteratively generate theory-informed candidate personas, optimize their parameters, and systematically evaluate their generalizability across samples.
One can also imagine just starting with a novel setting, and then having an AI system actually search for the relevant training and validation data to execute our approach.
Recent research supports the feasibility of these ideas \citep{Jackson2025Mixture,zhu2025ExplPredicting}.
In particular, \cite{Manning2024Automated} demonstrate AI systems capable of automating the full social scientific workflow---from hypothesis generation to experimental simulation and data analysis.

More generally, the results in this paper linking theory-grounded \aisub{s} to robust generalizability in novel settings, and atheoretical \aisub{s} to failure, are notable for reasons beyond prediction \citep{hofman2021integrating}.
They suggest that the underlying LLM has correctly learned the relevant relationships between the \aisub{s} and human responses to the given setting.
This is even more notable given that it is highly unlikely that these mappings were explicitly specified during training.  
Such a finding aligns with recent evidence that LLMs form rich internal representations of their human-generated training corpus rather than merely memorizing it \citep{lindsey2025biology, ameisen2025circuit}.
If true for LLMs and human behavior more generally, our \persona{}-alignment-generalizability exercises may offer more than improved predictive power.
If an LLM armed with a particular theory-grounded \persona{} matches human data particularly well across a wide range of related settings, it might be evidence that the theory has a lot of explanatory power for the underlying human sample.
Building on a young social scientific literature \citep{peterson2021lottery,hirasawa2022using,EnkeComplexity2023,Si2024NovelResearch,MLhypothesis2024,Ashesh2024Anomaly,batista2024words,movva2025sparse}, this could, in turn, provide researchers with robust machine learning methods to rapidly and efficiently inform promising new hypotheses.

\newpage \clearpage

\bibliographystyle{ecta}
\bibliography{optimize}

\newpage \clearpage

\appendix

\renewcommand{\thefigure}{A\arabic{figure}} 
\setcounter{figure}{0}  

\renewcommand{\thetable}{A\arabic{table}} 
\setcounter{table}{0}

\section{Predicting behavior in novel allocation games}
\label{app:predict_new_CR}

We further test the robustness of our approach by predicting human behavior on a set of novel allocation games.
Unlike the strategic reasoning games studied in Sections~\ref{sec:predict_new_AR} and \ref{sec:ext-valid}, these games---adapted from \cite{charness2002understanding}'s (CR) experiments on social preferences---require individuals to balance their own monetary payoffs against those of others. 
They offer a distinct theoretical and empirical context for validating the generalizability of \persona{s} identified using our approach.
In addition to this new context, there are two key technical differences from the previous section.
First, samples of agents are optimized over several training settings simultaneously.
This further decreases the likelihood of overfitting on idiosyncratic features of any particular setting.
Second, we employ the novel construction method and parameterize a prompt template to optimize the agents in-sample.

We follow the same analytical structure and use similar notation as in Section~\ref{sec:predict_new_AR}.
We first briefly describe the dictator settings originally explored by CR, as these form our training dataset. 
Next, we detail our procedure for optimizing samples of \aisub{s}, where the motivating theory is drawn directly from the social-preference models in CR. 
We then validate these subjects on a distinct set of two-player games studied by CR and demonstrate the pitfalls of optimizing over atheoretical \persona{s}.
Finally, we introduce a new series of structurally distinct three-player allocation games and use them to illustrate the empirical efficacy of our approach in novel games that were not in the LLM's training corpus.

\subsection{\cite{charness2002understanding}'s unilateral dictator games}
\label{sec:cr_app}

CR study a set of simple allocation decisions in which one player (the dictator) chooses between two ways of splitting money with a passive recipient. 
In one version of these games, for example, the dictator (Person B) unilaterally decides between the options ``Left'' and ``Right'':
\begin{quote}
\centering
$\underbrace{(\underbrace{400}_{\mbox{To A}}\:,\: \underbrace{600}_{\mbox{To B}})}_{\mbox{``Left''}}$ $\quad$ vs. $\quad$ $\underbrace{(\underbrace{700}_{\mbox{To A}}\:,\: \underbrace{300}_{\mbox{To B}})}_{\mbox{``Right''}}$.
\end{quote}
CR collected human responses for six variations of this basic dictator setting, each featuring different payoff distributions.
These six settings constitute our training dataset from which we derive the joint empirical distribution $P$ of choosing Left.

\begin{figure}[h!]
\begin{center}
\begin{minipage}{\textwidth}
\caption{Distribution of responses for the single-stage training dictator games} 
\label{fig:cr_train}
\vspace*{-0.15in}
\centering
\includegraphics[width=.9\textwidth]{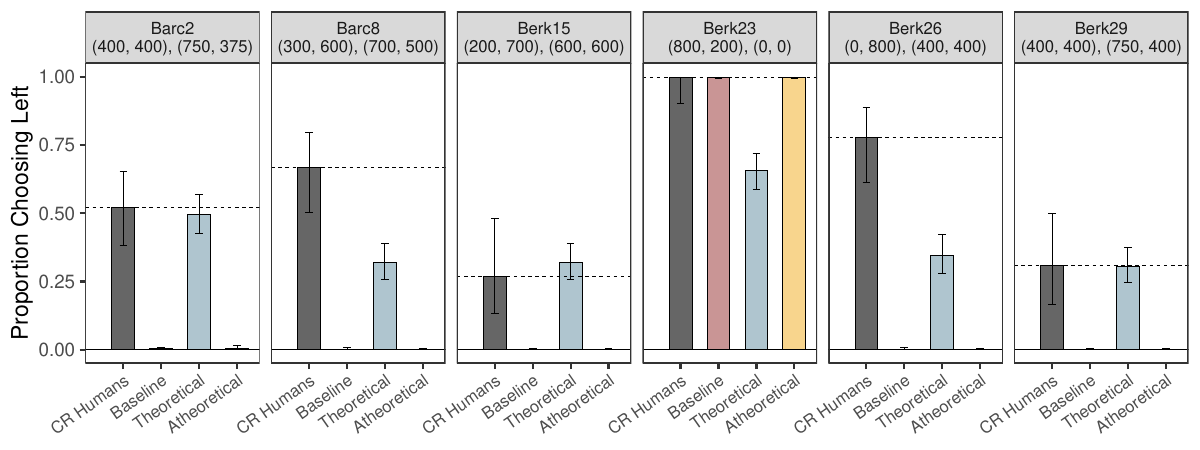} 
\end{minipage}
\end{center}
\begin{footnotesize}
\begin{singlespace}
\vspace*{-0.05in}
\emph{Notes:}
This figure reports the results of replications of the unilateral dictator games from \cite{charness2002understanding}. 
Columns each represent a different game, and the x-axis corresponds to different samples of subjects playing each game.
The y-axis shows the proportion of that sample choosing the option ``Left.'' The black bars (and the dashed black lines) are the human responses from the original paper, red is the baseline \aisub{s}, blue are the agents optimized using efficiency, self-interest, inequity aversion as parameters, and the yellow are atheoretical agents with preferences for the TV show new girl, taxidermy, and swimming. 
Error bars report 95\% Wilson confidence intervals.
\end{singlespace}
\end{footnotesize}
\end{figure}

Figure~\ref{fig:cr_train} shows the original results from CR.
The columns represent different settings and show the payoffs for each player depending on the dictator's choice of ``Left'' or ``Right''.
The y-axis shows the proportion of the sample that chose ``Left'' for each setting, and the black bars correspond to the distribution of human responses from CR.
Besides the Pareto-dominated Berk23 setting, where everyone chooses ``Right,'' the human data is balanced across the two options.

To establish the baseline ($\hat P_0$), we elicited 1,000 responses per setting from \textsc{Gpt-4o}, without any additional instructions.\footnote{\cite{horton2023large} also explore the baseline for the same games. Although their goal is to provide an early demonstration of AI simulation more generally.}
The red bars in Figure~\ref{fig:cr_train} represent these baseline AI responses. Notably, the baseline AI strongly favors choosing ``Right'' in nearly every setting, diverging sharply from the balanced human distributions. 
Quantitatively, this mismatch is substantial: using mean absolute error (MAE) as our distance metric, we find $\frac{1}{6}\sum_{s \in S} d(P_s,\hat P_0)=\CRBaselineTrainMAE$. 
Given that the maximum possible MAE is 1, this is poor baseline predictive accuracy.

\subsection{Constructing the sample of \aisub{s}}
\label{sec:apply_prompt}

CR hypothesize that a combination of efficiency concerns, inequity aversion, and self-interest is a key determinant of dictators' choices.
To construct the sample of \aisub{s} to better match the human data from these six settings simultaneously, we build a prompt template that incorporates these three traits as our theoretical motivation for the agents.
Specifically, we parameterize each trait in the following prompt: 
\begin{quote}
\vspace{-.5cm}
\singlespacing
$\theta(\phi_{eff}, \phi_{self}, \phi_{ineq})$ = \emph{On a scale from 1 to 10, your efficiency level is: \{\(\phi_{eff}\)\}. 
10 means you strongly prioritize maximizing combined payoffs, and 1 means you don't care.
On a scale from 1 to 10, your self-interest level is: \{\(\phi_{self}\)\}. 
10 means you strongly prioritize your own payoffs, and 1 means you don't care.
On a scale from 1 to 10, your inequity aversion level is: \{\(\phi_{ineq}\)\}. 
10 means you strongly prioritize fairness between players, and 1 means you don't care.}
\end{quote}

Our goal is to identify the parameter vector (or combination of vectors) that generates AI response distributions closely matching the observed human data.
To do this, we create sets of $k=3$ agents, each with a distinct parameter vector $\bm{\phi}$. 
Thus, each agent's \persona{} is $\theta(\bm{\phi})$, where
$\bm{\phi} \;=\; (\phi_{eff},\,\phi_{self},\,\phi_{ineq})$.
We begin by randomly sampling 5 triples from the feasible space $\Phi = \{1,\dots,10\}^3$.
For each sampled combination, we query the model 30 times per agent, producing the empirical distribution of responses $P$.

We then employ Bayesian optimization to iteratively search the parameter space, evaluating an additional 15 sets of parameter combinations (for a total of 20). 
Using mean absolute error to measure divergence from human data, this optimization identifies the optimal parameter vectors as:
$\bigl(\bm{\phi}_1^*,\,\bm{\phi}_2^*,\,\bm{\phi}_3^*\bigr) \;=\; \bigl(\,(7,\,10,\,10),\;(3,\,1,\,3),\;(1,\,10,\,2)\bigr)$.
Assigning these parameters to three \aisub{s} forms the optimized sample $\bm{\theta}^*$. 
As shown by the blue bars in Figure~\ref{fig:cr_train}, the resulting distribution aligns much closer with the human responses: $\frac{1}{6}\sum_{s \in S} d(P_s,\hat P_{\bm{\theta}^*})=\CROptTrainMAE$.
This divergence represents a significant improvement, more than halving the baseline AI's error (MAE = \CRBaselineTrainMAE).

\subsection{Validation using two-stage games from \cite{charness2002understanding}}
\label{sec:validate_prompt}

To validate whether $\bm{\theta}^*$ generalizes to new games, we apply the same prompt template and the values to a new set of more complicated sequential two-stage games from CR---the test set.
Like the validation variants from AR (costless and cycle), these games are plausibly driven by similar underlying mechanisms as the training games, but are still different enough to provide a nontrivial test of generalization.
In the first stage, Person A chooses either a given allocation or lets Person B choose one of two other known allocations.
Person B chooses an allocation but is not informed of Person A's choice---until the payoffs are realized.
For example, in one game, players are shown the following options:
\begin{quote}
Stage 1 (Person A chooses): $\underbrace{(\underbrace{500}_{\mbox{To A}}\:, \:\underbrace{500}_{\mbox{To B}})}_{\mbox{``Left''}}$ $\quad$ vs. $\quad$  $\underbrace{\underbrace{(400\:\:,\:600)\:\text{    vs.  }(700\:\:,\: 300)}_{\mbox{Let Person B choose}}}_{\mbox{``Right''}}$.
\end{quote}
\begin{quote}
Stage 2  (Person B chooses): $\underbrace{(\underbrace{400}_{\mbox{To A}}\:,\: \underbrace{600}_{\mbox{To B}})}_{\mbox{``Left''}}$ $\quad$ vs. $\quad$ $\underbrace{(\underbrace{700}_{\mbox{To A}}\:,\: \underbrace{300}_{\mbox{To B}})}_{\mbox{``Right''}}$.
\end{quote}
Table~\ref{tab:cr_games_results} in Appendix~\ref{app:figs} provides all 20 versions of these two-stage games (each with a different set of payoffs), along with the human results from CR.

As a baseline, we elicit \textsc{Gpt-4o's} responses to these 20 games 150 times each with the temperature set to 1.
We then do the same for the theory-grounded sample $\bm{\theta}^*$---each of the three agents in the mixture plays each game 50 times.

\begin{figure}[h!]
\begin{center}
\begin{minipage}{\textwidth}
\caption{Distances between human and \aisub{s} for the two-stage dictator games} 
\label{fig:cr_test}
\vspace*{-0.15in}
\centering
\includegraphics[width=.9\textwidth]{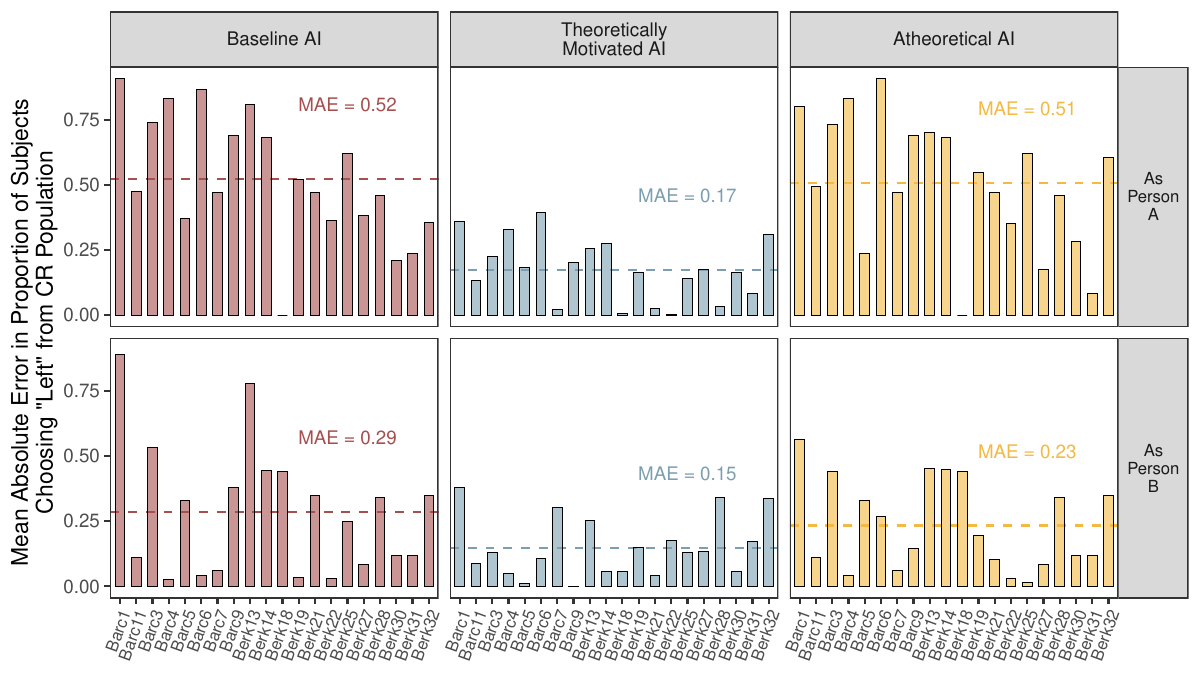} 
\end{minipage}
\end{center}
\begin{footnotesize}
\begin{singlespace}
\vspace*{-0.05in}
\emph{Notes:}
This figure reports the results of replications of the sequential two-player games from \cite{charness2002understanding} with \aisub{s}.
Each row shows responses from either Person A (left) or Person B (right), while each column corresponds to a different set of subjects. 
The x-axis shows the game, and the y-axis shows the mean absolute difference between the fraction of \aisub{s} choosing ``Left'' and the fraction of human subjects choosing ``Left'' in \citeauthor{charness2002understanding}.
The left column displays the baseline \aisub{s} (red), the middle column is the selected \aisub{s} (blue), and the right column shows the atheoretical \aisub{s} (yellow).  
The horizontal dashed lines show the mean absolute error in each pane.
\end{singlespace}
\end{footnotesize}
\end{figure}

Figure~\ref{fig:cr_test} shows the results. 
The top row shows the responses for the \aisub{s} as Person A, and the bottom row for Person B.
Each column corresponds to a different sample of subjects.
The x-axis shows the setting name and the y-axis shows the mean absolute difference between the fraction of \aisub{s} choosing ``Left'' and the fraction of human subjects choosing ``Left'' in \citeauthor{charness2002understanding}.
The difference between the baseline (red) and selected \aisub{s} (blue) is substantial.
The MAE between the baseline and the human subjects as Player A (\CRTestBaselineGameAMAE) is three times larger than that for the optimized agents relative to the humans (\CRTestOptGameAMAE).
The difference in MAE is twice as large for Player B (\CRTestBaselineGameBMAE{} vs. \CRTestOptGameBMAE).

This predictive improvement is robust across settings.
In $\CRTestNOptBetterThanBaseline$ of the 40 total decisions (20 settings, each played as Person A and Person B), the optimized theory-grounded agents more accurately matched human behavior than the baseline AI. 
Never was the absolute error in a game larger than 0.50 for the optimized theory-grounded agents, which is less than the MAE for the baseline AI as player A.

\subsection{Optimizing among atheoretical \persona{s}}
\label{sec:atheory_prompt}

Similarly to our study of AR, grounding a \persona{} template in theory is important for generalization.
We do this via negative example.
Specifically, without a theoretical grounding, the optimization procedure may fail to find a \persona{} that even fits in-sample.

Unlike in the Section~\ref{sec:predict_new_AR} with the Always Pick `N' agents, we do not offer an analogous overfitting example.
Because human data from multiple games is used for optimization, finding a sample of \aisub{s} which overfits requires finding \persona{s} which overfit to all six settings.
This is a much more difficult task than finding a \persona{} which overfits to a single game.
Indeed, this is an attractive feature of using multiple training samples to construct and validate agent samples.

To generate samples of arbitrary agents, we repeat the entire process from Section~\ref{sec:apply_prompt} but replace the theory-grounded attributes (i.e., efficiency, inequity aversion, and self-interest) with wholly unscientific ones: a self-reported fondness for the TV show \emph{New Girl}, an enthusiasm for taxidermy, and swimming ability.  
This new prompt template is:
\begin{quote}
\vspace{-.5cm}
\singlespacing
$\theta(\phi_{ng}, \phi_{tax}, \phi_{swim})$ = \textit{On a scale from 1 to 10, you think the show New Girl is: \{$\phi_{ng}$\}.
10 means you love New Girl, and 1 means you hate it.
On a scale from 1 to 10, your passion for taxidermy is: \{\(\phi_{tax}\)\}.
10 means you love taxidermy, and 1 means you hate it.
On a scale from 1 to 10, your ability to swim is: \{\(\phi_{swim}\)\}.
10 means you are a great swimmer, and 1 means you can't swim.}
\end{quote}

Using identical hyperparameters and the same Bayesian optimization procedure, we search over this atheoretical space to see if any combination of $(\phi_{ng}, \phi_{tax}, \phi_{swim})$ (each a sample of 3 agents with their own parameter vector) could even match the original single-stage dictator games in-sample.  
The resulting atheoretical parameter vector was: $\bigl(\bm{\phi}_{ath-1}^*,\bm{\phi}_{ath-2}^*,\bm{\phi}_{ath-3}^*\bigr) \;=\; \bigl(\,(5,\,7,\,1),\;(9,\,9,\,5),\;(7,\,6,\,8)\bigr)$.
As shown in Figure~\ref{fig:cr_train} (yellow), $\bm{\theta}_{ath}^*$ constructed using these parameters and template failed to beat even the baseline \aisub{s}' performance.  
In fact, throughout the search, no parameter combination for ``loving \emph{New Girl},'' ``passion for taxidermy,'' or ``swimming skill'' ever produced a distribution of choices that aligned more closely with real humans.
This result demonstrates the importance of grounding \aisub{s} in theoretical constructs.

This lack of improvement persisted in the two-stage validation games as well (Figure~\ref{fig:cr_test}; rightmost column).  
The atheoretical \aisub{s} and the baseline were effectively indistinguishable in their distribution of responses as Player A, and the atheoretical subjects were only a little better as Player B.
Overall, the atheoretical \aisub{s} were far less aligned than the theory-grounded \aisub{s}. 
They were closer to the human data than the baseline in $\CRTestAtheoryTotalProp$\% of the settings, worse than the baseline in $\CRTestBaselineTotalPropAtheo$\% of the settings, and identical in the remaining games.
The only way these arbitrary \persona{s} generalize is that their poor performance is consistent across settings.

\subsection{Predicting the novel three-player games}
\label{sec:cr_novel}

We conclude this section by introducing a set of 8 novel three-player allocation games to evaluate $\bm{\theta}^*$ and $\bm{\theta}_{ath}^*$ in new settings with a new participant pool.
We recruited $n = \NNCRovel$ participants from Prolific to make three allocation decisions drawn from eight distinct settings, each involving a choice between two monetary allocations.
Participants were paid \$1.00 and could earn a bonus of up to an additional \$1.00, depending on their own or others' choices. 
A representative setting is:
$$
\begin{array}{cc}
\mathbf{\textbf{Option A: }} \begin{cases}
\$1.00 & \text{To Selected} \\
\$0.75 & \text{To Each Other Player}
\end{cases}
&
\quad\quad
\mathbf{\textbf{Option B: }} \begin{cases}
\$0.50 & \text{To Selected} \\
\$1.00 & \text{To Each Other Player}
\end{cases}
\end{array}
$$
After completion, one of the three decisions was randomly selected for payment. 
Participants were then randomly grouped into triads, with one member randomly chosen as the Selected Player. 
All three members received bonuses according to the allocation chosen by their group's Selected Player.\footnote{Suppose you, the reader, are completing this task and choose option A in the above setting.
If after the survey is completed, the decision above is selected for payment and you are randomly chosen as the Selected Player, you will receive a \$1.00 bonus, and the other two players each receive an extra \$0.75.
However, if another player was chosen as the Selected Player and they had picked Option A, then you would receive a \$0.75 bonus payment.}
Multiple attention checks confirmed participants understood the instructions and the payoffs.
Participants' decisions only determined payments if they were the Selected Player.
Uncertainty over selection aimed to ensure that choices reflected genuine social preferences.

The entire experimental design---including settings, procedures, and the optimized \aisub{} parameters---was preregistered prior to data collection with human participants.
Figure~\ref{fig:novel_allocation_pic} in Appendix~\ref{app:instruct} shows the full instructions for an example setting.
To the best of our knowledge, games with these exact payoffs have never been used in an experiment with publicly available data.\footnote{CR tested some 3-player games, but these had different payoffs, bonus rules, and involved imperfect information.}

\begin{figure}[h!]
\begin{center}
\caption{Results from the novel three-player allocation games}
\vspace*{-0.12in}
\label{fig:novel_cr}

\begin{subfigure}{\textwidth}  
\refstepcounter{subfigure}\label{fig:novel_cr_a}
\centering
\begin{overpic}[width=.95\textwidth]{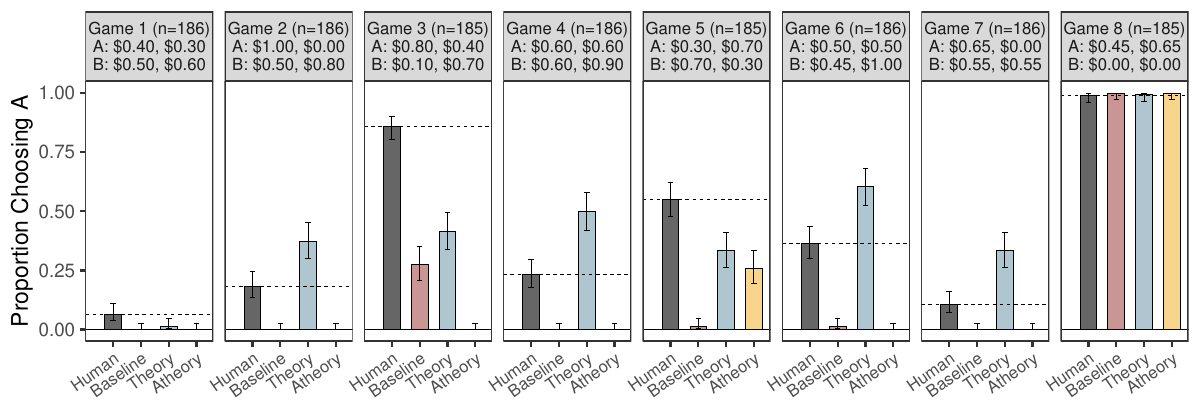}
\put(-1,35){\large\bfseries (a)}
\end{overpic}
\end{subfigure}

\begin{subfigure}{\textwidth}
\refstepcounter{subfigure}\label{fig:novel_cr_b}
\centering
\begin{overpic}[width=.95\textwidth]{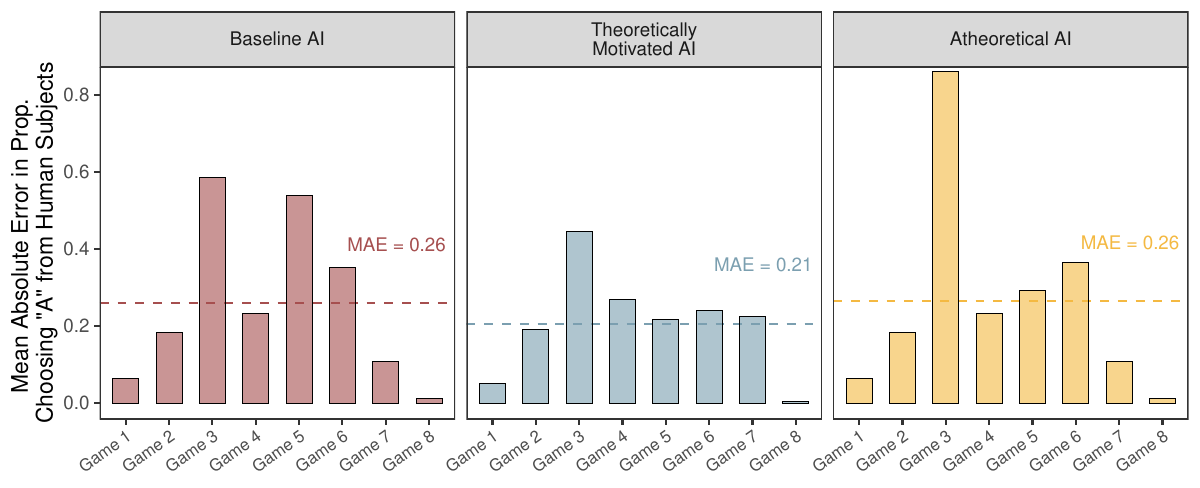}
\put(-1,40){\large\bfseries (b)}
\end{overpic}
\end{subfigure}

\end{center}
\vspace*{-0.2in}
\begin{footnotesize}
\begin{singlespace}
\textit{Notes:} Panel (a) shows the proportion of choices for Option A across eight novel three-player allocation settings. Human responses are depicted in black (with the dashed lines), baseline AI in red, theoretically motivated AI in blue, and atheoretical AI in yellow. Error bars indicate 95\% Wilson confidence intervals. Panel (b) presents the absolute error between human and AI choices across the settings, with dashed lines marking the MAE.
\end{singlespace}
\end{footnotesize}
\end{figure}

Figure~\ref{fig:novel_cr_a} shows the results for all four subject samples: human participants (black), baseline \aisub{s} (red), theory-grounded \aisub{s} ($\bm{\theta}^*$ in blue), and atheoretical \aisub{s} ($\bm{\theta}_{ath}^*$ in yellow). 
Each column corresponds to a setting with the relevant options indicated, with the y-axis indicating the proportion of subjects choosing Option A. 

Human responses generally reflect a fairly even split between options, except for the extreme setting 8, where participants unanimously select Option A.
The baseline AI consistently diverges from human behavior, disproportionately favoring Option B in nearly every setting (MAE = \CRBaselineNovelMAE).
Atheoretical \aisub{s} offer no relative improvement, with a slightly worse fit (MAE = \CRAtheoryNovelMAE).

On average, $\bm{\theta}^*$ better approximates human choices across settings (MAE = \CRTheoryNovelMAE)---about \CRPercImprove\% less than the baseline.
This improvement is emphasized in Figure~\ref{fig:novel_cr_b}, showing the per-game absolute error along with the MAE.
Importantly, this performance improvement is not driven by a few outliers: the theoretically motivated sample matches or exceeds both baseline and atheoretical \aisub{s} in five settings.
And in the games where the baseline AI and atheoretical \aisub{s} are better, the difference is not large.

As with the results in Section~\ref{sec:predict_new_AR}, these findings demonstrate that theoretically grounded \aisub{s}, optimized and then validated on data from related but distinct data-generating processes, can significantly improve the predictive power of AI simulations in novel settings.

\begin{table}[h!]
\caption{Human subjects results for two-person response games in \cite{charness2002understanding}}
\label{tab:cr_games_results}
\begin{center}
\footnotesize
\begin{tabular}{llcccc}
\hline
& & \multicolumn{4}{c}{Human Subject Responses} \\
Game & Description & Out & Enter & Left & Right \\
\hline
\\[-1.8ex]
\multicolumn{6}{l}{\textit{Panel A: B's payoffs identical}} \\[0.5ex]
Barc7 & A chooses (750,0) or lets B choose & .47 & .53 & .06 & .94 \\
& (400,400) vs. (750,400) & & & & \\
Barc5 & A chooses (550,550) or lets B choose & .39 & .61 & .33 & .67 \\
& (400,400) vs. (750,400) & & & & \\
Berk28 & A chooses (100,1000) or lets B choose & .50 & .50 & .34 & .66 \\
& (75,125) vs. (125,125) & & & & \\
Berk32 & A chooses (450,900) or lets B choose & .85 & .15 & .35 & .65 \\
& (200,400) vs. (400,400) & & & & \\[1ex]

\multicolumn{6}{l}{\textit{Panel B: B's sacrifice helps A}} \\[0.5ex]
Barc3 & A chooses (725,0) or lets B choose & .74 & .26 & .62 & .38 \\
& (400,400) vs. (750,375) & & & & \\
Barc4 & A chooses (800,0) or lets B choose & .83 & .17 & .62 & .38 \\
& (400,400) vs. (750,375) & & & & \\
Berk21 & A chooses (750,0) or lets B choose & .47 & .53 & .61 & .39 \\
& (400,400) vs. (750,375) & & & & \\
Barc6 & A chooses (750,100) or lets B choose & .92 & .08 & .75 & .25 \\
& (300,600) vs. (700,500) & & & & \\
Barc9 & A chooses (450,0) or lets B choose & .69 & .31 & .94 & .06 \\
& (350,450) vs. (450,350) & & & & \\
Berk25 & A chooses (450,0) or lets B choose & .62 & .38 & .81 & .19 \\
& (350,450) vs. (450,350) & & & & \\
Berk19 & A chooses (700,200) or lets B choose & .56 & .44 & .22 & .78 \\
& (200,700) vs. (600,600) & & & & \\
Berk14 & A chooses (800,0) or lets B choose & .68 & .32 & .45 & .55 \\
& (0,800) vs. (400,400) & & & & \\
Barc1 & A chooses (550,550) or lets B choose & .96 & .04 & .93 & .07 \\
& (400,400) vs. (750,375) & & & & \\
Berk13 & A chooses (550,550) or lets B choose & .86 & .14 & .82 & .18 \\
& (400,400) vs. (750,375) & & & & \\
Berk18 & A chooses (0,800) or lets B choose & .00 & 1.00 & .44 & .56 \\
& (0,800) vs. (400,400) & & & & \\[1ex]

\multicolumn{6}{l}{\textit{Panel C: B's sacrifice hurts A}} \\[0.5ex]
Barc11 & A chooses (375,1000) or lets B choose & .54 & .46 & .89 & .11 \\
& (400,400) vs. (350,350) & & & & \\
Berk22 & A chooses (375,1000) or lets B choose & .39 & .61 & .97 & .03 \\
& (400,400) vs. (250,350) & & & & \\
Berk27 & A chooses (500,500) or lets B choose & .41 & .59 & .91 & .09 \\
& (800,200) vs. (0,0) & & & & \\
Berk31 & A chooses (750,750) or lets B choose & .73 & .27 & .88 & .12 \\
& (800,200) vs. (0,0) & & & & \\
Berk30 & A chooses (400,1200) or lets B choose & .77 & .23 & .88 & .12 \\
& (400,200) vs. (0,0) & & & & \\
\hline
\end{tabular}
\end{center}
\begin{footnotesize}
\begin{singlespace}
\vspace*{-0.05in}
\textit{Notes:} This table presents the complete set of two-person response games from \citeauthor{charness2002understanding} along with human subject responses. 
This figure is identical to the one they show in the original paper.
For each game, we show the proportion of subjects choosing each option. ''Out'' and ''Enter'' refer to Person A's initial choice, while ''Left'' and ''Right'' refer to Person B's choice if given the opportunity. All payoff values are in experimental currency units.
\end{singlespace}
\end{footnotesize}
\end{table}
  
\newpage \clearpage

\renewcommand{\thefigure}{B\arabic{figure}} 
\setcounter{figure}{0}  

\renewcommand{\thetable}{B\arabic{table}} 
\setcounter{table}{0}

\section{All game instructions}
\label{app:instruct}

\noindent\textbf{Basic 11-20 Game}
\begin{quote}
\emph{You and another player are playing a game in which each player requests an amount of money. The amount must be (an integer) between 11 and 20 shekels.  Each player will receive the amount he requests. A player will receive an additional amount of 20 shekels if he asks for exactly one shekel less than the other player. What amount of money would you request?}
\end{quote}

\noindent\textbf{Cycle 11-20 Game}
\begin{quote}
\emph{You and another player are playing a game in which each player requests an amount of money.
The amount must be (an integer) between 11 and 20 shekels. 
Each player will receive the amount of money he requests.
A player will receive an additional amount of 20 shekels if: (i)  he asks for exactly one shekel less than the other player or  (ii) he asks for 20 shekels and the other player asks for 11 shekels.
What amount of money would you request?}
\end{quote}

\noindent\textbf{Costless 11-20 Game}
\begin{quote}
\emph{You and another player are playing a game in which each player chooses an integer in the range 11-20.
A player who chooses 20 will receive 20 shekels (regardless of the other player's choice). 
A player who chooses any other number in this range will receive three shekels less than in the case where he chooses 20. 
However, he will receive an additional amount of 20 shekels if he chooses a number that is one less than that chosen by the other player. 
Which number would you choose?}
\end{quote}

\noindent\textbf{Basic 1-10 Game}
\begin{quote}
\emph{You are going to play a game where you must select a whole number between 1 and 10.
You will receive a number of points equivalent to that number. For example, if you select 3, you will get 3 points. 
If you select 7, you will get 7 points, etc.
After you tell us your number, we will randomly pair you with another Prolific worker who is also playing this game. 
They will also have chosen a number between 1 and 10.
If either of you select a number exactly one less than the other player's number, than the player with the lower number will receive an additional 10 points.
Please choose a number between 1 and 10. }
\end{quote}

\noindent\textbf{Cycle 1-10 Game}
\begin{quote}
\emph{You are going to play a game where you must select a whole number between 1 and 10.
You will receive a number of points equivalent to that number. 
For example, if you select 3, you will get 3 points. 
If you select 7, you will get 7 points, etc.
After you tell us your number, we will randomly pair you with another Prolific worker who is also playing this game. 
They will also have chosen a number between 1 and 10.
There are 2 ways to win an additional 10 points based on both yours and the other player's choice:
1. If either of you select a number exactly one less than the other player's number, then the player with the lower number will receive an additional 10 points.
2. If either of you select 10 and the other selects 1, then the player who chose 10 will receive an additional 10 points.
Please choose a number between 1 and 10.}
\end{quote}

\noindent\textbf{Costless 1-10 Game}
\begin{quote}
\emph{You are going to play a game where you must select a whole number between 1 and 10.
You will receive 10 points if you select the number 10 and you will receive 7 points for selecting any other number. 
After you tell us your number, we will randomly pair you with another Prolific worker who is also playing this game. 
They will also have chosen a number between 1 and 10.
If either of you select a number exactly one less than the other player's number, than the player with the lower number will receive an additional 10 points.
Please choose a number between 1 and 10.}
\end{quote}

\noindent\textbf{1-7 Game}
\begin{quote}
\emph{You are going to play a game where you must select a whole number between 1 and 7.
You will receive a number of points equivalent to that number. For example, if you select 3, you will get 3 points. 
If you select 6, you will get 6 points, etc.
After you tell us your number, we will randomly pair you with another Prolific worker who is also playing this game. 
They will also have chosen a number between 1 and 7.
If either of you select a number exactly one less than the other player's number, than the player with the lower number will receive an additional 10 points.
Please choose a number between 1 and 7.}
\end{quote}

\begin{figure}[h!]
\begin{center}
\begin{minipage}{1 \linewidth}
\caption{Screenshot of the three-player game instructions} 
\centering
\vspace*{-0.15in}
\label{fig:novel_allocation_pic}
\includegraphics[width=\textwidth]{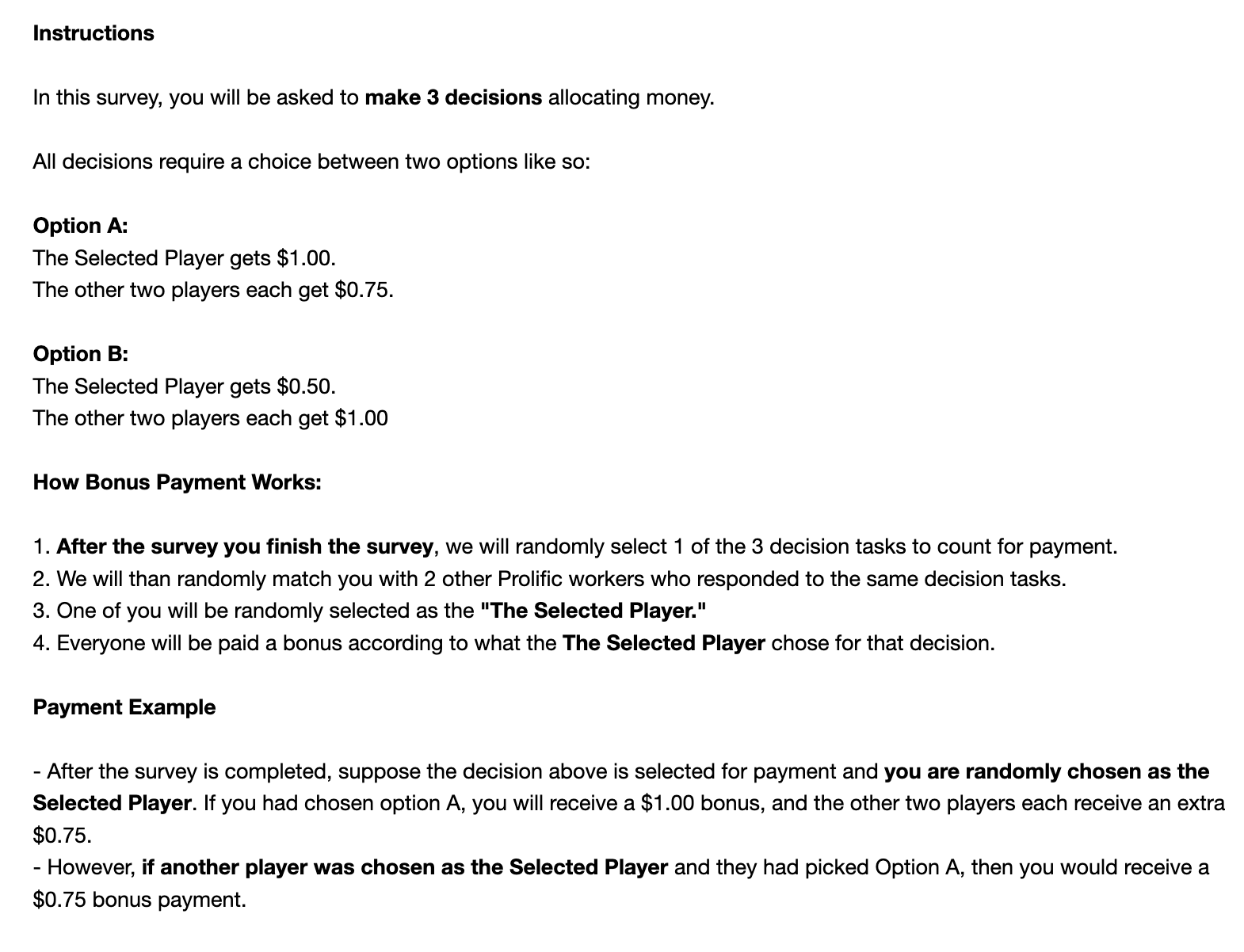}
\end{minipage}
\end{center}
\begin{footnotesize}
\begin{singlespace}
\vspace*{-0.15in}
\emph{Notes:} This figure shows the instructions for the novel three-player allocation game presented to participants.
\end{singlespace}
\end{footnotesize}
\end{figure}

\begin{figure}[h!]
\begin{center}
\begin{minipage}{1 \linewidth}
\caption{Bonus opportunity for the games in Section~\ref{sec:ext-valid}} 
\centering
\vspace*{-0.15in}
\label{fig:bonus_bounds_pic}
\includegraphics[width=.75\textwidth]{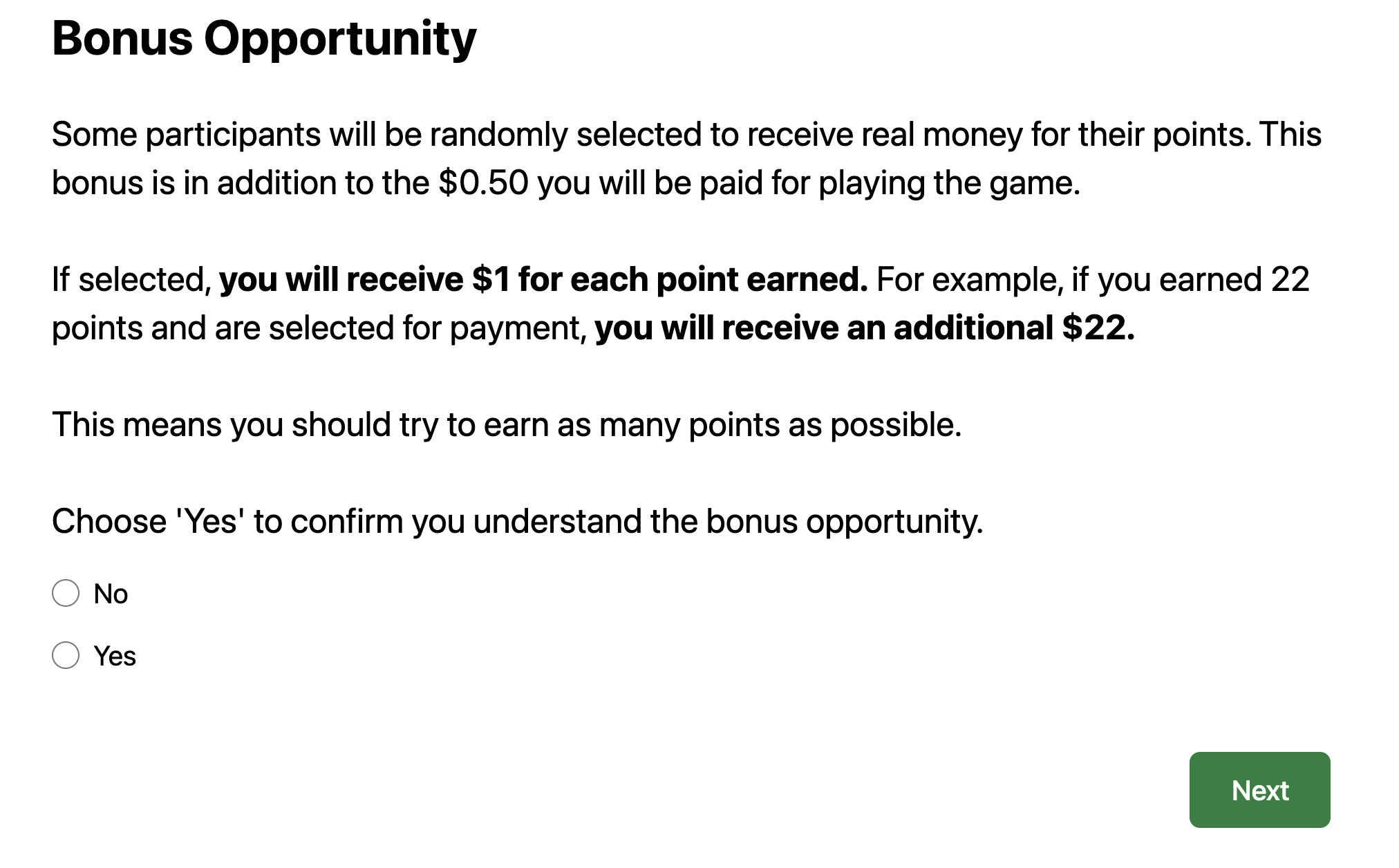}
\end{minipage}
\end{center}
\begin{footnotesize}
\begin{singlespace}
\vspace*{-0.15in}
\emph{Notes:} This shows the instructions for the bonus opportunity presented to participants for the novel sample of 1,500 games.
\end{singlespace}
\end{footnotesize}
\end{figure}

\newpage \clearpage

\begin{figure}[h!]
\begin{center}
\begin{minipage}{1 \linewidth}
\caption{Choosing a number for the assigned game in Section~\ref{sec:ext-valid}} 
\centering
\vspace*{-0.15in}
\label{fig:choice_bounds_pic}
\includegraphics[width=0.75\textwidth]{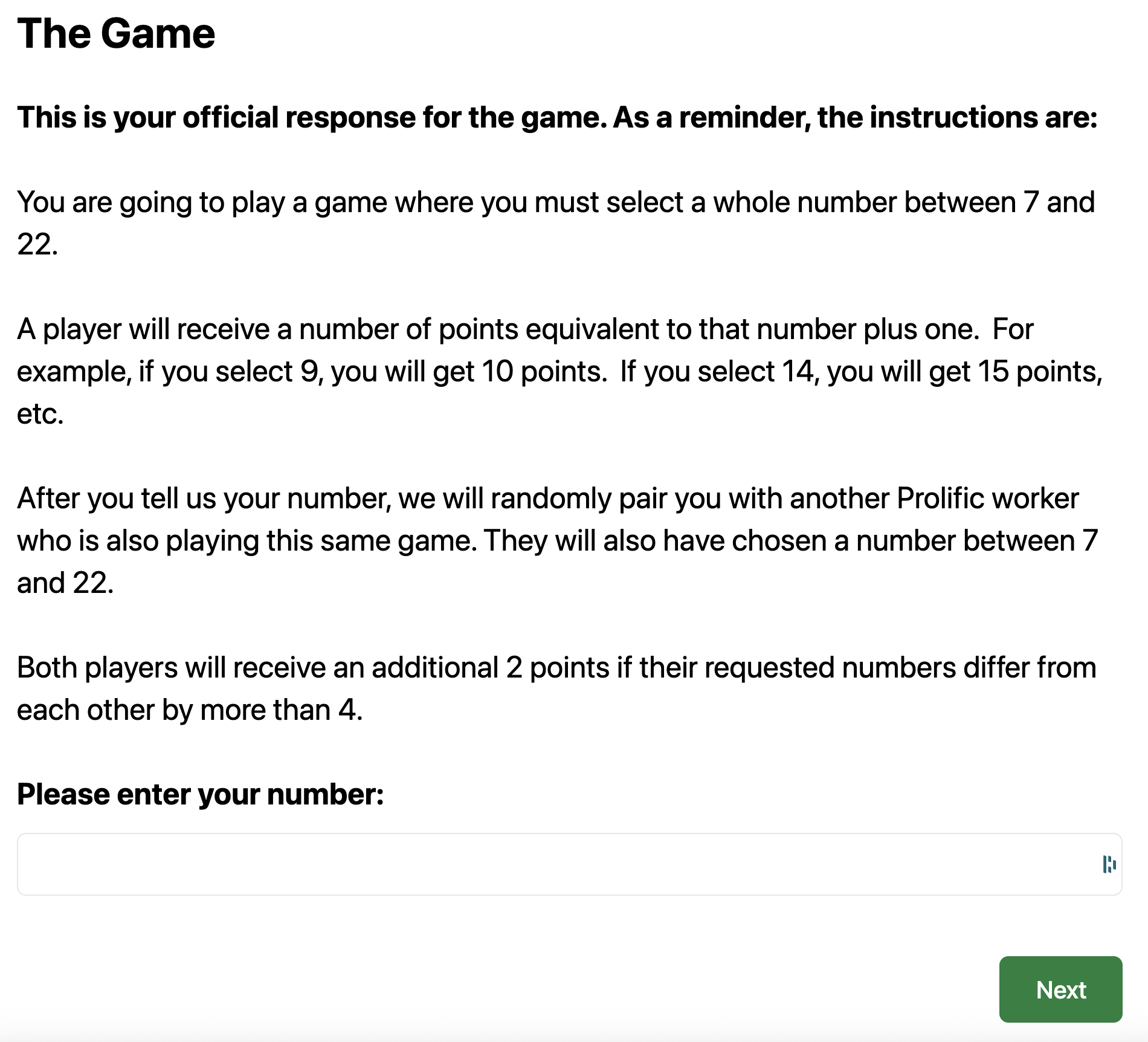}
\end{minipage}
\end{center}
\begin{footnotesize}
\begin{singlespace}
\vspace*{-0.15in}
\emph{Notes:} This figure shows an example screenshot of participants selecting their number for their assigned game from the set of 1,500 games.
\end{singlespace}
\end{footnotesize}
\end{figure}

\renewcommand{\thefigure}{C\arabic{figure}} 
\setcounter{figure}{0}  

\renewcommand{\thetable}{C\arabic{table}} 
\setcounter{table}{0}  

\section{Harsanyi-Selten selector implementation details}
\label{app:nash-selector}

Let $E$ be the finite set of Nash equilibria of a simultaneous, two-player normal-form game that is symmetric, so that the row player's payoff matrix is $U$ and the column player's payoff matrix is $U^\top$.
\cite{HarsanyiSelten1988}'s four-stage procedure deterministically selects a single equilibrium.
For the vast majority of games in our setting, an equilibrium is selected in one of the first three steps.
The fourth is barely used and is more a formality.

Our implementation follows that blueprint with two minimal deviations that (1) protect symmetric components in the Pareto filter and (2) enforce symmetry in the reported profile after tracing.
The procedure deterministically returns a single selection; in symmetric games, and when the tracing routine returns normally, the selected profile is symmetric.
The code is or will soon be available at \url{https://benjaminmanning.io/}.
The following pseudocode broadly outlines the procedure.

\paragraph{Step 1 (component decomposition).}
Two equilibria $e=(\sigma^r,\sigma^c)$ and $e'=(\tau^r,\tau^c)$ are adjacent when they differ in exactly one player's strategy.
The connected components of the resulting graph---call them $C_1,\dots,C_K$---are the equilibrium components.

\paragraph{Step 2 (Pareto filter with symmetry safeguard).}
For each component $C_k$ compute its security vector $v(C_k)=\bigl(\min_{e\in C_k}u_1(e),\,\min_{e\in C_k}u_2(e)\bigr)$, where $u_1(e)=\sigma^{r\top}U\sigma^c$ and $u_2(e)=\sigma^{r\top}U^\top\sigma^c$.
Delete $C_k$ if some $C_\ell$ is strictly better in both coordinates. 

\paragraph{Step 3 (symmetry filter and risk dominance).}
Discard all remaining components that contain no symmetric equilibrium.
If multiple components survive, choose the one whose representative symmetric equilibrium (the first symmetric equilibrium encountered when iterating the component) minimizes the risk-dominance index
\[
R(\sigma)=\sum_{i\neq j}\sigma_i\sigma_j\,
\bigl[U_{ii}-U_{ji}\bigr]\bigl[U_{ii}-U_{ij}\bigr].
\]
If two or more symmetric components attain exactly the same minimal value, we keep the first one encountered in iteration order.
If no symmetric components remain after the symmetry filter, we select the first Pareto-surviving component as a fallback before Step~4.

\paragraph{Step 4 (alpha-tracing).}
Let the winning component be the one selected in Step~3 (or the
first Pareto-surviving component if no symmetric component remains).
\begin{itemize}
\item If the winning component is a singleton that already contains a
symmetric equilibrium, we return it directly (no tracing).
\item Otherwise, we run Gambit's logit $\alpha$-tracing procedure on the full game---not restricted to the winning component---starting from the uniform prior. 
We follow the path to $\alpha=1$ and take the resulting profile as the candidate equilibrium. 
To guard against numerical asymmetries, we then enforce symmetry in the reported profile by setting $\sigma^{r}=\sigma^{c}$ equal to the traced row strategy.
Because the prior and the game are symmetric, the traced profile is generically symmetric; the coercion is a safeguard.
\item \emph{Deviation~2 (singleton asymmetric case).} If the winning
component is a singleton asymmetric equilibrium, we run the same
$\alpha$-tracing procedure and then report the coerced symmetric profile as above. 
If the tracing routine raises an exception, the unique equilibrium is returned unchanged.
\end{itemize}

\renewcommand{\thefigure}{D\arabic{figure}} 
\setcounter{figure}{0}  

\renewcommand{\thetable}{D\arabic{table}} 
\setcounter{table}{0}  

\section{Additional Tables and Figures}
\label{app:figs}

\subsection{Permutations of the 11-20 game in Section~\ref{sec:predict_new_AR}}

\begin{figure}[h!]
\begin{center}
\begin{minipage}{1 \linewidth}
\caption{Response Distributions for the Basic Version for the 11-20 Game with raw candidate responses} 
\centering
\vspace*{-0.15in}
\label{fig:no_mix}
\includegraphics[width=\textwidth]{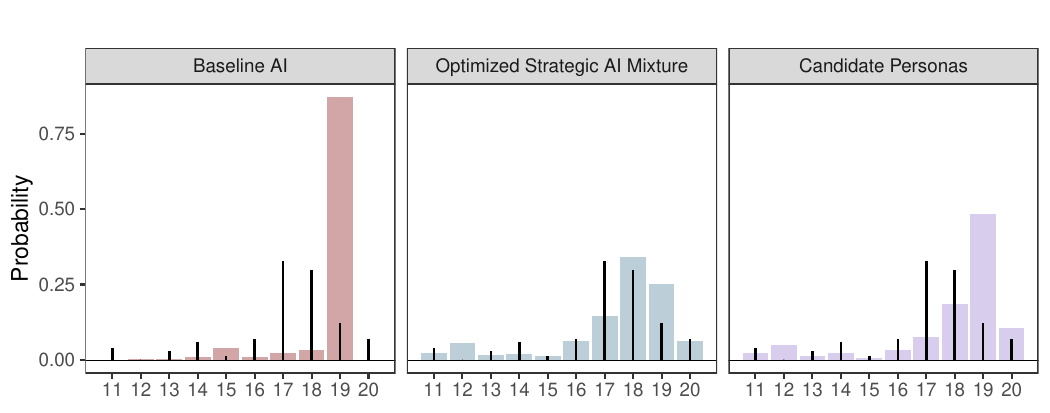}
\end{minipage}
\end{center}
\begin{footnotesize}
\begin{singlespace}
\vspace*{-0.15in}
\emph{Notes:} This figure displays empirical PMFs for three samples playing the basic 11-20 money request game: human subjects from \citeauthor{1120Arad2012} (left panel), the naive baseline (center-left panel), responses from our selected \aisub{s} based on the weights in Table~\ref{tab:persona} (center-right panel), and responses based on the unweighted and evenly distributed \persona{s} in Table~\ref{tab:persona} (right panel).
\end{singlespace}
\end{footnotesize}
\end{figure}

\begin{table}[h!]
\begin{center}
\caption{Atheoretical \aisub{s} and resulting mixture weights} \label{tab:bad_personas}
\scriptsize
\begin{tabular}{l cl c}
\hline \addlinespace[4pt]
\multicolumn{4}{c}{\textbf{Historical Figures}}\\\addlinespace[5pt]
\hline\addlinespace[4pt]
Persona & Weight & Persona & Weight \\
\hline\addlinespace[4pt]
Cleopatra & 0.000 & Genghis Khan & 0.000 \\
Julius Caesar & 0.891 & Mother Teresa & 0.000 \\
Confucius & 0.109 & Martin Luther King & 0.000 \\
Joan of Arc & 0.000 & Frida Kahlo & 0.000 \\
Nelson Mandela & 0.000 & George Washington & 0.000 \\
Mahatma Gandhi & 0.000 & Winston Churchill & 0.000 \\
Harriet Tubman & 0.000 & Mansa Musa & 0.000 \\
Leonardo da Vinci & 0.000 & Sacagawea & 0.000 \\
Albert Einstein & 0.000 & Emmeline Pankhurst & 0.000 \\
Marie Curie & 0.000 & Socrates & 0.000 \\
\hline\addlinespace[5pt]
\multicolumn{4}{c}{\textbf{MBTI Types}} \\\addlinespace[4pt]
\hline\addlinespace[4pt]
Type & Weight & Type & Weight \\
\hline\addlinespace[4pt]
You are an ESTJ  & 0.000 & You are an ISTJ  & 0.000 \\
You are an ESTP  & 0.000 & You are an ISTP  & 0.000 \\
You are an ESFJ  & 0.000 & You are an ISFJ  & 0.000 \\
You are an ESFP  & 0.000 & You are an ISFP  & 0.000 \\
You are an ENTJ  & 0.000 & You are an INTJ  & 0.000 \\
You are an ENTP  & 0.000 & You are an INTP  & 0.000 \\
You are an ENFJ  & 0.000 & You are an INFJ  & 0.000 \\
You are an ENFP  & 1.000 & You are an INFP  & 0.000 \\
\hline\addlinespace[5pt]
\multicolumn{4}{c}{\textbf{Always Pick `N`}} \\\addlinespace[4pt]
\hline\addlinespace[4pt]
Number & Weight & Number & Weight \\
\hline\addlinespace[4pt]
You always like to pick 11 & 0.037 & You always like to pick 16 & 0.065 \\
You always like to pick 12 & 0.000 & You always like to pick 17 & 0.324 \\
You always like to pick 13 & 0.028 & You always like to pick 18 & 0.296 \\
You always like to pick 14 & 0.056 & You always like to pick 19 & 0.120 \\
You always like to pick 15 & 0.009 & You always like to pick 20 & 0.065 \\
\hline
\end{tabular}

\end{center}
\begin{footnotesize}
\begin{singlespace}
\vspace*{-0.05in}
\emph{Notes:} This table displays three sets of arbitrary \persona{s}---each a different $\Theta$. 
The weights columns display the optimized weights $\bm{w^*}$ when performing the selection method on the basic version of the 11-20 game.
Weights sum to 1 within each set. 
For the historical figures, each \persona{} is told \emph{``You are X''} where X is a historical figure.
For the Myers-Briggs set, each \persona{} is also told that the four letters are in reference to the Myers-Briggs personality type indicator.
\end{singlespace}
\end{footnotesize}
\end{table}

\begin{figure}[h!]
\begin{center}
\begin{minipage}{1 \linewidth}
\caption{Comparison of novel 1-10 for alternative distance metrics} 
\centering
\vspace*{-0.15in}
\label{fig:dist_compare_1_10}
\includegraphics[width=\textwidth]{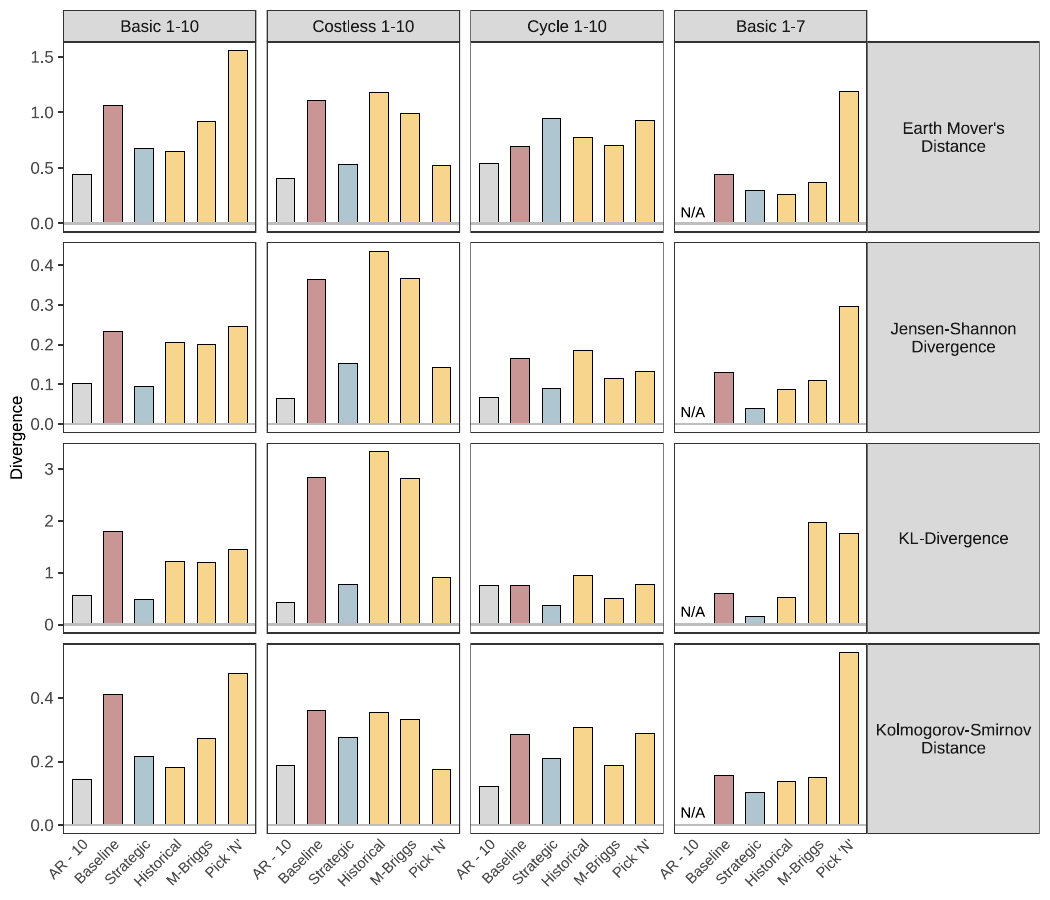}
\end{minipage}
\end{center}
\begin{footnotesize}
\begin{singlespace}
\vspace*{-0.15in}
\emph{Notes:} Reports the divergence between human and each AI distribution for the novel games across various additional distance metrics. 
For three of the four metrics, the optimized strategic agents outperform the baseline. 
Only in the costless version of the game, with the Earth Mover’s distance as the metric, is the baseline slightly better.
\end{singlespace}
\end{footnotesize}
\end{figure}

\newpage \clearpage

\subsection{Additional Tables and Figures for Section~\ref{sec:ext-valid}}

\begin{figure}[h!]
\begin{center}
\begin{minipage}{\textwidth}
\caption{Equilibrium PMFs across all games in $S$ with mixed strategy Harsanyi-Selten solutions} 
\label{fig:eq_plot}
\vspace*{-0.1in}
\includegraphics[width=\textwidth]{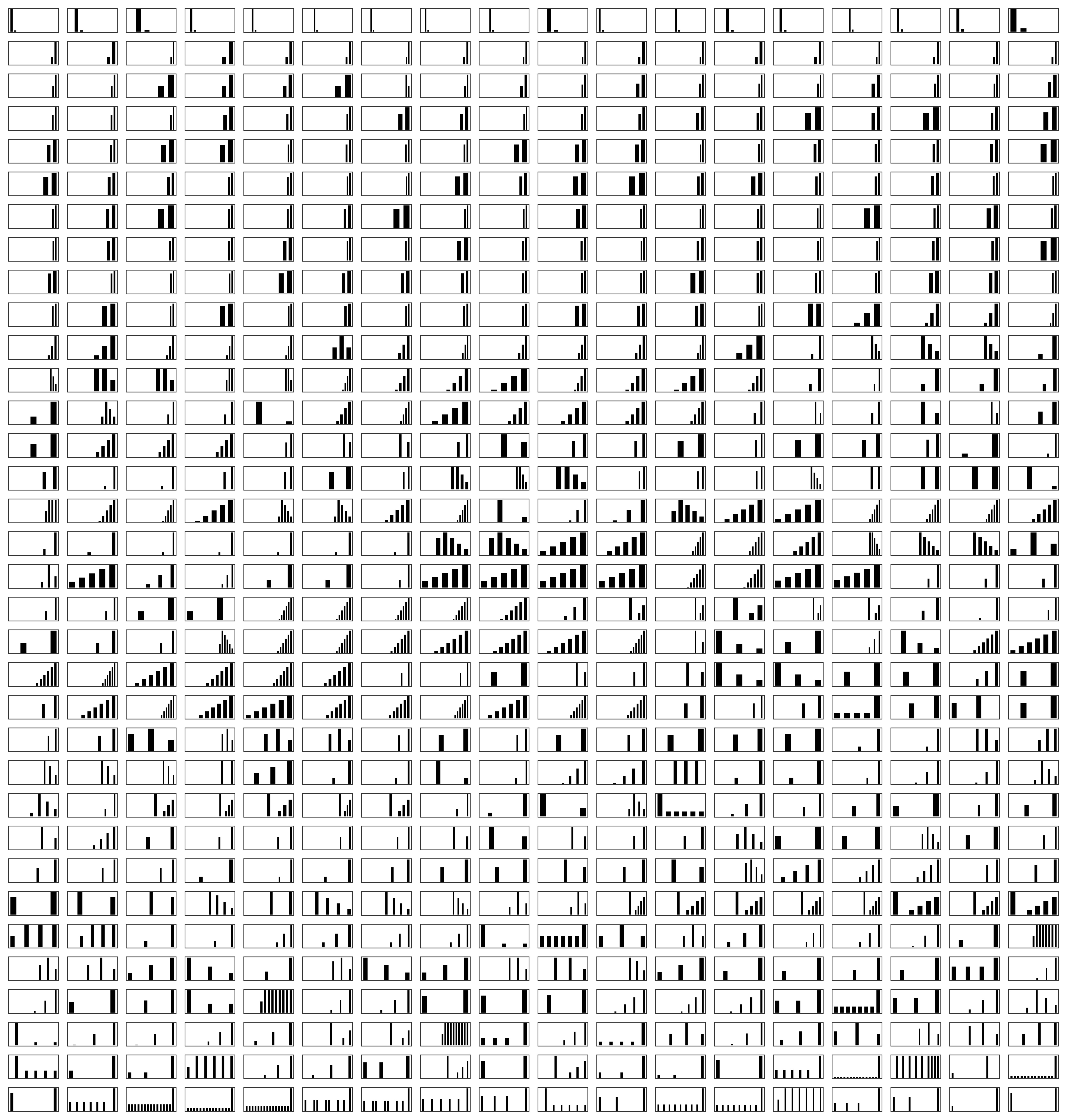} 
\end{minipage}
\end{center}
\begin{footnotesize}
\begin{singlespace}
\vspace*{-0.05in}
\emph{Notes:} Each panel shows the PMF of a mixed strategy Harsanyi-Selten equilibrium for a game in $S$. 
Panels are ordered by the variance of the equilibrium distribution (left to right, top to bottom). 
The x-axes are scaled freely, so games with different action spaces (e.g., 5, 11, or 20 options) span the same width. All y-axes use probability units but are scaled freely in height to better visualize different distributions. 
612 games with mixed strategy equilibria are shown. 
The rest of $S$ admit pure-strategy equilibria, collapsing to a single vertical bar (not shown).
\end{singlespace}
\end{footnotesize}
\end{figure}

\newpage \clearpage

\subsection{Robustness checks for the family of pre-committed games in Section~\ref{sec:ext-valid}}
\label{app:robustness_checks}

\begin{table}[h!]
\begin{center}
\caption{Summary statistics for absolute predictive accuracy of different models ($\varepsilon = 0$).} 
\label{tab:match_rates}
\footnotesize
\renewcommand{\arraystretch}{1.25}
\begin{tabular}{lcccc}
\toprule
 & Strategic AI & Baseline AI & HS Eq. & Cog. Hierarchy\\
\midrule
\% Humans Choose Max Prob. Strategy & 24.3 & 16.8 & 30.4 & 28.5\\
\% Humans Choose Top 3 Prob. Strategy & 52.9 & 39.1 & 49.6 & 49.5\\
\% Humans Choose Pos. Prob. Strategy & 94.3 & 81.9 & 46.4 & 100.0\\
\% Games Any Human Chooses Pos. Prob. Strategy & 99.3 & 93.7 & 74.7 & 100.0\\
\% Games All Humans Choose Pos. Prob. Strategy & 86.3 & 65.3 & 17.7 & 100.0\\
\bottomrule
\end{tabular}

\end{center}
\begin{footnotesize}
\begin{singlespace}
\vspace*{-0.05in}
\emph{Notes:} This table reports summary statistics for the absolute predictive accuracy of different models in Section~\ref{sec:ext-valid}.
These are the raw results without any smoothing.
The first row corresponds to the proportion of human subjects who chose the most likely strategy for the given model in the columns.
The second row corresponds to the proportion of human subjects who chose the top 3 most likely strategies for the given model in the columns.
The third is the proportion of human subjects who chose a strategy with a positive probability for the given model in the columns.
The fourth is the proportion of games in which any human chose a strategy with a positive probability for the given model in the columns.
The fifth is the proportion of games in which all humans chose a strategy with a positive probability for the given model in the columns.
Note that the cognitive hierarchy model, by design, provides positive probability for all strategies (even without smoothing) because the level-$0$ player chooses uniformly at random.
None of the other models has this feature.
\end{singlespace}
\end{footnotesize}
\end{table}

\begin{table}[h!]
\begin{center}
\caption{Statistical tests comparing strategic \aisub{s} vs other models ($\varepsilon = 0.05$)} \label{tab:stat_bounds_05}
\scriptsize
\begin{tabular}{l c c c c}
    \hline \addlinespace[4pt]
    \textbf{Comparison ($n$ Games)} & \multicolumn{1}{c}{\textbf{$\mathbf{\bar{\Lambda}_S}$}} & \multicolumn{1}{c}{\textbf{Wilcoxon}} & \multicolumn{1}{c}{\textbf{Permutation Test}} & \multicolumn{1}{c}{\textbf{$\sum_{s \in S} \mathbf 1\{\hat \Lambda_s > 0\} / |S|$}} \\
\addlinespace[6pt]\hline\addlinespace[6pt]
Baseline AI & \phantom{(}\phantom{....}0.668*** & $p<.001$*** & $p<.001$*** & \phantom{(}\phantom{....}0.726*** \\
 &  (0.027) & & &  (0.012) \\
\addlinespace[2pt]
\addlinespace[6pt]
Cognitive Hierarchy & \phantom{(}\phantom{....}0.302*** & $p<.001$*** & $p<.001$*** & \phantom{(}\phantom{....}0.596*** \\
 &  (0.030) & & &  (0.013) \\
\addlinespace[2pt]
\addlinespace[6pt]
Harsanyi-Selten Nash & \phantom{(}\phantom{....}0.924*** & $p<.001$*** & $p<.001$*** & \phantom{(}\phantom{....}0.730*** \\
 &  (0.039) & & &  (0.012) \\
\addlinespace[2pt]
\quad\quad \textit{Mixed} & \phantom{(}\phantom{....}0.790*** & $p<.001$*** & $p<.001$*** & \phantom{(}\phantom{....}0.732*** \\
 &  (0.048) & & &  (0.018) \\
\addlinespace[2pt]
\quad\quad \textit{Pure} & \phantom{(}\phantom{....}1.018*** & $p<.001$*** & $p<.001$*** & \phantom{(}\phantom{....}0.728*** \\
 &  (0.054) & & &  (0.015) \\
\addlinespace[2pt]
\addlinespace[6pt]
Random Pure Strategy & \phantom{(}\phantom{....}2.574*** & $p<.001$*** & $p<.001$*** & \phantom{(}\phantom{....}0.942*** \\
 &  (0.036) & & &  (0.006) \\
\addlinespace[2pt]
\addlinespace[6pt]
Uniform & \phantom{(}\phantom{....}0.106*** & $p<.001$*** & $p<.001$*** & \phantom{(}\phantom{....}0.598*** \\
 &  (0.021) & & &  (0.013) \\
\addlinespace[2pt]
\hline
\end{tabular}

\end{center}
\begin{footnotesize}
\begin{singlespace}
\vspace*{-0.05in}
\emph{Notes:} This table shows the results of the statistical tests comparing the strategic \aisub{s} to the other models for $\varepsilon = 0.05$.
No smoothing is applied to the cognitive hierarchy model.
The first column shows the comparison model.
The second presents $\bar \Lambda_S$ with bootstrap confidence intervals comparing the strategic \aisub{s} to the other models.
The third and fourth columns present p-values for the Wilcoxon signed-rank test and random-sign permutation test, respectively.
The fifth column presents the proportion of games for which the strategic \aisub{s} is the best predictor with its 95\% Clopper-Pearson interval. \starlanguage
\end{singlespace}
\end{footnotesize}
\end{table}

\begin{table}[h!]
\begin{center}
\caption{Statistical tests comparing optimized vs other models ($\varepsilon = 0.1$)} \label{tab:stat_bounds_10}
\scriptsize
\begin{tabular}{l c c c c}
    \hline \addlinespace[4pt]
    \textbf{Comparison ($n$ Games)} & \multicolumn{1}{c}{\textbf{$\mathbf{\bar{\Lambda}_S}$}} & \multicolumn{1}{c}{\textbf{Wilcoxon}} & \multicolumn{1}{c}{\textbf{Permutation Test}} & \multicolumn{1}{c}{\textbf{$\sum_{s \in S} \mathbf 1\{\hat \Lambda_s > 0\} / |S|$}} \\
\addlinespace[6pt]\hline\addlinespace[6pt]
Baseline AI & \phantom{(}\phantom{....}0.556*** & $p<.001$*** & $p<.001$*** & \phantom{(}\phantom{....}0.724*** \\
 &  (0.023) & & &  (0.012) \\
\addlinespace[2pt]
\addlinespace[6pt]
Cognitive Hierarchy & \phantom{(}\phantom{....}0.350*** & $p<.001$*** & $p<.001$*** & \phantom{(}\phantom{....}0.619*** \\
 &  (0.029) & & &  (0.013) \\
\addlinespace[2pt]
\addlinespace[6pt]
Harsanyi-Selten Nash & \phantom{(}\phantom{....}0.614*** & $p<.001$*** & $p<.001$*** & \phantom{(}\phantom{....}0.679*** \\
 &  (0.033) & & &  (0.012) \\
\addlinespace[2pt]
\quad\quad \textit{Mixed} & \phantom{(}\phantom{....}0.534*** & $p<.001$*** & $p<.001$*** & \phantom{(}\phantom{....}0.691*** \\
 &  (0.040) & & &  (0.019) \\
\addlinespace[2pt]
\quad\quad \textit{Pure} & \phantom{(}\phantom{....}0.670*** & $p<.001$*** & $p<.001$*** & \phantom{(}\phantom{....}0.671*** \\
 &  (0.046) & & &  (0.016) \\
\addlinespace[2pt]
\addlinespace[6pt]
Random Pure Strategy & \phantom{(}\phantom{....}2.004*** & $p<.001$*** & $p<.001$*** & \phantom{(}\phantom{....}0.928*** \\
 &  (0.032) & & &  (0.007) \\
\addlinespace[2pt]
\addlinespace[6pt]
Uniform & \phantom{(}\phantom{....}0.153*** & $p<.001$*** & $p<.001$*** & \phantom{(}\phantom{....}0.611*** \\
 &  (0.019) & & &  (0.013) \\
\addlinespace[2pt]
\hline
\end{tabular}

\end{center}
\begin{footnotesize}
\begin{singlespace}
\vspace*{-0.05in}
\emph{Notes:} This table shows the results of the statistical tests comparing the strategic \aisub{s} to the other models for $\varepsilon = 0.1$.
No smoothing is applied to the cognitive hierarchy model.
The first column shows the comparison model.
The second presents $\bar \Lambda_S$ with bootstrap confidence intervals comparing the strategic \aisub{s} to the other models.
The third and fourth columns present p-values for the Wilcoxon signed-rank test and random-sign permutation test, respectively.
The fifth column presents the proportion of games for which the strategic \aisub{s} is the best predictor with its 95\% Clopper-Pearson interval. \starlanguage
\end{singlespace}
\end{footnotesize}
\end{table}

\begin{table}[h!]
\begin{center}
\caption{Statistical tests comparing optimized vs other models ($\varepsilon = 0.2$)} \label{tab:stat_bounds_20}
\scriptsize
\begin{tabular}{l c c c c}
    \hline \addlinespace[4pt]
    \textbf{Comparison ($n$ Games)} & \multicolumn{1}{c}{\textbf{$\mathbf{\bar{\Lambda}_S}$}} & \multicolumn{1}{c}{\textbf{Wilcoxon}} & \multicolumn{1}{c}{\textbf{Permutation Test}} & \multicolumn{1}{c}{\textbf{$\sum_{s \in S} \mathbf 1\{\hat \Lambda_s > 0\} / |S|$}} \\
\addlinespace[6pt]\hline\addlinespace[6pt]
Baseline AI & \phantom{(}\phantom{....}0.429*** & $p<.001$*** & $p<.001$*** & \phantom{(}\phantom{....}0.715*** \\
 &  (0.018) & & &  (0.012) \\
\addlinespace[2pt]
\addlinespace[6pt]
Cognitive Hierarchy & \phantom{(}\phantom{....}0.395*** & $p<.001$*** & $p<.001$*** & \phantom{(}\phantom{....}0.640*** \\
 &  (0.029) & & &  (0.012) \\
\addlinespace[2pt]
\addlinespace[6pt]
Harsanyi-Selten Nash & \phantom{(}\phantom{....}0.323*** & $p<.001$*** & $p<.001$*** & \phantom{(}\phantom{....}0.622*** \\
 &  (0.027) & & &  (0.013) \\
\addlinespace[2pt]
\quad\quad \textit{Mixed} & \phantom{(}\phantom{....}0.299*** & $p<.001$*** & $p<.001$*** & \phantom{(}\phantom{....}0.647*** \\
 &  (0.031) & & &  (0.019) \\
\addlinespace[2pt]
\quad\quad \textit{Pure} & \phantom{(}\phantom{....}0.339*** & $p<.001$*** & $p<.001$*** & \phantom{(}\phantom{....}0.604*** \\
 &  (0.038) & & &  (0.017) \\
\addlinespace[2pt]
\addlinespace[6pt]
Random Pure Strategy & \phantom{(}\phantom{....}1.436*** & $p<.001$*** & $p<.001$*** & \phantom{(}\phantom{....}0.902*** \\
 &  (0.026) & & &  (0.008) \\
\addlinespace[2pt]
\addlinespace[6pt]
Uniform & \phantom{(}\phantom{....}0.198*** & $p<.001$*** & $p<.001$*** & \phantom{(}\phantom{....}0.643*** \\
 &  (0.016) & & &  (0.012) \\
\addlinespace[2pt]
\hline
\end{tabular}

\end{center}
\begin{footnotesize}
\begin{singlespace}
\vspace*{-0.05in}
\emph{Notes:} This table shows the results of the statistical tests comparing the strategic \aisub{s} to the other models for $\varepsilon = 0.2$.
No smoothing is applied to the cognitive hierarchy model.
The first column shows the comparison model.
The second presents $\bar \Lambda_S$ with bootstrap confidence intervals comparing the strategic \aisub{s} to the other models.
The third and fourth columns present p-values for the Wilcoxon signed-rank test and random-sign permutation test, respectively.
The fifth column presents the proportion of games for which the strategic \aisub{s} is the best predictor with its 95\% Clopper-Pearson interval. \starlanguage
\end{singlespace}
\end{footnotesize}
\end{table}

\begin{table}[h!]
\begin{center}
\caption{Statistical tests comparing optimized vs other models ($\varepsilon = 0.3$)} \label{tab:stat_bounds_30}
\scriptsize
\begin{tabular}{l c c c c}
    \hline \addlinespace[4pt]
    \textbf{Comparison ($n$ Games)} & \multicolumn{1}{c}{\textbf{$\mathbf{\bar{\Lambda}_S}$}} & \multicolumn{1}{c}{\textbf{Wilcoxon}} & \multicolumn{1}{c}{\textbf{Permutation Test}} & \multicolumn{1}{c}{\textbf{$\sum_{s \in S} \mathbf 1\{\hat \Lambda_s > 0\} / |S|$}} \\
\addlinespace[6pt]\hline\addlinespace[6pt]
Baseline AI & \phantom{(}\phantom{....}0.344*** & $p<.001$*** & $p<.001$*** & \phantom{(}\phantom{....}0.705*** \\
 &  (0.015) & & &  (0.012) \\
\addlinespace[2pt]
\addlinespace[6pt]
Cognitive Hierarchy & \phantom{(}\phantom{....}0.414*** & $p<.001$*** & $p<.001$*** & \phantom{(}\phantom{....}0.647*** \\
 &  (0.028) & & &  (0.012) \\
\addlinespace[2pt]
\addlinespace[6pt]
Harsanyi-Selten Nash & \phantom{(}\phantom{....}0.165*** & $p<.001$*** & $p<.001$*** & \phantom{(}\phantom{....}0.575*** \\
 &  (0.023) & & &  (0.013) \\
\addlinespace[2pt]
\quad\quad \textit{Mixed} & \phantom{(}\phantom{....}0.175*** & $p<.001$*** & $p<.001$*** & \phantom{(}\phantom{....}0.611*** \\
 &  (0.026) & & &  (0.020) \\
\addlinespace[2pt]
\quad\quad \textit{Pure} & \phantom{(}\phantom{....}0.158*** & $p<.001$*** & $p<.001$*** & \phantom{(}\phantom{..}0.551** \\
 &  (0.033) & & &  (0.017) \\
\addlinespace[2pt]
\addlinespace[6pt]
Random Pure Strategy & \phantom{(}\phantom{....}1.102*** & $p<.001$*** & $p<.001$*** & \phantom{(}\phantom{....}0.884*** \\
 &  (0.023) & & &  (0.008) \\
\addlinespace[2pt]
\addlinespace[6pt]
Uniform & \phantom{(}\phantom{....}0.217*** & $p<.001$*** & $p<.001$*** & \phantom{(}\phantom{....}0.666*** \\
 &  (0.014) & & &  (0.012) \\
\addlinespace[2pt]
\hline
\end{tabular}

\end{center}
\begin{footnotesize}
\begin{singlespace}
\vspace*{-0.05in}
\emph{Notes:} This table shows the results of the statistical tests comparing the strategic \aisub{s} to the other models for $\varepsilon = 0.3$.
No smoothing is applied to the cognitive hierarchy model.
The first column shows the comparison model.
The second presents $\bar \Lambda_S$ with bootstrap confidence intervals comparing the strategic \aisub{s} to the other models.
The third and fourth columns present p-values for the Wilcoxon signed-rank test and random-sign permutation test, respectively.
The fifth column presents the proportion of games for which the strategic \aisub{s} is the best predictor with its 95\% Clopper-Pearson interval. \starlanguage
\end{singlespace}
\end{footnotesize}
\end{table}

\begin{figure}[h!]
\begin{center}
\begin{minipage}{\textwidth}
\caption{Relative predictive accuracy of the strategic \aisub{} vs other models for different game types ($\varepsilon = 0.05$)} 
\label{fig:game_level_plots_eps_005}
\vspace*{-0.1in}
\includegraphics[width=\textwidth]{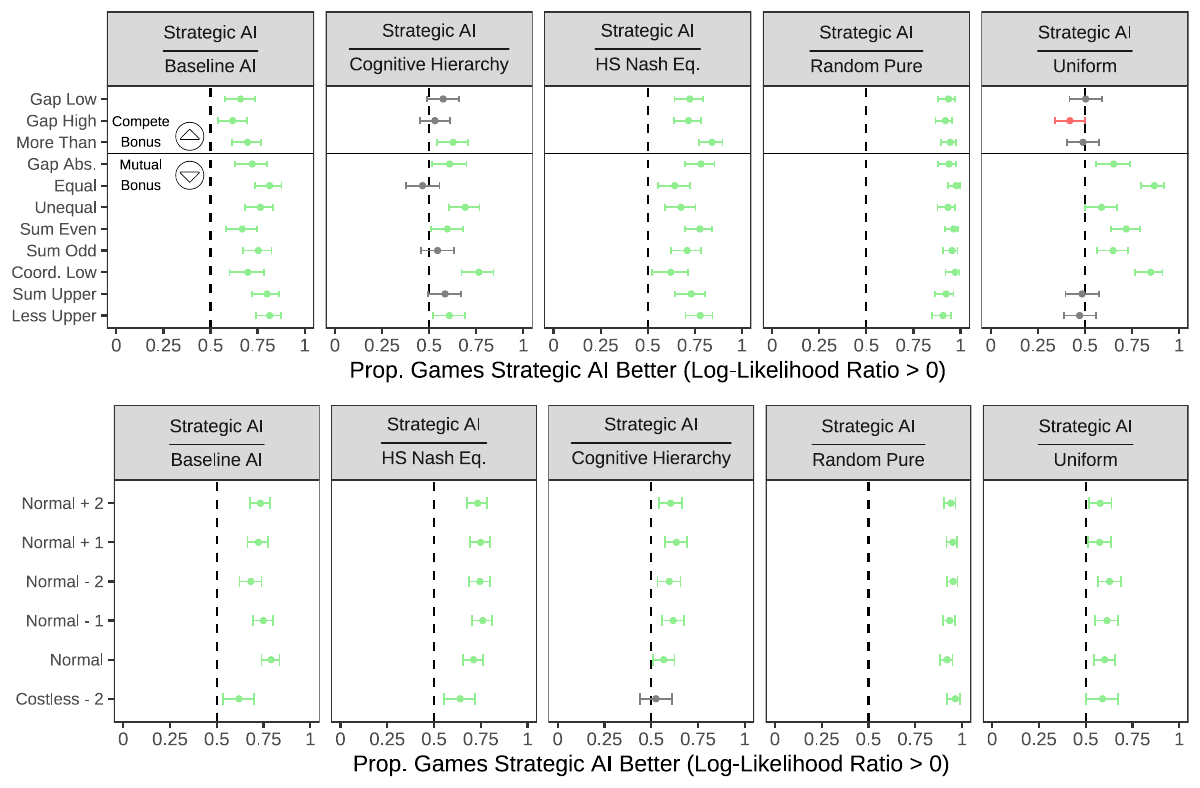} 
\end{minipage}
\end{center}
\begin{footnotesize}
\begin{singlespace}
\vspace*{-0.05in}
\emph{Notes:} This figure shows the proportion of games for which strategic \aisub{s} are the best predictor of initial play, separated by bonus rule (top panel) and points rule (bottom panel). 
All distributions except the cognitive hierarchy model are smoothed with $\varepsilon = 0.05$. 
The vertical dashed line represents equal performance (50-50 split). 
Green indicates that strategic \aisub{} significantly outperforms the reference model in more than 50\% of games, red indicates significantly worse performance in more than 50\% of games, and grey indicates no significant difference. 
Error bars show 95\% Clopper-Pearson confidence intervals.
\end{singlespace}
\end{footnotesize}
\end{figure}

\begin{figure}[h!]
\begin{center}
\begin{minipage}{\textwidth}
\caption{Relative predictive accuracy of the strategic \aisub{} vs other models for different game types ($\varepsilon = 0.1$)} 
\label{fig:game_level_plots_eps_01}
\vspace*{-0.1in}
\includegraphics[width=\textwidth]{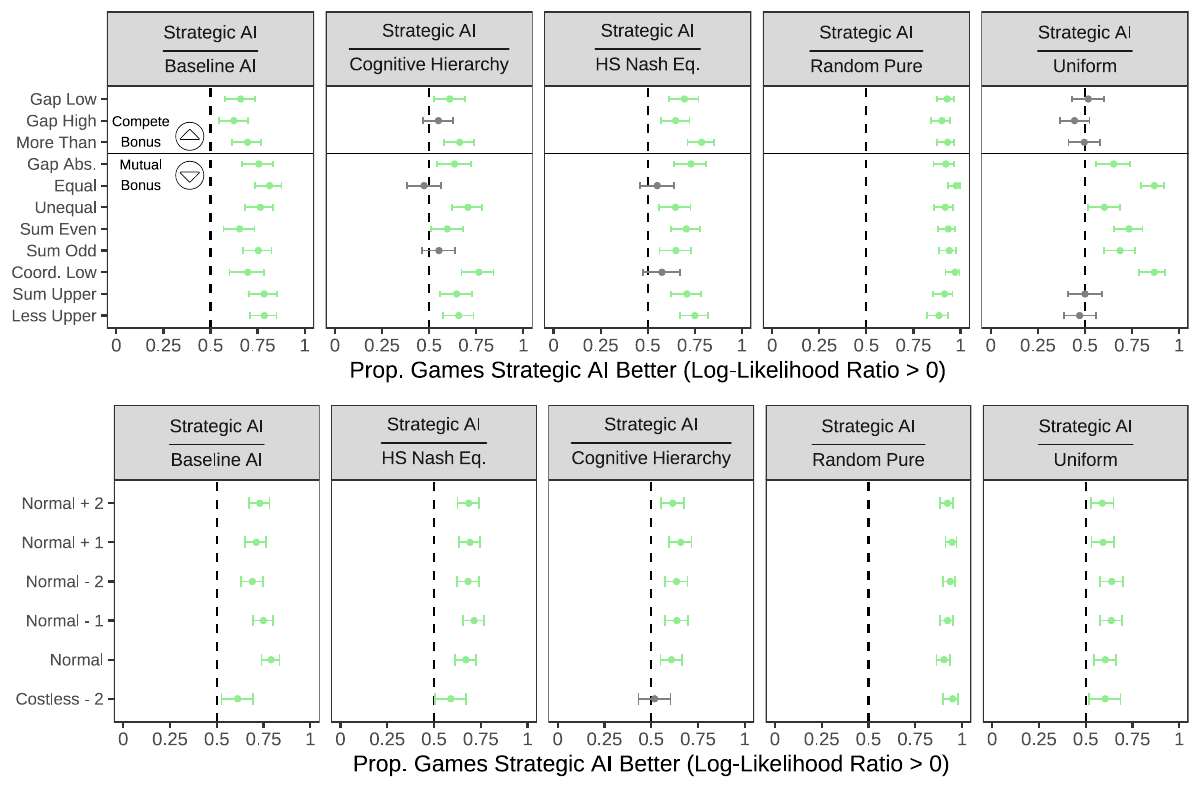} 
\end{minipage}
\end{center}
\begin{footnotesize}
\begin{singlespace}
\vspace*{-0.05in}
\emph{Notes:} This figure shows the proportion of games for which strategic \aisub{s} are the best predictor of initial play, separated by bonus rule (top panel) and points rule (bottom panel). 
All distributions except the cognitive hierarchy model are smoothed with $\varepsilon = 0.1$.
The vertical dashed line represents equal performance (50-50 split). 
Green indicates that strategic \aisub{} significantly outperforms the reference model in more than 50\% of games, red indicates significantly worse performance in more than 50\% of games, and grey indicates no significant difference. 
Error bars show 95\% Clopper-Pearson confidence intervals.
\end{singlespace}
\end{footnotesize}
\end{figure}

\begin{figure}[h!]
\begin{center}
\begin{minipage}{\textwidth}
\caption{Relative predictive accuracy of the strategic \aisub{} vs other models for different game types ($\varepsilon = 0.2$)} 
\label{fig:game_level_plots_eps_02}
\vspace*{-0.1in}
\includegraphics[width=\textwidth]{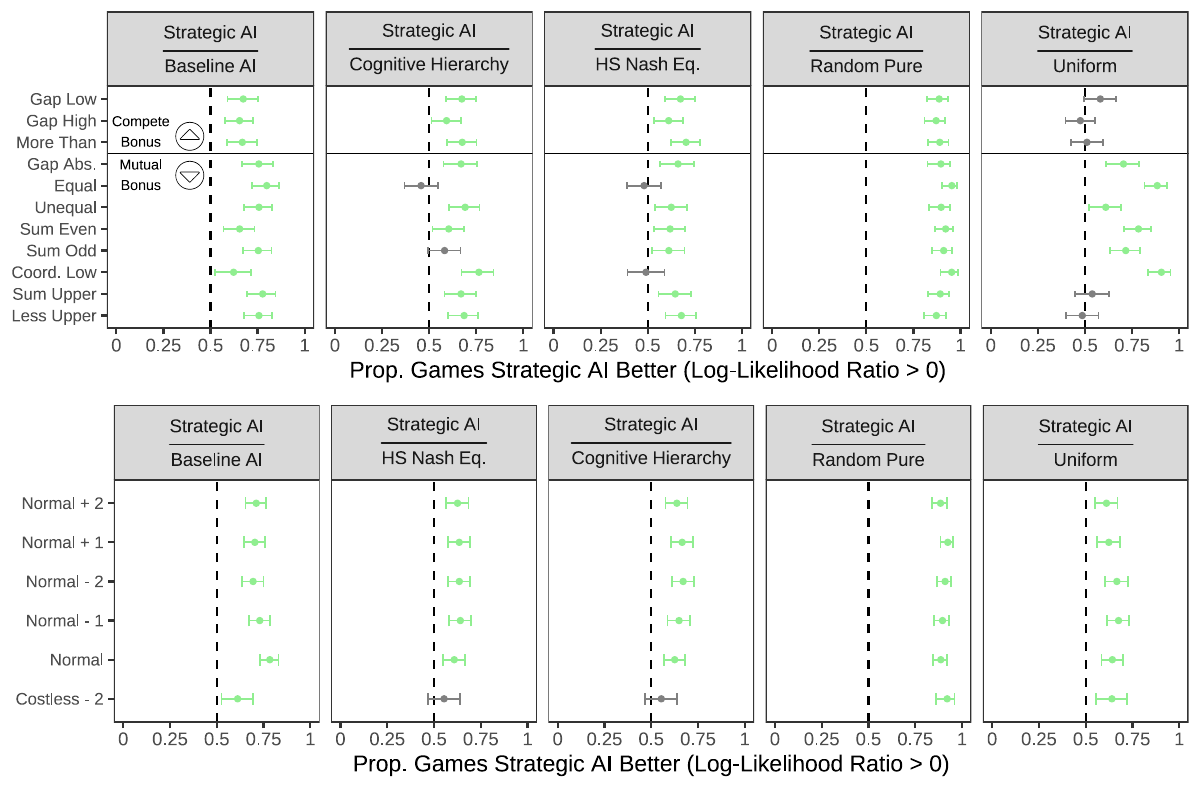} 
\end{minipage}
\end{center}
\begin{footnotesize}
\begin{singlespace}
\vspace*{-0.05in}
\emph{Notes:} This figure shows the proportion of games for which strategic \aisub{s} are the best predictor of initial play, separated by bonus rule (top panel) and points rule (bottom panel). 
All distributions except the cognitive hierarchy model are smoothed with $\varepsilon = 0.2$.
The vertical dashed line represents equal performance (50-50 split). Green indicates that strategic \aisub{} significantly outperforms the reference model in more than 50\% of games, red indicates significantly worse performance in more than 50\% of games, and grey indicates no significant difference. 
Error bars show 95\% Clopper-Pearson confidence intervals.
\end{singlespace}
\end{footnotesize}
\end{figure}

\begin{figure}[h!]
\begin{center}
\begin{minipage}{\textwidth}
\caption{Relative predictive accuracy of the strategic \aisub{} vs other models for different game types ($\varepsilon = 0.3$)} 
\label{fig:game_level_plots_eps_03}
\vspace*{-0.1in}
\includegraphics[width=\textwidth]{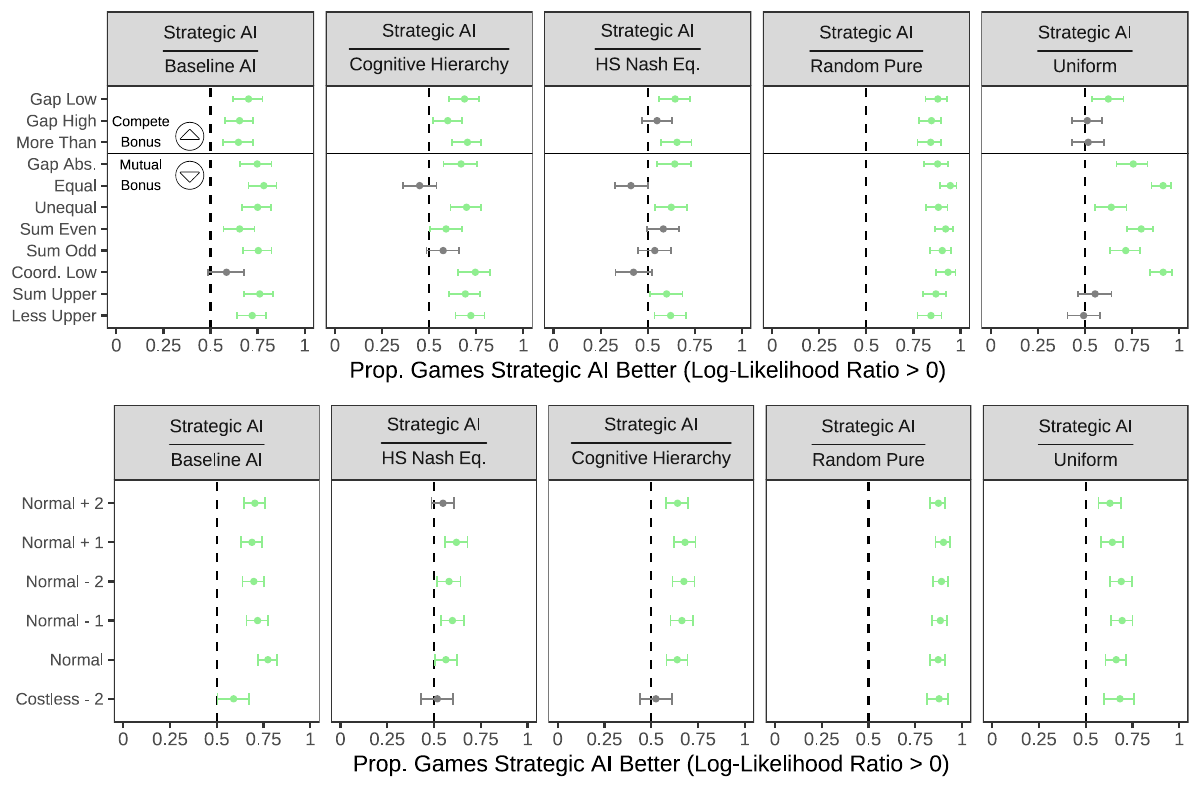} 
\end{minipage}
\end{center}
\begin{footnotesize}
\begin{singlespace}
\vspace*{-0.05in}
\emph{Notes:} This figure shows the proportion of games for which strategic \aisub{s} are the best predictor of initial play, separated by bonus rule (top panel) and points rule (bottom panel). 
All distributions except the cognitive hierarchy model are smoothed with $\varepsilon = 0.3$.
The vertical dashed line represents equal performance (50-50 split). Green indicates that strategic \aisub{} significantly outperforms the reference model in more than 50\% of games, red indicates significantly worse performance in more than 50\% of games, and grey indicates no significant difference. 
Error bars show 95\% Clopper-Pearson confidence intervals.
\end{singlespace}
\end{footnotesize}
\end{figure}

\newpage \clearpage

\begin{table}[h!] 
\begin{center}
\begin{minipage}{1 \linewidth}                          
\caption{Log-likelihood ratio regressions across game types ($\varepsilon = 0.05$)}  
\label{tab:game_level_eps_005}     
\centering                                                  
\scriptsize
\renewcommand{\arraystretch}{1.2}
\begin{tabular}{@{\extracolsep{5pt}}lccccc} 
\\[-1.8ex]\hline 
\hline \\[-1.8ex] 
 & \multicolumn{5}{c}{Log-Likelihood Ratio (Strategic AI / \{...\})} \\ 
\cline{2-6} 
 & \{Baseline AI\} & \{HS Nash Eq.\} & \{Cognitive Hierarchy\} & \{Random Pure\} & \{Uniform\} \\ 
\\[-1.8ex] & (1) & (2) & (3) & (4) & (5)\\ 
\hline \\[-1.8ex] 
 Bonus Size & 0.006 & $-$0.020$^{**}$ & 0.009 & 0.014$^{*}$ & 0.010$^{**}$ \\ 
  & (0.005) & (0.007) & (0.005) & (0.006) & (0.004) \\ 
  & & & & & \\ 
 Lower Bound & 0.015$^{**}$ & $-$0.001 & $-$0.009 & $-$0.001 & $-$0.006 \\ 
  & (0.005) & (0.007) & (0.005) & (0.006) & (0.004) \\ 
  & & & & & \\ 
 Action Space Size & 0.030$^{***}$ & 0.020$^{*}$ & $-$0.004 & 0.049$^{***}$ & 0.011$^{*}$ \\ 
  & (0.006) & (0.008) & (0.007) & (0.008) & (0.005) \\ 
  & & & & & \\ 
 Gap & $-$0.019 & 0.060 & 0.017 & $-$0.026 & $-$0.017 \\ 
  & (0.025) & (0.034) & (0.028) & (0.031) & (0.019) \\ 
  & & & & & \\ 
 Constant & 0.127 & 0.745$^{***}$ & 0.315$^{*}$ & 1.901$^{***}$ & $-$0.025 \\ 
  & (0.117) & (0.158) & (0.128) & (0.159) & (0.091) \\ 
  & & & & & \\ 
\hline \\[-1.8ex] 
Observations & 1,477 & 1,477 & 1,477 & 1,477 & 1,477 \\ 
R$^{2}$ & 0.024 & 0.013 & 0.005 & 0.030 & 0.011 \\ 
\hline 
\hline \\[-1.8ex] 
\end{tabular}

\end{minipage}
\end{center}
\begin{footnotesize}
\begin{singlespace}
\vspace*{-0.15in}
\emph{Notes:} Each column reports regression estimates where the dependent variable is the log-likelihood ratio of the strategic \aisub{} to the other models.
The independent variables are the game features from Table~\ref{tab:game_parameters} (excluding the bonus rule and points rule).
We report heteroskedasticity-robust standard errors.
\starlanguage{}
\end{singlespace}
\end{footnotesize}
\end{table}

\begin{table}[h!] 
\begin{center}
\begin{minipage}{1 \linewidth}                          
\caption{Log-likelihood ratio regressions across game types ($\varepsilon = 0.1$)}  
\label{tab:game_level_eps_01}     
\centering                                                
\scriptsize
\renewcommand{\arraystretch}{1.2}
\begin{tabular}{@{\extracolsep{5pt}}lccccc} 
\\[-1.8ex]\hline 
\hline \\[-1.8ex] 
 & \multicolumn{5}{c}{Log-Likelihood Ratio (Strategic AI / \{...\})} \\ 
\cline{2-6} 
 & \{Baseline AI\} & \{HS Nash Eq.\} & \{Cognitive Hierarchy\} & \{Random Pure\} & \{Uniform\} \\ 
\\[-1.8ex] & (1) & (2) & (3) & (4) & (5)\\ 
\hline \\[-1.8ex] 
 Bonus Size & 0.006 & $-$0.015$^{**}$ & 0.009 & 0.013$^{*}$ & 0.009$^{**}$ \\ 
  & (0.004) & (0.006) & (0.005) & (0.005) & (0.003) \\ 
  & & & & & \\ 
 Lower Bound & 0.012$^{**}$ & $-$0.001 & $-$0.009 & $-$0.001 & $-$0.006 \\ 
  & (0.004) & (0.006) & (0.005) & (0.005) & (0.003) \\ 
  & & & & & \\ 
 Action Space Size & 0.027$^{***}$ & 0.016$^{*}$ & $-$0.001 & 0.046$^{***}$ & 0.015$^{***}$ \\ 
  & (0.005) & (0.007) & (0.006) & (0.007) & (0.004) \\ 
  & & & & & \\ 
 Gap & $-$0.016 & 0.048 & 0.022 & $-$0.021 & $-$0.013 \\ 
  & (0.021) & (0.029) & (0.027) & (0.027) & (0.017) \\ 
  & & & & & \\ 
 Constant & 0.078 & 0.461$^{***}$ & 0.310$^{*}$ & 1.371$^{***}$ & $-$0.030 \\ 
  & (0.098) & (0.134) & (0.125) & (0.137) & (0.081) \\ 
  & & & & & \\ 
\hline \\[-1.8ex] 
Observations & 1,477 & 1,477 & 1,477 & 1,477 & 1,477 \\ 
R$^{2}$ & 0.026 & 0.011 & 0.005 & 0.035 & 0.016 \\ 
\hline 
\hline \\[-1.8ex] 
\end{tabular}

\end{minipage}
\end{center}
\begin{footnotesize}
\begin{singlespace}
\vspace*{-0.15in}
\emph{Notes:} Each column reports regression estimates where the dependent variable is the log-likelihood ratio of the strategic \aisub{} to the other models.
The independent variables are the game features from Table~\ref{tab:game_parameters} (excluding the bonus rule and points rule).
We report heteroskedasticity-robust standard errors.
\starlanguage{}
\end{singlespace}
\end{footnotesize}
\end{table}

\begin{table}[h!] 
\begin{center}
\begin{minipage}{1 \linewidth}                          
\caption{Log-likelihood ratio regressions across game types ($\varepsilon = 0.2$)}  
\label{tab:game_level_eps_02}     
\centering                                            
\scriptsize
\renewcommand{\arraystretch}{1.2}
\begin{tabular}{@{\extracolsep{5pt}}lccccc} 
\\[-1.8ex]\hline 
\hline \\[-1.8ex] 
 & \multicolumn{5}{c}{Log-Likelihood Ratio (Strategic AI / \{...\})} \\ 
\cline{2-6} 
 & \{Baseline AI\} & \{HS Nash Eq.\} & \{Cognitive Hierarchy\} & \{Random Pure\} & \{Uniform\} \\ 
\\[-1.8ex] & (1) & (2) & (3) & (4) & (5)\\ 
\hline \\[-1.8ex] 
 Bonus Size & 0.005 & $-$0.010$^{*}$ & 0.008 & 0.012$^{**}$ & 0.008$^{**}$ \\ 
  & (0.003) & (0.005) & (0.005) & (0.004) & (0.003) \\ 
  & & & & & \\ 
 Lower Bound & 0.009$^{**}$ & $-$0.002 & $-$0.008 & $-$0.001 & $-$0.005 \\ 
  & (0.003) & (0.005) & (0.005) & (0.005) & (0.003) \\ 
  & & & & & \\ 
 Action Space Size & 0.024$^{***}$ & 0.012$^{*}$ & 0.002 & 0.041$^{***}$ & 0.017$^{***}$ \\ 
  & (0.004) & (0.006) & (0.006) & (0.006) & (0.003) \\ 
  & & & & & \\ 
 Gap & $-$0.013 & 0.036 & 0.026 & $-$0.015 & $-$0.008 \\ 
  & (0.017) & (0.023) & (0.026) & (0.023) & (0.014) \\ 
  & & & & & \\ 
 Constant & 0.024 & 0.207 & 0.317$^{**}$ & 0.867$^{***}$ & $-$0.023 \\ 
  & (0.078) & (0.109) & (0.122) & (0.114) & (0.069) \\ 
  & & & & & \\ 
\hline \\[-1.8ex] 
Observations & 1,477 & 1,477 & 1,477 & 1,477 & 1,477 \\ 
R$^{2}$ & 0.030 & 0.008 & 0.004 & 0.040 & 0.023 \\ 
\hline 
\hline \\[-1.8ex] 
\end{tabular}

\end{minipage}
\end{center}
\begin{footnotesize}
\begin{singlespace}
\vspace*{-0.15in}
\emph{Notes:} Each column reports regression estimates where the dependent variable is the log-likelihood ratio of the strategic \aisub{} to the other models.
The independent variables are the game features from Table~\ref{tab:game_parameters} (excluding the bonus rule and points rule).
We report heteroskedasticity-robust standard errors.
\starlanguage{}
\end{singlespace}
\end{footnotesize}
\end{table}

\begin{table}[h!] 
\begin{center}
\begin{minipage}{1 \linewidth}                          
\caption{Log-likelihood ratio regressions across game types ($\varepsilon = 0.3$)}  
\label{tab:game_level_eps_03}     
\centering                                           
\scriptsize
\renewcommand{\arraystretch}{1.2}
\begin{tabular}{@{\extracolsep{5pt}}lccccc} 
\\[-1.8ex]\hline 
\hline \\[-1.8ex] 
 & \multicolumn{5}{c}{Log-Likelihood Ratio (Strategic AI / \{...\})} \\ 
\cline{2-6} 
 & \{Baseline AI\} & \{HS Nash Eq.\} & \{Cognitive Hierarchy\} & \{Random Pure\} & \{Uniform\} \\ 
\\[-1.8ex] & (1) & (2) & (3) & (4) & (5)\\ 
\hline \\[-1.8ex] 
 Bonus Size & 0.004 & $-$0.007 & 0.007 & 0.011$^{**}$ & 0.007$^{**}$ \\ 
  & (0.003) & (0.004) & (0.005) & (0.004) & (0.002) \\ 
  & & & & & \\ 
 Lower Bound & 0.007$^{*}$ & $-$0.002 & $-$0.008 & $-$0.001 & $-$0.004 \\ 
  & (0.003) & (0.004) & (0.005) & (0.004) & (0.003) \\ 
  & & & & & \\ 
 Action Space Size & 0.021$^{***}$ & 0.008 & 0.002 & 0.037$^{***}$ & 0.017$^{***}$ \\ 
  & (0.003) & (0.005) & (0.006) & (0.005) & (0.003) \\ 
  & & & & & \\ 
 Gap & $-$0.011 & 0.029 & 0.029 & $-$0.012 & $-$0.006 \\ 
  & (0.014) & (0.020) & (0.026) & (0.020) & (0.013) \\ 
  & & & & & \\ 
 Constant & $-$0.005 & 0.081 & 0.327$^{**}$ & 0.588$^{***}$ & $-$0.013 \\ 
  & (0.065) & (0.094) & (0.121) & (0.098) & (0.060) \\ 
  & & & & & \\ 
\hline \\[-1.8ex] 
Observations & 1,477 & 1,477 & 1,477 & 1,477 & 1,477 \\ 
R$^{2}$ & 0.032 & 0.006 & 0.004 & 0.043 & 0.030 \\ 
\hline 
\hline \\[-1.8ex] 
\end{tabular}

\end{minipage}
\end{center}
\begin{footnotesize}
\begin{singlespace}
\vspace*{-0.15in}
\emph{Notes:} Each column reports regression estimates where the dependent variable is the log-likelihood ratio of the strategic \aisub{} to the other models.
The independent variables are the game features from Table~\ref{tab:game_parameters} (excluding the bonus rule and points rule).
We report heteroskedasticity-robust standard errors.
\starlanguage{}
\end{singlespace}
\end{footnotesize}
\end{table}

\end{document}